\documentclass[fleqn,usenatbib]{mnras}
\usepackage{newtxtext,newtxmath}
\usepackage{cancel}
\usepackage[T1]{fontenc}
\usepackage{ae,aecompl}

\usepackage{graphicx,epstopdf}
\DeclareGraphicsExtensions{.ps}
\graphicspath{ {./figures/} }
\usepackage{amsmath}	
\usepackage{mathtools}
\usepackage{longtable}
\usepackage{subfigure}
\usepackage{natbib}
\usepackage{adjustbox}
\usepackage{multirow}
\usepackage{tikz}
\usetikzlibrary{shapes.geometric, arrows}

\tikzstyle{startstop} = [rectangle, rounded corners, minimum width=3cm, minimum height=1cm,text centered, draw=black, fill=red!30]

\tikzstyle{io} = [trapezium, trapezium left angle=70, trapezium right angle=110, minimum width=3cm, minimum height=1cm, text centered, draw=black, fill=blue!30]

\tikzstyle{process} = [rectangle, minimum width=3cm, minimum height=1cm, text centered, draw=black, fill=orange!30]
\tikzstyle{decision} = [diamond, minimum width=3cm, minimum height=1cm, text centered, draw=black, fill=green!30]

\tikzstyle{arrow} = [thick,->,>=stealth]

\newcommand{\tsne}{\emph{t}-SNE}
\newcommand{\lps}{{\it LPlcv}}
\newcommand{\cleans}{{\it Cleanlcv}}
\newcommand{\ie}{\emph{i.e.}}
\newcommand{\eg}{\emph{e.g.}}

\title[Anomaly detection]{A method for finding anomalous astronomical light curves and their analogs} 

\author[Mart\'{\i}nez-Galarza et al.]
{J.~Rafael Mart\'{\i}nez-Galarza$^{1}$,
{Federica B. Bianco}$^{2,3,4,5}$
Dennis Crake$^{1,7,8}$, 
\newauthor
Kushal Tirumala$^{6}$,
Ashish~A.~Mahabal$^{6}$, 
Matthew J. Graham$^{6}$,
Daniel Giles$^{9}$\\
$^{1}$Center for Astrophysics | Harvard \& Smithsonian\\
$^{2}${University of Delaware
Department of Physics and Astronomy
217 Sharp Lab
Newark, DE 19716 USA}\\
$^{3}${University of Delaware
Joseph R. Biden, Jr. School of Public Policy and Administration, 
184 Academy St, Newark, DE 19716 USA}\\
$^4${University of Delaware
Data Science Institute}\\
$^5${Center for Urban Science and Progress, New York University, 
370 Jay St, Brooklyn, NY 11201, USA}\\
$^{6}$Division of Physics, Mathematics and Astronomy, California Institute of Technology, Pasadena, CA 91125, USA\\
$^{7}$School of Physics and Astronomy, University of Southampton, Hampshire, SO17 1BJ, UK\\
$^{8}$Institute for Astronomy, University of Edinburgh, Royal Observatory, Blackford Hill, Edinburgh, EH9 3HJ, UK\\
$^{9}$The SETI Institute. 189 Bernardo Ave, Suite 200, Mountain View, CA 94043\\
}

\date{Accepted XXX. Received YYY; in original form ZZZ}

\pubyear{2020}

\begin{document}
\label{firstpage}
\pagerange{\pageref{firstpage}--\pageref{lastpage}}
\maketitle

\begin{abstract}
Our understanding of the Universe has profited from deliberate, targeted studies of known phenomena, as well as from serendipitous, unexpected discoveries, such as the discovery of a complex variability pattern in the direction of KIC 8462852 (Boyajian's star). Upcoming surveys, such as the Vera C. Rubin Observatory Legacy Survey of Space and Time (LSST),  will explore the parameter space of astrophysical transients at all time scales, and offer the opportunity to discover even more extreme examples of unexpected phenomena. We investigate strategies to identify novel objects and to contextualize them within large time-series data sets in order to facilitate the discovery of new classes of objects, as well as the physical interpretation of their anomalous nature. We develop a method that combines tree-based and manifold-learning algorithms for anomaly detection in order to perform two tasks: 1) identify and rank anomalous objects in a time-domain dataset; and 2) group those anomalies according to their similarity in order to identify analogs. We achieve the latter by combining an anomaly score from a tree-based method with a dimensionality manifold-learning reduction strategy. Clustering in the reduced space allows for the successful identification of anomalies and analogs. We also assess the impact of pre-processing and feature engineering schemes and investigate the astrophysical nature of the objects that our models identify as anomalous by augmenting the Kepler data with Gaia color and luminosity information. We find that multiple models, used in combination, are a promising strategy to identify novel light curves and light curve families.
\end{abstract}

\begin{keywords}
methods: data analysis, methods: statistical, stars: flare, stars: peculiar (except chemically peculiar)
\end{keywords}


\newcommand{\red}[1]{{\color{red} #1}}

\section{Introduction}

Scientific discovery in astronomy can be thought of as happening via two complementary approaches. The \emph{question-driven} approach attempts to provide answers to questions that have already been conceived based on our present knowledge of existing theories, models, or observed phenomena. Cosmological supernovae (SNe type Ia) are a good example. They are relatively well-known objects to us in terms of their energetic output, redshift distribution, and spectral properties, and we design surveys with observational parameters fine-tuned to find them on the basis of those known properties, to improve our understanding of cosmology as well as stellar physics. The \emph{exploration-driven} approach, on the other hand, attempts to enable us with the capability to find objects that are unknown, unexpected, or extremely rare, by either expanding the space of observational parameters (for example by increasing the spatial resolution available to us with a larger telescope) or by employing novel ways to dissect the increasingly complex data sets that are becoming available to us \citep{Li21}. These unexpected discoveries often require additions or modifications to our theoretical apparatus, and, on occasion, force us to formulate new hypotheses.

This approach has led to serendipitous discoveries, which are prevalent in astronomy and which have produced significant breakthroughs. Recent examples include the discovery of peculiar light curves in \emph{Kepler} data, such as KIC~8462852, commonly know as Boyajian's star \citep{Boyajian16}, the discovery of the interstellar object 1I/'Oumuamua \citep{Meech17}, and the detection of quasi-periodic oscillations in the X-ray light curve of galaxy G 159 \citep{Miniutti19}, among many others. The question then becomes, how do we make serendipity ``systematic'' \citep{Giles19} in order to increase the chance of discovery in the era of large astronomical data sets? A related question that is at the core of this paper is whether it is possible to find analogs of a particular anomaly of interest.

Anomalies are traditionally identified as data points that lie beyond some threshold distance from the bulk of the population of other data points in some representative space \citep[see, for example the comparative study in][]{goldstein16}. This distance is usually determined in terms of the population scatter, often measured by the standard deviation. For example, assuming a Gaussian distribution, one can flag one in 370 objects $(3 \sigma)$ or one in 1.7 million objects ($5 \sigma)$ as ``outliers''. However, there is no reason to think that an arbitrary data set in an arbitrary data space should be described by a Gaussian distribution. Furthermore, \cite{aggarwal2001outlier} have shown that as the number of dimensions increases, the proximity-related measures of similarity between objects become less meaningful, since in a high-dimensional space objects are more sparse, and more likely to show up as proximity-based outliers. In addition, anomaly detectors can also be affected by false alarms, either false positives or false negatives.

Time domain-based studies of astrophysical phenomena date back centuries, but in the context of modern astrophysical surveys an example can be found in \citep{Eyer19} where the authors also emphasize the pivotal role of time-domain surveys in the future of astrophysics. Reviews of anomaly detection methods in time-series data can be found in \citet{goldstein16} and \citet{blazquez20}.  Several unsupervised and semi-supervised algorithms for astrophysical anomaly detection have been used, including approaches that use Euclidean proximity/clustering information to isolate anomalies \citep{Giles19, Dutta07, Henrion13}, and approaches that use more complex representations of the data that do not involve their projection into an Euclidean space, such as neural networks, ensemble methods, and active learning \citep{Baron17,  Druetto19, Skoda20, Margalef20}, as well as Gaussian processes \citep{Chen18}. Significant effort has been put into the problem of anomaly detection in supernova surveys, in particular by the Supernova Anomaly Detection (SNAD) group, which have used Isolation Forest and active learning to boost the discovery of unusual objects \citep{Pruzhinskaya19, Ishida19, Aleo20}. There are also significant differences in the feature engineering aspects of those algorithms. While some of them require a feature extraction step that produces a set of synthetic features to represent the data, others work directly on the data points, either light curves, spectra, or images. While each of these methods can perform well for specific data sets, one could ask whether different methods applied to the same data set would find the same anomalies.

One potential caveat of anomaly detection algorithms based on a similarity score calculated using a certain set of features, is that the selected features might not fully characterize the anomalous nature of the objects. That is, the objects can be anomalous in a certain space, but not in all possible representations. This makes it hard to efficiently identify objects that are ``like'' a particular object of interest when dealing with a large dataset, because the anomaly score based on those features is usually a one-dimensional quantity. Our hypothesis in the present work is that by combining a manifold-learning proximity method with an independently derived anomaly score from a tree-based method, one can effectively break the degeneracies in the one-dimensional space of the anomaly score, and make useful inferences regarding which anomalies are similar to each other, simplifying the discovery of analogs and new classes. This approach also offers the advantage that a different set of features can be used for the anomaly scoring and for the clustering, offering additional flexibility.

Previous work has made significant contributions towards the goal of finding analogs to Boyajian's star and other anomalies. For example \citet{Giles19} and \citet{Giles20} apply a clustering method to a set of synthetic features derived for Kepler light curves and demonstrate that their method is capable of identifying anomalies such as Boyajian's star, as well as cataclysmic variables. \cite{Schmidt19} use a photometric selection method to isolate analogs of KIC~8462852 by looking for light curve dips in All Sky Automated Survey for Supernovae (ASAS-SN, \citealt{asassn}) data, and they find about 20 similar objects that deserve follow-up studies (see also \citealt{2021A&C....3600481L}). Yet, to the best of our knowledge, no reproducible methods have been proposed with the specific goal of finding analogs to a light curve of interest, be it Boyajian's star or other type of anomalous light curve. In this paper we aim to provide a recipe not only for finding the most compelling objects in a time domain survey, but also for finding any analogs (\ie, objects with similar light curves) of those objects. The method combines:

\begin{itemize}
    \item A tree-based anomaly detection algorithm, the Unsupervised Random Forest \citep{Shi06} that operates on the joint space of light curve points and power spectrum.
    \item Two manifold-learning algorithms: \tsne\  \citep{Maaten08} and UMAP \citep{McInnes18}, that operate on an image representation of the light curves and finds low-dimensional embedded representations of these images. 
\end{itemize} 

We apply this combined method to the full set of Quarter 16 light curves from the \emph{Kepler} Space Observatory, and test it against a set of previously identified anomalies in the \emph{Kepler} dataset \citep{Giles20}. We rank the anomalies according to their URF-based similarity score and compare them with the scores for the general population in order to assess the ability of the method to identify \emph{bona-fide} anomalous objects. In order to test our ability to find analogs, we investigate the location and clustering properties of these anomalies in the space of embedded features derived from the manifold methods, and whether other similar and previously unknown anomalies are identified. By using \emph{Gaia} observations of the sources to construct their Hertzprung-Russell diagram, we find that the method is able to identify anomalies that share astrophysical properties, either intrinsic or extrinsic.

The Kepler telescope generated evenly sampled time series. This is possible with a space mission, but impossible with ground-based surveys, due to weather, moon phase, visibility, and, generally, optimization of survey strategies \citep{Bianco21}. Surveys like the Catalina Realtime-Transient Survey \citep{catalina} and the Zwicky Transient Facility \citep{ztf} are an invaluable reservoir of transients, all measured with unevenly sampled light curves. Vera C. Rubin Observatory's Legacy Survey of Space and Time (LSST, \citealt{Ivezic19}) will generate an unprecedentedly large data set including tens of billions of stars measured with over $\sim800$ unevenly sampled data points. To test if our results generalize to unevenly sampled data sets, such as the LSST data, we generate a second data set from the Kepler data by sub-sampling the original time series, and repeat our analysis on these uneven, sparse light curves.

This paper is organized as follows. In \autoref{sec:data} we describe the data set used to test the performance of our method, and discuss the impact of different pre-processing in the process of finding anomalies. In \autoref{sec:methods} we describe the methods used in the paper for anomaly detection, dimensionality reduction, and clustering. In \autoref{sec:results} we present our set of \emph{bona-fide} anomalies, show that they can be identified with our method, and identify additional anomalies in the dataset. We also show how the manifold methods can be used to group similar anomalies together, and how those similarities relate to shared astrophysical properties. Finally, in \autoref{sec:discussion}, we discuss and summarize our findings. The code used in this work is available at \url{https://github.com/kushaltirumala/WaldoInSky}

\section{Data set and feature extraction}\label{sec:data}

In this section, we describe the data set used in the subsequent analysis, as well as the feature extraction approaches that we employed to create the input to the anomaly detection and dimensionality reduction methods.

\subsection{Kepler light curves}
\begin{figure}
\includegraphics[scale=0.415,angle=0,trim=0.3cm 0cm 0cm 0cm,clip=true]{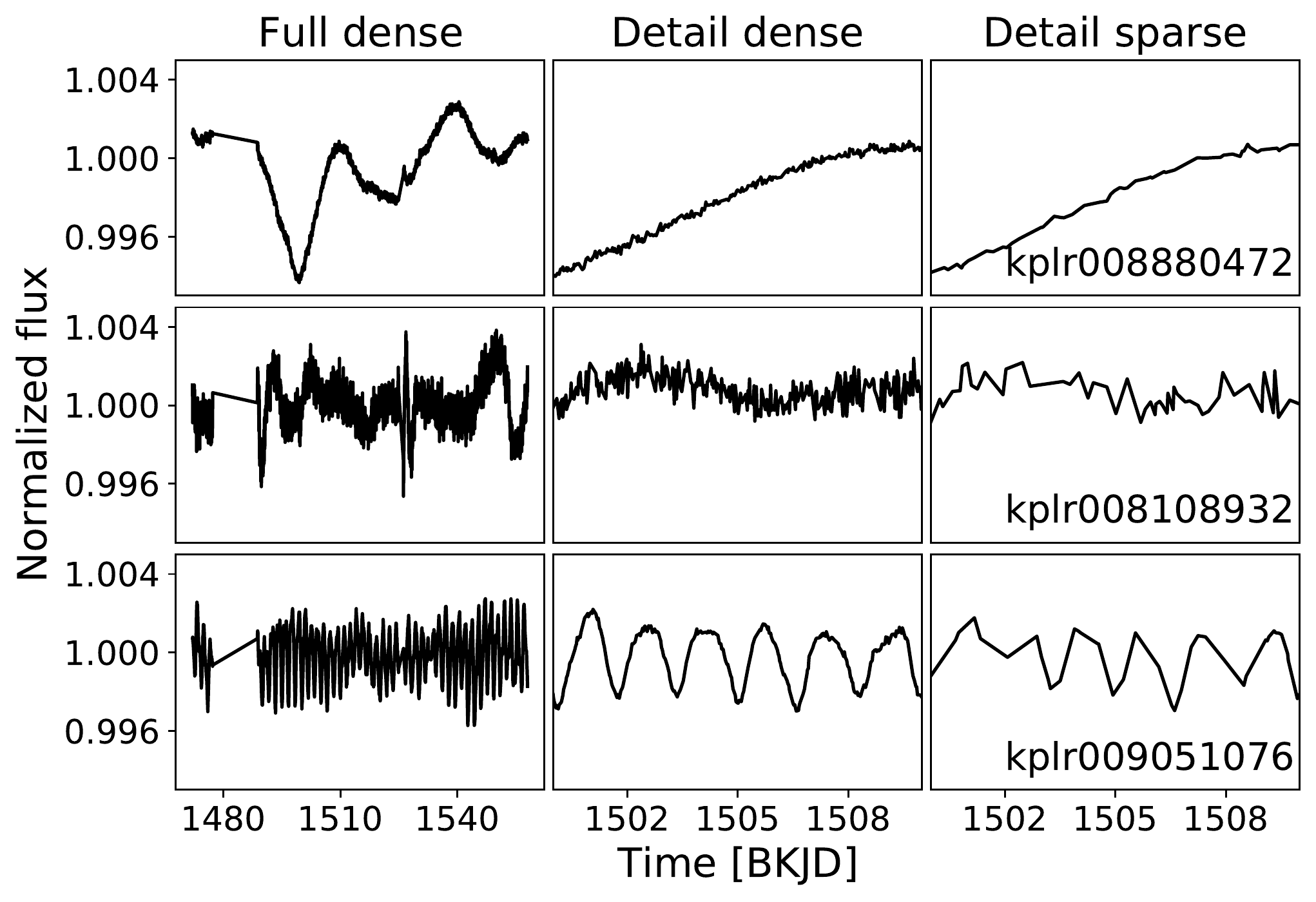}
\caption{Examples of normalized Kepler light curves used in this paper. The left panels show the full 85 day long Q16 dense light curves for three different Kepler targets. The middle panels show a detail of the same light curves plotted over a period of 10 Julian days. The right panels show the same detail but this time for the corresponding sparse version of the light curves. The Kepler source IDs are shown. Note the data gap early on in each light curve due to missing observations.}
\label{fig:LC_examples}
\end{figure}

The data set considered here is comprised of 147,036 light curves from the \emph{Kepler} telescope archive. These light curves were obtained by searching the Mikulski Archive for Space Telescopes (MAST) for all light curves in Quarter 16. From the total of over 160,000 light curves, we then excluded those that we were not be able to pre-process with the methods of \autoref{sec:pre-processing}. The downloaded light curves are de-trended from spacecraft effects and have a uniform cadence, with one photometry point obtained every 30 minutes. Our pre-processed version on the light curves (see \autoref{sec:pre-processing}) covers a total span of approximately 85 days, starting at Barycentric Kepler Julian Date (BKJD) 1472 and ending at BKJD 1558, for a total of 3520 measurements per light curve. We used the most up-to-date pipeline processed data (PDCsap flux) according to the Kepler Data Release Notes and the Kepler Data Processing Handobook \citep{Jenkins17}. Fluxes are given in relative flux units as provided in the standard Kepler de-trended light curves delivered by the MAST archive. We normalized the fluxes to have a mean value of 1, \ie, we divided them by their mean. These normalized fluxes are not formally standardized, \ie, their standard deviations are those resulting from this normalization process, without any further adjustments. We chose this approach because we do not want the anomaly detection being dominated by the noise level. In what follows, however, we use standardized and normalized indistinctly. In \autoref{fig:LC_examples} we show a small sample of the light curves studied here, in order to illustrate the wide range of variability properties in the Kepler targets.

In order to assess the effect that uneven, sparse sampling has on our ability to find anomalies, we have also produced a separate data set by uniform random sub-sampling of the full light curves, selecting only $\sim$10\% of the original data points time-stamps (the same time stamps for all light curves). We refer to this second set of light curves as the \emph{sparse} set, as opposed to the \emph{dense} set of the original light curves. Three sparse light curves are shown in \autoref{fig:LC_examples}, right panel.



\subsection{Feature engineering}
\label{sec:features}
Selecting and engineering features from astronomical light curves (and more generally from time series and other one-dimensional data) is a crucial step in setting up a successful anomaly detection algorithm, and, in general, any machine learning model, either supervised or unsupervised \citep{Fulcher17}. The decision regarding which features to use should be based upon the predictive power of these features and on how efficiently they can be used to split the data set into multiple classes, or as is the case here, on how well this set of features represent the different types of variability of the original data set.  

Feature selection is crucial not only to maximize the efficiency of our methods, but especially if we intend to use the features to physically characterize the objects.
In the case of supervised tree methods, on which some of the models we use are based, since at each step the model deals with only a single feature and features are never combined in a mathematical sense, the predictions are generally robust against covariance \citep{Breiman01, Biau12}. Consider, however, the case of two features that are completely co-linear: the trees in the forest will use one or the other based on an initial random selection. When encountering the second feature the trees will disregard it since the information carried by the feature has already been used in splits based on the other, co-linear feature. However, this affects the feature importance evaluation. Trees assess the importance of features based on how much each feature contributes to splitting. In the covariant example, the second co-linear feature does not ultimately contribute and therefore its covariance with other features has suppressed its importance. 

Feature engineering is particularly challenging when the sampling of a time series is sparse and uneven, specially if the sampled epochs are different for different light curves. In that case, phase information may be lost, and the individual photometric measurements are rendered much less useful for the analysis \citep{Che18}. Yet, due to technical and astronomical constraints, this is necessarily the case for ground-based astronomical time domain surveys (\eg, SDSS, \citealt{SDSS}, ASASSN, \citealt{asassn}, etc.), and will be the case of upcoming surveys such as the LSST \citep{Ivezic19}. Additionally, depending on the number of light curves to analyze and the specific features to be extracted, the process of feature engineering can be computationally expensive. It is therefore desirable to use feature extraction methods that do not require heavy processing.

It is common to use statistical parameters of the time series as the set of features \citep{Dubath11, Nun16, Richards11, Johnston17}. A number of feature extraction packages are available for time series. One such codes is the Feature Analysis for Time Series (FATS) code \citep{Nun15}, that is able to evaluate over 40 features in time series with the epochs, magnitudes, magnitude errors, and filters as an input, and extracts statistical features such as means, standard deviations, linear trends, variability index, skewness, kurtosis, etc (see also \citealt{2021MNRAS.502.5147M}). One issue with this approach of feature extraction is that not all of the features included are properly defined for all light curves, since there might be missing data, different number of elements, etc. As a result, these features may be undefined or unreliable for possibly a significant fraction of the light curves.  

Rather than relying on second-order features, it seems appropriate to remain as close as possible to the data themselves, therefore avoiding biases introduced by the extraction process. In this paper, we have chosen to use a combination of the light curve points themselves and their power spectrum as the input features for the anomaly detection, and an image representation of the light curves as the input for the dimensionality reduction algorithms. The latter method falls within the representation learning approach \citep{Bengio14}, in which a different representation of the data (as opposed to engineered features) is used in order to extract useful information for classification and/or anomaly detection tasks. Representation learning has been used for astronomical analysis, for example in \citep{jamal20, szklenar20} for classification purposes, but not as extensively for anomaly detection \cite[but see][for the use of generative models for anomaly detection]{storey21}.

Using the power spectrum (or rather the periodogram since the data are unevenly sampled due to the presence of a gap in the data) is desirable especially for light curves that do not share a common phase reference. Periodogram analysis can also be used to find the best period for periodic variables. However, period-finding itself is a significant challenge \citep{Graham13, 2018ApJS..236...16V}, and therefore we do not use estimated periods, but the periodogram values themselves, as the features.

\subsubsection{Periodograms}

The periodogram of a light curve is based on its Fourier transform and it measures the signal's power as a function of angular frequency. \citet{Lomb76} and \citet{Scargle82} generalized the concept of periodogram for the case of unevenly sampled data. Their normalized periodogram can be written as:

\begin{equation}
\label{eq:LS_periodogram}
\begin{multlined}
P_N(\omega) = \frac{1}{2\sigma_y}\left[\frac{\left[\sum_k(y_k-\hat{y})\cos\omega(t_k-\tau)\right]^2}{\sum_k\cos^2\omega(t_k-\tau)}\right. \\ +\left.\frac{\left[\sum_k(y_k-\hat{y})\sin\omega(t_k-\tau)\right]^2}{\sum_k\sin^2\omega(t_k-\tau)}\right],
\end{multlined}
\end{equation}
where $\sigma_y$ is the variance of the photometry ${y_k}$ and $\tau$ is a time offset that orthogonalizes the model and makes the expression independent on a time translation. As demonstrated in \citet{Lomb76}, this expression is fundamentally equivalent to estimating the harmonic content given a least-squares fit to a sinusoidal model consisting of a single component. It can therefore easily be computed using a fast Fourier transform.

\begin{figure*}
\includegraphics[scale=0.47,angle=0,trim=0cm 0.6cm 0cm 0.1cm,clip=true]{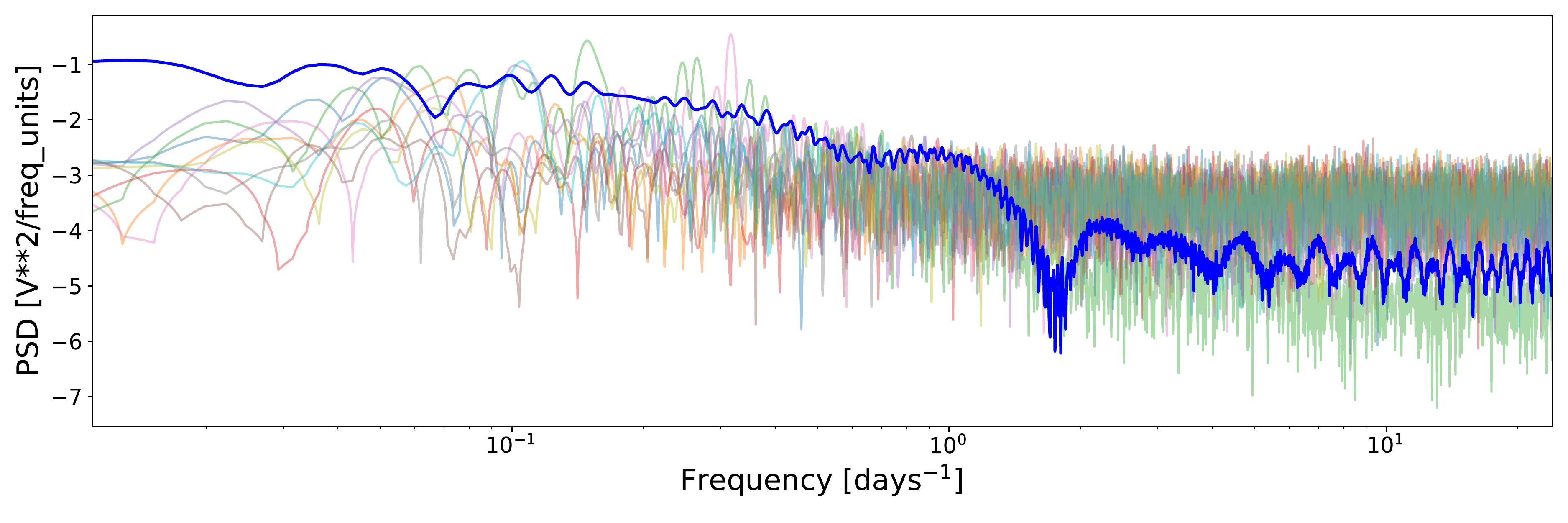}
\caption{Example periodograms for ten Kepler Q16 light curves, with Boyajian's star indicated in thick blue. Note the particular structure of Boyajian's periodogram, indicating the pronounced dips in the light curve. The logarithmic frequency range covers periods from 1 hour to about 90 days. All periodograms are obtained over the same array of frequencies.}
\label{fig:periodograms}
\end{figure*}

Given a light curve, we can evaluate the periodogram in \autoref{eq:LS_periodogram} for a discrete number of frequencies, and use the values of the periodogram evaluated at each of these frequencies as features for our algorithm. This is conceptually similar to using the individual pixels in a spectrum as the features, except that using the same array of frequencies for all light curves, we can compare them using the same absolute reference. This approach can also be used for multi-band, non simultaneous light curves, since a different periodogram can be calculated for each filter, and a final array of features can be obtained by concatenating the single-band periodograms. 

For the dense light curves, we have extracted periodograms with 3000 single points covering a logarithmic range of frequencies corresponding to periods between one hour (twice the Kepler cadence) and 90 days (the approximate duration of the observations). This range does not only cover the periods sampled by the observations, but also the typical timescales of different stellar variability phenomena. For example, an HST survey of the variability properties of luminous ($M_I<-5$) stars in M51 \citep{Conroy18} finds that the variability fraction for these is $\sim 50\%$, with many stars showing typical timescales between 1 and 100 days.  More in general, the most common pulsating variable stars have characteristic timescales that range from a few minutes to a couple of years \citep{Eyer08}. For the sparse light curves we construct periodograms of 300 points, covering a range of frequencies between 4 hours and 90 days. In \autoref{fig:periodograms} we show examples of the dense light curve periodograms computed for a sub-sample of relatively normal light curves in the Kepler data set, together with the periodogram of Boyajian's star, one of our bona-fide anomalies. Note that, with respect to the periodogram of Boyajian's star, ``normal'' light curves have a more evenly distributed spectral power as a function of frequency.

In terms of transferability, we note that the approach adopted here requires the generation of the power spectrum prior to the anomaly detection analysis. The properties of this power spectrum will be affected by the cadence and total number of points in the light curve, which implies that, in principle, it would not be possible to transfer a model trained on \emph{Kepler} data and apply it to a different survey, such as LSST. However, as long as all the cadence has been the same for all the objects belonging to a survey, the Lomb-Scargle algorithm provides a method to obtain a periodogram that is consistent for all objects in that survey. Our investigations on simualted LSST-like datasets, such as the one resulting from the PLAsTiCC data challenge \citep{Kessler19} have shown that a similar approach as the one adopted here can be used to study the much more sparse and irregular time series of ground-based surveys.

For anomaly detection with the tree-based method, we have constructed the feature vector for each light curve as the concatenation of the light curve points and the periodogram points. This yields a total of 6520 features for the dense light curves, and 632 features for the sparse light curves.

\subsubsection{DMDT maps}
\label{sec:dmdtmaps}

One additional feature extraction method, the \emph{DMDT} approach proposed in \citealt{Mahabal17}, is adopted to obtain the input features for the manifold learning-based dimensionality reduction, and is described below.

\begin{figure*}
\includegraphics[scale=0.445,angle=0,trim={10mm 10mm 10mm 10mm},clip]{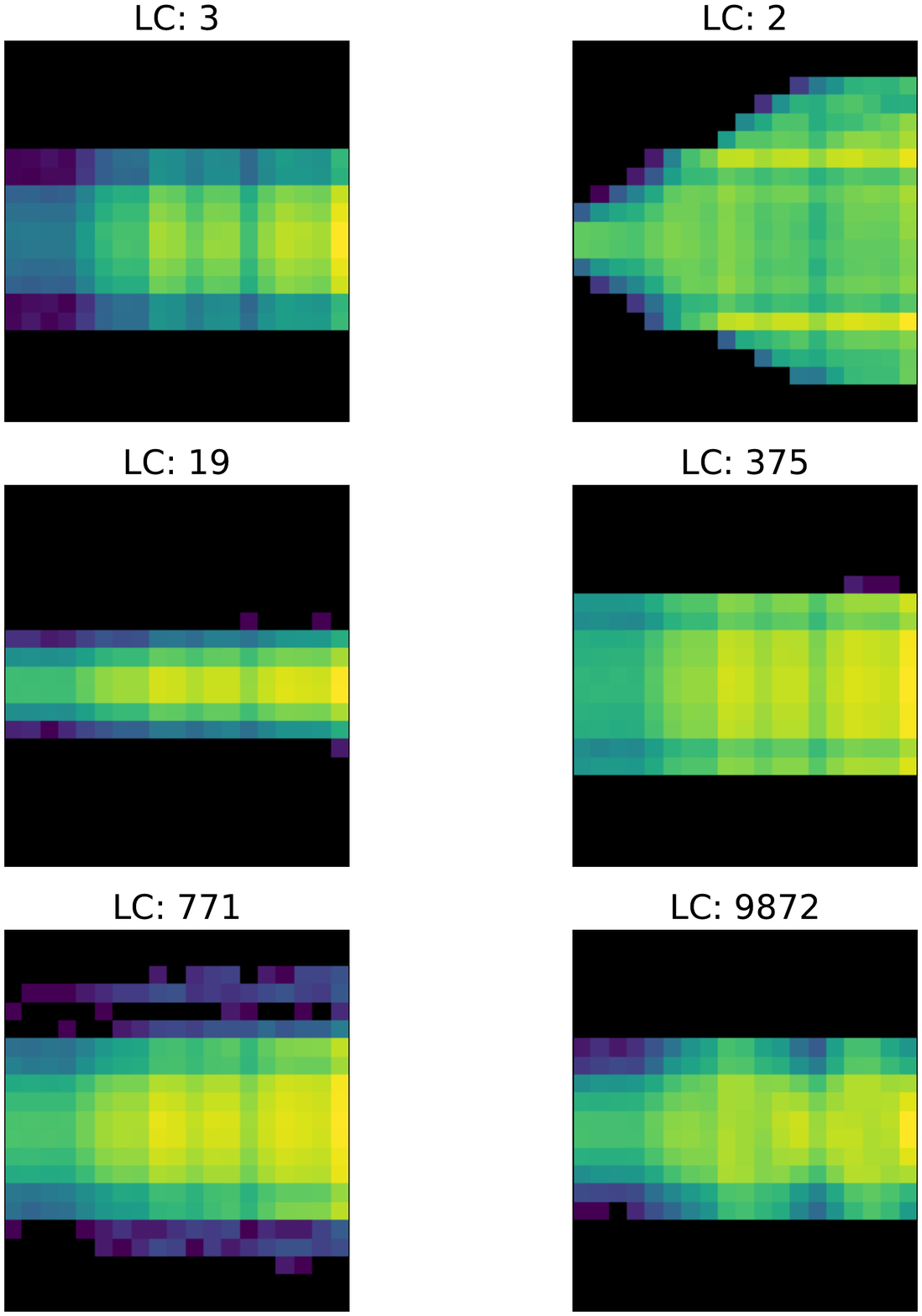}
\includegraphics[scale=0.445
,angle=0,trim={10mm 10mm 10mm 10mm},clip]{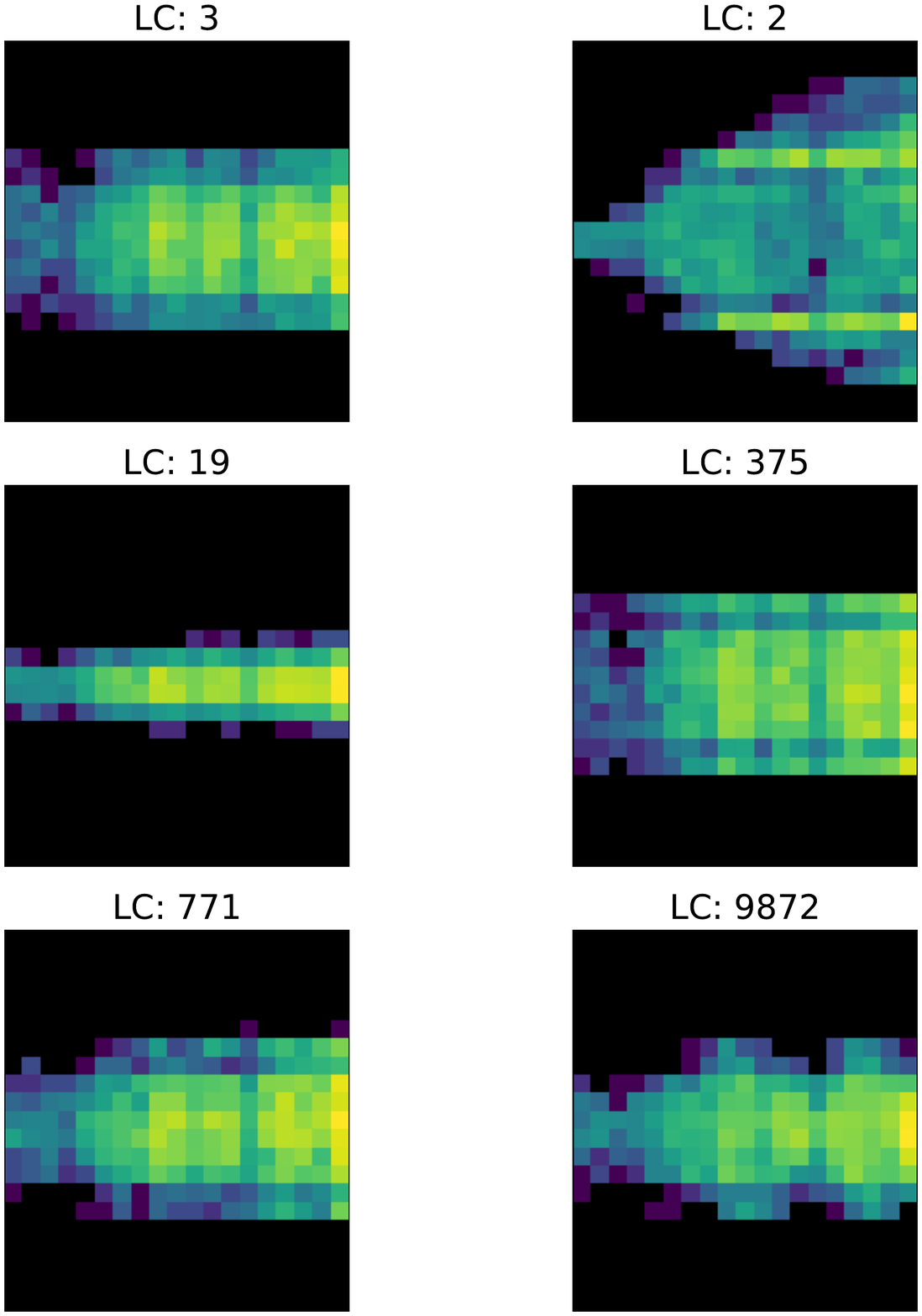}
\caption{A plot of raw \emph{DMDT} representations of dense (six left panels) and sparse (right six panels) light curves, for a random selection of 6 objects from our dataset.}
\label{fig:tsne_dense_dmdt_images}
\end{figure*}

Light curves are sequences of stellar brightness as a function of time, and the \emph{DMDT} map is a light curve image representation that simply considers differences in magnitude --- \emph{DM} --- against differences in time --- \emph{DT}. The values of magnitude differences and time differences are then binned, resulting in a 2D image representation of the light curve \citep[we refer the reader to][for more information on the specifics on the \emph{DMDT} representation of light curves]{Mahabal17}. The output feature space is a  $\mathbb{R}^{21 \times 19}$ matrix (its dimension relates to the desired dimensions of the \emph{DMDT} mapping) representing the pixel matrix of the \emph{DMDT} images. We show examples of \emph{DMDT} representations for dense and sparse light curves  in \autoref{fig:tsne_dense_dmdt_images}.

After producing the \emph{DMDT} map for each light curve, we then flatten the matrices into $399$-long arrays, each array representing a light curve. We consider this our starting high dimensional space for the dimensionality reduction algorithms. While keeping the 2D information of the matrix would be ideal to reduce the amount of information lost, to the best of our knowledge there are currently no codes available that are robust, tested and scalable that can handle distances between 2D datasets. One possible approach is proposed in \citet{Johnston19}, where the authors represent light curves of variable stars as matrices of features, and then define a metric that defines distances directly between matrices. The authors acknowledge that the method requires some improvements between it can be fully exploited. We have therefore opted for flattening the matrices, as is common practice for the treatment of 2D images. Distances can be directly estimated in the multi-dimensional space of the vector elements.

\subsection{The influence of pre-processing choices}
\label{sec:pre-processing}
Anomaly detection methodologies identify sources whose properties stand out with respect to 
all other objects in a data set. These sources might represent examples of unknown types of objects, in which case we may require the formulation of new physical hypotheses to explain their properties, or instances of rare evolutionary stages of known types. In both cases, they represent an expansion of our discovery space. But anomaly detection can also reveal instrumental or processing artifacts. 

In the specific case of light curves, such artifacts can include spurious trends in the light curve baselines or bad pixels values that result in artificial spikes or dips in the time series. Therefore, in order to characterize the effect of data artifacts in our analysis, we explore how different pre-processing approaches affect our ability to find anomalies.

We consider the results of applying an anomaly detection algorithm to our set of \emph{normalized} light curves both before and after further pre-processing. For demonstration purposes, we use here the Isolation Forest (IF) method, which will be explained in detail in the next section (\autoref{sec:methods}). For now, it is sufficient to say that the IF generates a score such that the lower the score value, the more unlike-the-others the object is, and, formally, having decided what percentage of objects are expected to be anomalous through the hyperparmeter \texttt{contamination}, anomalies are associated with negative scores. We note that the IF leads to the identification of similar anomalies as the URF methods (\autoref{sec:URF}) which we extensively use in our analysis.  By comparing the anomaly scores of the pre-processed light curves with those of the original light curves, we expect to learn to what extent pre-processing (or the lack of pre-processing) influences our ability to detect anomalies.

\begin{figure*}
\includegraphics[width=\textwidth,angle=0,trim={5.5cm 2cm 4.5cm 22cm},clip]{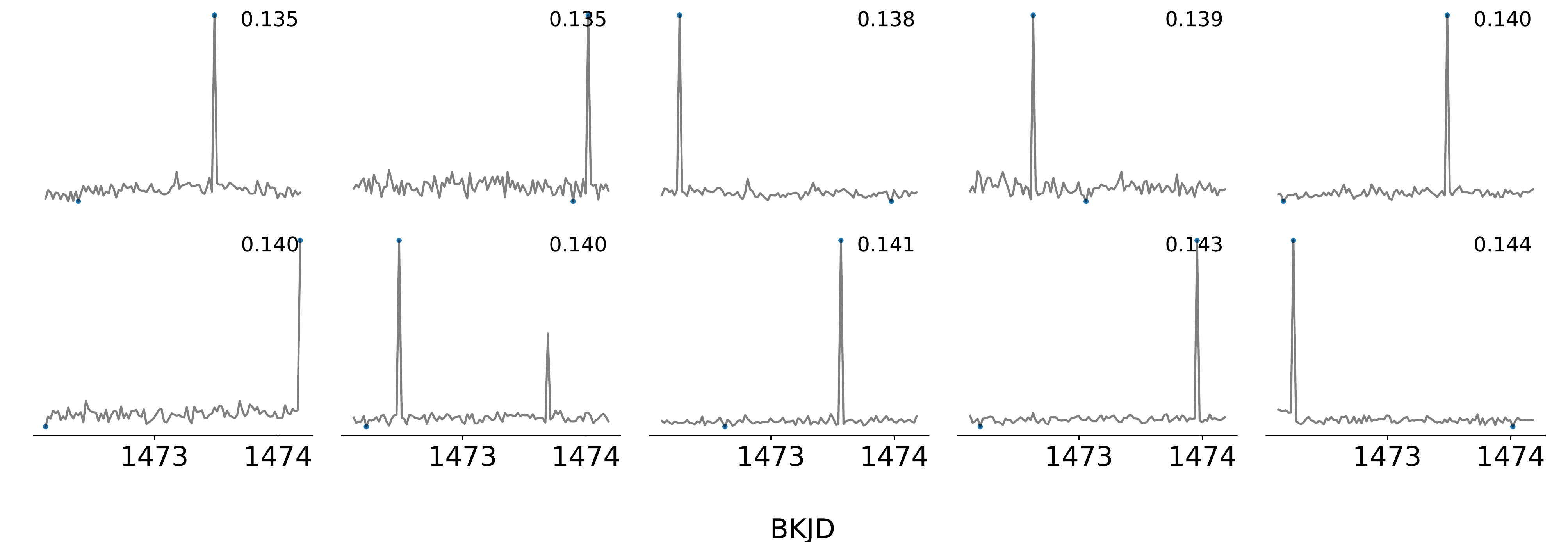}
\caption{Normalized, dense, partial (5-days) Kepler Q16 light curves with corresponding normality scores derived from the Isolation Forest method for 10 light curves with low anomaly score. The anomaly score is dominated by single-point events which may be due to artifacts or extremely high energy and short duration events, such as flares, which are rare, and therefore arguably should be contributing to the anomaly estimate. The brightest and dimmest point in each light curve are indicated by a blue point.}
\label{fig:IFweirdos}
\end{figure*}

In \autoref{fig:IFweirdos} we show the 10 least anomalous light curves, according to a similarity score derived from the IF method. The fact that prominent spikes consisting of a single bright data point are present in all of them deserves some attention. These spikes can be due to events unrelated to the source (\eg, cosmic rays), or to instrumental artifacts, such as hot pixels. They could also be very high energy astrophysical events, such as star flares, but we note that these spikes are unlikely to be high energy flares simply due to how commonly they appear in our data compared to flare statistics  \citep[see for example][]{Paudel19}. 
These large features dominate the information-content in the normalized light curves, and therefore the similarity score of objects, potentially hiding less prominent, anomalous features. 
Here we are interested in finding unique, anomalous light curves, so we cannot let either data artifacts or flares 
to dominate the anomaly score.  
when the methods are applied to the normalized light curves. 

We investigate the sensitivity of the anomaly score to pre-processing by removing these spikes using two different approaches and then comparing the anomaly scores obtained for the normalized light curves, and both sets of pre-processed light curves. The first spike-removal approach involves low-pass filtering: we take the light curve's rolling average over a window of 10 points, henceforth suppressing any high-frequency variations. We call these the low-pass filtered light curves set \lps 's. The second approach involves directly removing the outlying datapoints by replacing every point outside of a 3$\sigma$ deviation range from the light curve mean with the corresponding value of a \lps 's generated with a window of size 15 data points (7.5 hours). We refer to this last set of pre-processed time series as \cleans 's.

\begin{figure*}
\centering
\includegraphics[width=1.0\textwidth]{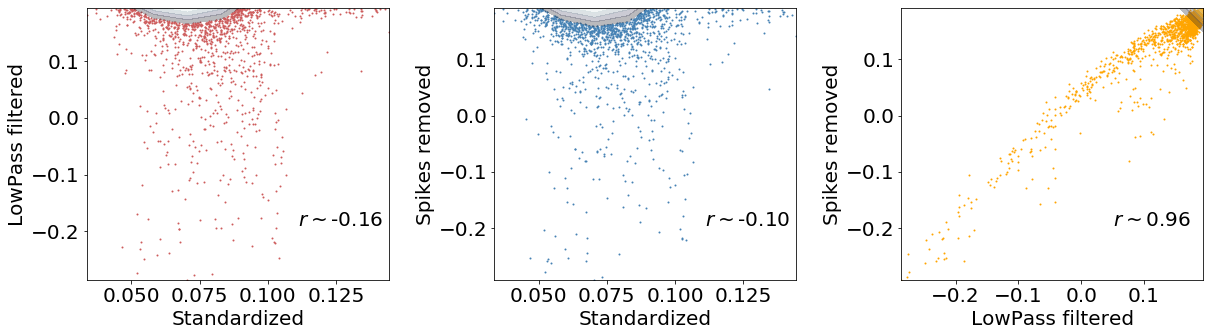}
  \caption{Comparison of anomaly scores produced by the Isolation Forest method for light curves pre-processed differently: standardized, low pass filters, and cleaned by replacing 3$\sigma$ outliers with a local mean value. The correlation in the scores generated for standardized lightcurves and lightcurves where sharp features are removed is low and inverse (Pearson's $r$ is reported in each panel) regardless of whether sharp features are low pass filtered or cleaned by replacing the values with a local mean. Conversely, the specific choice of pre-processing (by low-pass filtering or by replacement) has little influence on the anomaly score. (Regions of high point density are visualized as contours, instead of scatter points)}
 \label{fig:ITscorecorrelation}
\end{figure*}

The anomaly scores of the normalized light curves do not strongly correlate with the scores for either of the two pre-processing methods (left and center panels,  \autoref{fig:ITscorecorrelation}). The Pearson's $r$ correlation coefficients between each of the two sets of pre-processed light curves and the unprocessed version are $r\sim$-0.10 and $r\sim$-0.16: there is a weak \emph{inverse} proportionality, weak but statistically significant at a 3$\sigma$ level, for both \cleans~ and \lps\ (\emph{p}-value $< 0.01$). On the other hand, the two pre-processed light curve sets result in very similar anomaly scores (right panel  \autoref{fig:ITscorecorrelation}). In this case, the Pearson's $r$ value is 0.96. Although the IF reported ``anomaly'' threshold, where the anomaly score becomes negative, is somewhat arbitrary 
it is worth noting that all scores are positives for the unprocessed light curve set, while the pre-processed light curves find anomalies, consistently with the implicit request set up in the choice of parameters (\verb+contamination=0.1+).

Altogether, this demonstrates that the single spikes dominate the model's decision when assigning an anomaly score, and may hide important anomalous behavior of lesser flux amplitude by biasing the scores. Truly anomalous light curves (or at least not those with the most obvious data artifacts) are only found by these methods when such artifacts have been effectively removed and 
such features should be removed to reveal more subtle anomalies.

\begin{figure*}
\includegraphics[width=1.0\textwidth]{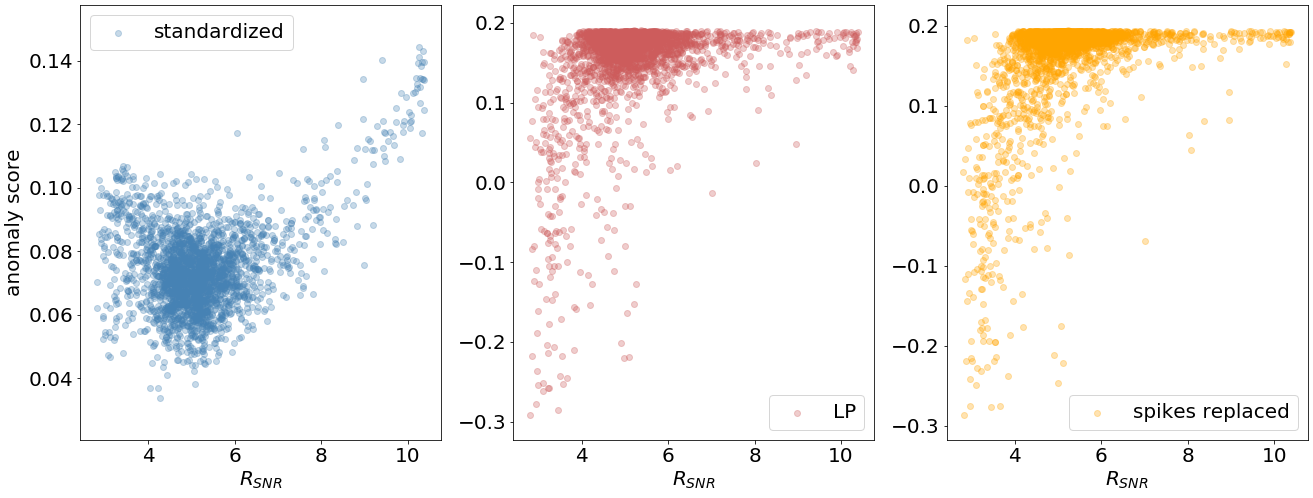}
\caption{The SNR of the light curves, measured as $R_{SNR} = \frac{max(\vec{t}) - min(\vec{t})}{\sigma_t}$, plotted against the anomaly score measured by the Isolation Forest model for standardized light curves (left), light curves that have been low-pass filtered (\lps, center), and light curves were spikes are replaced by the value of the local mean (\cleans, right). The dependency of the score on the $R_{SNR}$ is weak for the standardized light curve, and significant for the low-pass filtered and cleaned-by-replacement versions.}
\label{fig:LCSNR}
\end{figure*}

We further investigate the scores in order to understand why the presence of spikes is associated with a low anomaly score in the original light curve set. We look at the correlation of the anomaly scores for each pre-processing scheme with three simple light curve statistics: standard deviation (${\sigma_t}$) of the original light curve, standard deviation of the low pass filter version $\sigma_{tLP}$, and normalized flux range over the standard deviation $R_{SNR} = \frac{max(\vec{t}) - min(\vec{t})}{\sigma_t}$. This last statistics is significantly impacted by the presence of spikes. We find a strong correlation between the score obtained on the normalized time series and $R_{SNR} $ (Pearsons $r\sim0.5$, \emph{p}-value <0.001), see \autoref{fig:LCSNR}. Meanwhile, individually, neither the data range nor the standard deviation of the original time series is a good predictor of the normality score ($r \sim 10^{-3}$ and 0.04 respectively). Conversely, we find a strong anti-correlation of the score of the pre-processed time series, both \lps~  and \cleans, with the standard deviation of the original light curve, and of the pre-processed versions (all \emph{p}-values < 0.001), as well as with the $R_{SNR}$ metric; but for the latter, the distribution is far more evenly distributed, still with a significant lack of inliers at high $R_{SNR}$.

In the presence of light curves with significant spikes, therefore, less anomalous objects are largely dominated by these artifacts, as they are not easily isolated in the multi-dimensional space of all light curve points. Once we remove them, however, the pattern of the least anomalous objects changes to contain those light curves with smaller flux variances and a white noise spectrum. More severe perturbations of the light curves are then associated with anomalies. Pre-processing of light curves in order to remove non-astrophysical artifacts is, therefore, a necessary step in anomaly detection.

\begin{figure}
\includegraphics[scale=0.275,angle=0,trim=0cm 0cm 0cm 0.0cm,clip=true]{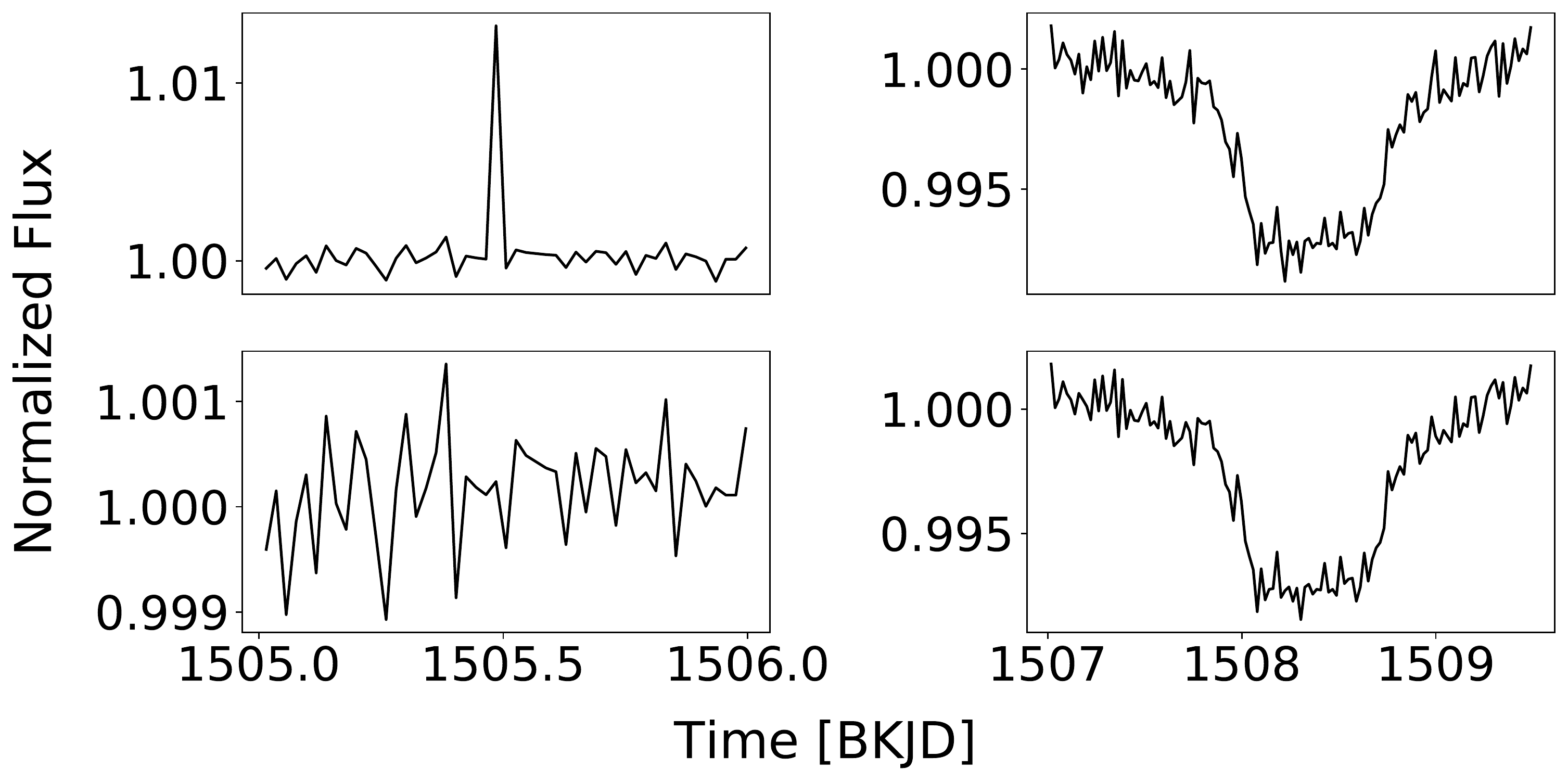}
\caption{Two light curve segments showing the effect of the \cleans\ pre-processing approach. The top panels show the original light curve points, whereas the bottom panels show the processed light curves. We note that the spike, potentially an artifact, seen in the top left panel is removed by our method, while the light curve remains unaffected for real astrophysical features, such as the transit-like dip seen in the right panels.}
\label{fig:spike_removal}
\end{figure}

For the remainder of this paper, we use light curves pre-processed according to the \cleans~ method. As illustrated in \autoref{fig:spike_removal}, this method removes the undesired spike artifacts without significantly affecting the frequency structure of the original data in other segments of the light curve. Only a very small fraction of true (non-spike) light curve points are affected by the $3\sigma$ threshold that we have imposed, and when they are affected, the effect is not dramatic, because most of the astrophysical dips and flares last for a few hours at least. The main effect of this pre-processing approach is that dips that deviate significantly from the mean value of the light curve are slightly less pronounced compared to the original light curve, as a result of the smoothing. We note that a similar normalization and spike-removal approach has been used in previous work as a pre-processing step for anomaly detection \citep{Rebbapragada09}.

The results in this section are reproducible and the code can be found in the \texttt{Github} repository that we have provided.

\section{Methods}\label{sec:methods}
In this section we introduce the anomaly detection and dimensionality reduction methods employed in our analysis. These methods are used with different purposes along the paper. The Unsupervised Random Forest method is used to estimate anomaly scores for the entire set of objects. The Isolation Forest method is used (see previous section) to support our analysis on pre-processing.  Finally, the manifold methods (\tsne~and UMAP) are used to reduce the dimensionality of the light curves and to group similar anomalies. For the latter, the addition of an Euclidean metric in the low-dimensionality space can be used as an additional anomaly measure, as explored in \autoref{sec:manifold_anomalies}.

\subsection{Unsupervised Random Forest}\label{sec:URF}

In order to understand how \emph{similarity} is defined in random forest (RF) methods, we ought to briefly describe the basics of their functionality as classifiers. A RF is an ensemble of decision trees. Each of these trees is a model in which the final prediction is based on a series of comparisons of the values of the predictors (features) against threshold values. Every tree, therefore, corresponds to a partition of the feature space by axis-aligned boundaries where the class of each final partition is given by the class of the majority of objects in that partition. These methods are inherently ``greedy'' and cannot exhaustively explore the parameter space (consider, for example, that a single continuous variable offers an infinite number of binary splits). As a result, decision trees have high variance (different trees give a different results) and are subject to overfitting. Ensemble approaches reduce the variance of these methods: each tree in a RF uses a random subset of objects from the training set and a subset of features. The forest then predicts the class of an object as the majority-vote prediction of all the trees in the ensemble.

The unsupervised random forest (URF) method for anomaly detection was first introduced by \citet{Shi06} in the context of tumor discovery using data sets comprising tumor marker expressions and has been adapted for the first time in astrophysics to look for anomalous objects in a large ($\sim$2 million sources) data set of SDSS spectra \citep{2017ascl.soft05015B, Baron17, Reis18}. 
URF is an adaptation of RF to unsupervised learning. The method works in two stages: the training of a supervised classifier, and the generation of an anomaly metric obtained by propagating the dataset of interest through the trained classifier. The classifier is trained to distinguish between members of the original dataset, and members of a \emph{synthetic} dataset that is generated by sampling the marginal distributions from that original dataset for to each of the features independently. That is, the synthetic dataset is identical to the original dataset in its marginal distributions, but it lacks the correlations between features that are present in the original data. The labels for the classifier are thus either \emph{original} or \emph{synthetic}, and the training set is the entire original dataset augmented with the synthetic dataset. The training is performed using all objects in the dataset for which an anomaly score wants to be determined.

The anomaly score is then computed by propagating the original dataset trough the trained classification model. The similarity score between two objects is measured as a normalized count of individual trees in which two given objects ended up in the same leaf node and with the same class. The \emph{weirdness} or anomaly score of an object can then be thought of as the average of the pair-wise dissimilarity between that particular object and all the other objects in the data set. Note that this dissimilarity measure is not necessarily associated to distance in an Euclidean space, because objects can be isolated in only a few of the many possible dimensions of the features space, and also because of the random nature of the forest. 

Formally, the similarity between the $i$-th and $j$-th objects in the data set can be calculated as:

\begin{equation}
\label{eq:weirdness}
D_{ij} = 1-N_{\rm{leaf}}/N_{\rm{tree}}
\end{equation}

Where $N_{\rm{leaf}}$ is the number of trees for which the $ij$ pair were both classified as \emph{real} in the same leaf node, and $N_{\rm{tree}}$ is the total number of trees. The \emph{weirdness} of object $i$ is then the average value of $D_{ij}$ for all possible pairings of $i$ with all other $j$ objects in the data set\footnote{For a graphical demonstration of how weirdness is defined in this algorithm, we refer the reader to the \texttt{GitHub} repository associated to \citet{Baron17}: \url{https://github.com/dalya/WeirdestGalaxies/blob/master/outlier_detection_RF_demo.ipynb}}.

In this work, we implement the URF using the \emph{scikit-learn} \citep{scikit-learn} Random Forest architecture, following the general recipe described in \citet{Baron17}. 
The URF has several important hyper-parameters that require fine-tuning in order to avoid over-fitting (and which of course depend on the number of $n_{\rm{feat}}$ features of each ligth curve, see \autoref{sec:features}). The most important hyperparameters are the number of trees in the forest ($N_{\rm{tree}}$); the maximum depth of the tree (the length of the longest path from the root of the tree to a leaf), \verb+max_depth+; and the maximum numbers of features that are considered when looking for the best split, \verb+max_features+. We performed the tuning of these parameters through a validation process that involved a gridsearch and a 80-20 training-test split. For the light curves and features described in \autoref{sec:data}, we found that achieving an optimal, non-overfitting validation accuracy for the classifier also results in a well defined peak of anomalous scores in the distribution of the URF scores. That is, if the classifier is less accurate, the peak of anomalies defined by the distribution of URF scores is less distinct from the less anomalous objects. The desired levels of accuracy are achieved with $N_{\rm{tree}}=700$, \verb+max_depth+$\: =100$ and \verb+max_features+$\: = \log_2(n_{\rm{feat}})$ in the case of the full dense light curves curves, and $N_{\rm{tree}}=150$, \verb+max_depth=default+, and \verb+max_features+$\: = \sqrt{n_{\rm{feat}}}$ in the case of the sparse light curves. 

\subsubsection{Isolation Forest}
A related method is the Isolation Forests (IF), that we have used in \autoref{sec:pre-processing} to evaluate the effect of pre-processing of the data. IFs are ensemble methods based on Isolation Trees \citep{Liu12}. In this method, which comprises a forest of isolation trees, each tree is partitioning a data set based on randomly selected features and randomly selected splitting points for each feature. The anomaly score is proportional to the number of random splits required to isolate an object (smaller score indicates more anomalous objects, negative scores indicate ``outliers''), averaged over all trees in the forest. Anomalies require fewer partitions, \ie, they are easy to isolate. The average number of partitions over a large number of random trees can therefore be considered as a measure of similarity to the bulk of the data. For the pre-processing analysis of \autoref{sec:pre-processing}, we use the \citep{scikit-learn} implementation \citep{Buitinck13} of this method and we embrace the default parameters: the number of samples used by each tree is set to \verb+max_samples=+ $2^8$ and the number of trees to 100. The ``contamination'' parameter sets the expectation for the fraction of objects in the set that are outliers, and it is set to \verb+contamination=0.1+. These parameters were demonstrated in \citealt{Liu12} to be effective under a large range of circumstances. The IF anomaly score is intuitively interpretable. For this, we chose it to guide our pre-processing choices. Yet IF is an extremely powerful methods for anomaly detection in spite of its simplicity (see for example the benchmarking study \citealt{10.1145/2500853.2500858}).

\subsection{Manifold learning methods}\label{sec:manifolds}
We now describe two related manifold learning methods: \tsne~and UMAP. The methods are designed to visualize high-dimensional data in a lower-, typically 2-, dimensional space where reciprocal distances in the high dimensional space are (optimally) preserved. 

The resulting embeddings can also be used for anomaly detection. As we detail in \autoref{sec:tsne_results}, an anomaly score can be defined by looking at the distribution of distances to the nearest neighbor in the converged distribution for both \tsne~and UMAP methods. This Euclidean anomaly score, which selects a very specific type of anomalies, can be used complementary to the score derived from the URF algorithm.

\subsubsection{\emph{t}-distributed Stochastic Neighbor Embedding}
\label{sec:tsne}
The \emph{t}-distributed Stochastic Neighbor Embedding (SNE) method, or \tsne, was introduced in \citet{Maaten08}, improving upon the well known nonlinear dimensionality reduction algorithm SNE \citep{Hinton03}. SNE works by embedding multidimensional Euclidean distances with conditional probabilities, which is what represents the similarities between datapoints. In other words, suppose we have a data point $x_i$ in the high dimensional space. Then consider a normal distribution of distances from $x_i$, wherein points near $x_i$ have a higher probability density under the distribution and further points have a lower probability density under the distribution. Then the similarity between $x_i$ and another data point $x_{i'}$ is the conditional probability $P_{x_{i'} | x_i}$ that $x_i$ will choose $x_{i'}$ as a neighbor under the normal distribution just described. 

Then we replicate the process for the lower dimensional space, for which we get another set of conditional probabilities. SNE then attempts to minimize the Kullback-Leibler (KL) divergence ~\citep[or relative entropy,][]{Kullback51}  between the two probability distributions using gradient descent.  However, SNE is computationally very expensive, largely because of the asymmetry imparted by the use of KL divergence as the distance metric;  \tsne~attempts to resolve this issue by looking at a ``symmetric'' SNE, specifically a symmetric version of the cost function with similarly simple gradients. \tsne~also redefines the lower dimensional distribution using a Student \emph{t}-distribution in place of the Gaussian distribution to solve the crowding problem, which stems from the fact that there is not enough area in a two-dimensional plot to accurately embed distances between points that are close, which leads to loss of information.

Here we use the Python \texttt{scikitlearn} implementation of \tsne~with hyperparameters (\verb+n_components+=2, \verb+perplexity+=200, \verb+learning_rate+=50.0, \verb+early_exaggeration+=5.0) for dense light curves, and similar hyperparameters for sparse light curves (with the only difference being \verb+early_exaggeration+=20.0); \verb+n_components+ represent the dimension of the space we want to map into. \verb+Perplexity+ is related to the number of neighbors to be considered when considering a certain data point (described in \citealt{McInnes18}), defining a notion of similarity; \verb+learning_rate+ affects the gradient descent portion of the tSNE algorithm, with too fast a learning rates resulting in ball like clusters where neighbors are equidistant, and too slow a learning rate creating dense cloud clusters; \verb+early_exaggeration+ relates to how tightly points are clumped in the embedding space, so that we can control the visualization of the high dimensional data (this affects all points similarly, so it mostly impacts visualization, and not the overall similarity results from the algorithm). We ranged the values of the three hyperparameters over different experiments, recording the KL divergence of each model after training on the full data set. We then set the hyperparameters based on the model with the lowest KL divergence.


\subsubsection{Uniform Manifold Approximation and Projection}
\label{sec:umap}
Uniform Manifold Approximation and Projection (UMAP) is a nonlinear dimensionality reduction technique introduced very recently in \citet{McInnes18}. Most dimensionality reduction techniques have a very similar structure: they aim to find some low dimensional representation of data that minimizes information loss between the same representation applied to the high dimensional data set. 
UMAP works as follows: consider data points $x_1, \dots, x_n$. We then create a \emph{k}-neighbor weighted graph by considering \emph{k}-neighbors of each $x_i$, and adding an edge in the graph with a defined weight $w$ that depends on the diameter of the \emph{k}-neighborhood of $x_i$, and the distance between $x_i$ and the closest neighbor.
Note that the weight, as defined in \citet{McInnes18}, is not symmetric. We handle this by, given $a = w(x_i, x_j), ~b = w(x_j, x_i)$, defining a new weight $w'(x_i, x_j) = a + b - ab$. Then the same process is repeated in the lower dimensional space, resulting in a new weight function for the lower dimensional space. Then UMAP minimizes the cross-entropy between the two weight functions as specified by the cost function so that the lower dimensional weights encapsulate (as closely as possible) the information from the higher dimensional weights. Like in \tsne, this optimization is done via stochastic gradient descent.

Here we use the UMAP implementation provided in \citet{McInnes18}, with hyperparameters (\verb+n_neighbors+=200, \verb+min_dist+=0.4, \verb+learning_rate+=0.25) for dense light curves. We used hyperparameters (\verb+n_neighbors+=200, \verb+min_dist+=0.1, \verb+learning_rate+=0.8) for sparse light curves. \verb+n_neighbors+ is similar to \verb+perplexity+ in the \tsne~algorithm, controlling the number of neighbors to be considered when defining the weights between data points in the weighted graph. \verb+min_dist+ is similar to the \verb+early_exaggeration+ parameter in the \tsne~algorithm, and controls how tightly points are clumped in the embedding space (to create more interpretable visualizations). Similar to the \tsne~method, we vary the hyperparameters for the model, and pick the model that gives the most stable embeddings.

\subsection{Processing requirements and scalability}
We now describe the computational performance of our algorithms in order to provide some isnight as to how our method scales to even larger datasets. Overall, processing times for the number of objects and the number of features involved in this project are not computationally prohibitive. The URF anomaly detection method took a few hours to process the entire set of 147,036 dense light curves, each containing about 6300 features, using a 2.4 GHz 8-Core Intel Core i9 Processor, typical of a Mac laptop. One advantage of ensemble methods such as the URF is that trees can be trained independently, and therefore parallelization is possible. The URF computing time is linear in the number of features and in the number of trees. The generation of the similarity matrix for the calculation of the anomaly score requires a total of $n_{\rm{obects}}\times n_{\rm{obects}-1}$ simple arithmetic operations, and it therefore scales roughly as $n_{\rm{objects}}^2$.

The machine that ran the \tsne/UMAP outlier pipeline had a GNU/Linux OS, x86-64 architecture (Linux kernel version 3.10.0-957.el7.x86-64) with 31 GB of free memory. DMDT generation for dense light curves took around 36.76 hours, and DMDT generation for sparse light curves took around 0.23 hours. Generating TSNE embeddings for dense light curves took about 1.04 hours and UMAP embeddings for dense light curves took around 2.36 hours. Generating \tsne/UMAP embeddings for sparse light curves took around 2 hours. Finding outliers given a specific embedding took around 1 hour. In terms of storage,  the dense and the sparse DMDT data sets occupy $\sim450$~MB each. Storing the \tsne/UMAP embeddings for dense and sparse light curves took around 2MB each. No batch processing was therefore required for our largest dataset.





\section{Results}\label{sec:results}
This section summarizes the results of implementing our method for the identification of anomalies in the Kepler data set, and for the identification of analogs to those light curves. We describe the list of \emph{bona-fide} anomalies that we have used as a ground truth to evaluate the performance of the method. We first apply the URF method to define a list of anomalies, and several diagnostics of their anomalous behavior. We then investigate how the 2D embeddings from the DMDT maps can be used to identify objects that are analogous to those identified with URF as anomalies. In the last subsection we examine some astrophysical implications of our findings.

\subsection{\emph{Bona-fide} anomalies}
To ensure our methods are effective, we use a control set of known anomalous objects as a first test of our method: rare objects should have a high anomalous score, as measured by our algorithm. In order to construct this set, we have relied on the  work of \citep{debosscher07}, who performed a comprehensive census of 35 variability classes and identified members of each class via an exhaustive literature search, that they later used to inform different supervised classifiers. Among the uncommon behaviors identified, they find:

\begin{itemize}
    \item Ellipsoidal variables
    \item $\gamma$-Doradus stars
    \item Slowly-pulsating B stars
    \item RR Lyrae stars
    \item RV-Tauri stars
    \item Classical cepheids
\end{itemize}

We constructed a list of 150 objects containing members of each of these classes. While these classes are not necessarily rare in the overall galactic field population, stellar variability is in fact rare in the \emph{Kepler} Input Catalog (KIC). Variability classes such as classical Cepheids, ellipsoidal variables, $\gamma$-Doradus stars, and pulsating B stars are have less than 1000 identified members in the KIC, out of a total of about 13 million targets. RR Lyrae stars and RV-Tauri stars are even rarer, with probably less than 100 identified objects. Even eclipsing binaries are below the 10,000 limit. For the purpose of this paper, thus, stellar variability is ``anomalous'', and our algorithm should also serve as a variability detector.

To the group of \emph{bona-fide} variables,  we have added Boyajian's star, which shows a truly unique behavior, unlike any other objects in the aforementioned classes. Some members of these classes, and Boyajian's star itself, have been successfully identified by existing anomaly detection methods,  \citep[notably in][]{Giles20}. However, no established mechanism exists to find analogs to those objects in a given dataset.

We present the full list of \emph{bona-fide} anomalies in \autoref{tab:bona_fide}.

\subsection{Finding anomalies with URF.}\label{sec:urf_anomalies}
To find anomalous light curves, we apply the URF algorithm with the hyper-parameters described in \autoref{sec:URF} to the set of features composed of the array of the pre-processed (normalized, spike-removed) light curve points concatenated with the periodogram points, for each of the 147,036 light curves in our dataset. The URF results are reproducible and presented at the \texttt{GitHub} repository that we have provided. The total number of features used for the dense light curves was 6520, whereas for the sparse light curves it was 632. The hyper-parameters of the URF were tuned to maximize accuracy during the classification step as described in \autoref{sec:methods}. We compute the URF weirdness scores for each light curve, and for both the dense and sparse datasets. In this section we present the resulting anomaly scores, highlight the important features for anomaly detection, and specifically search for the \emph{bona-fide} anomalies in the ranked list of weirdness scores, to evaluate the performance of our method in finding light curves of interest. We also provide a list of new anomalies identified with our method.

\begin{figure}
\includegraphics[scale=0.57,angle=0,trim=0.2cm 0.2cm 0cm 0.8cm,clip=true]{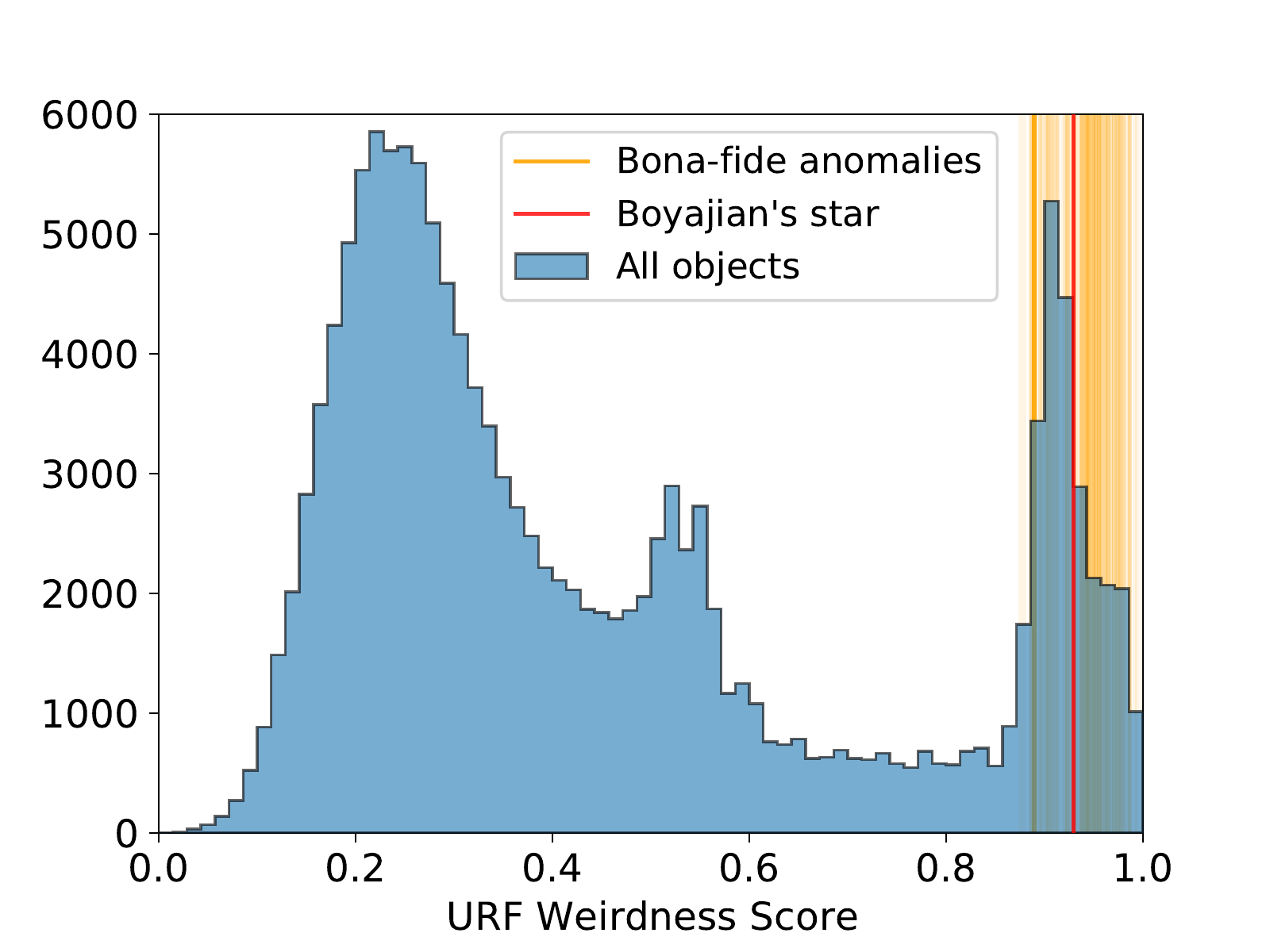}
\includegraphics[scale=0.57,angle=0,trim=0.2cm 0.2cm 0cm 0.8cm,clip=true]{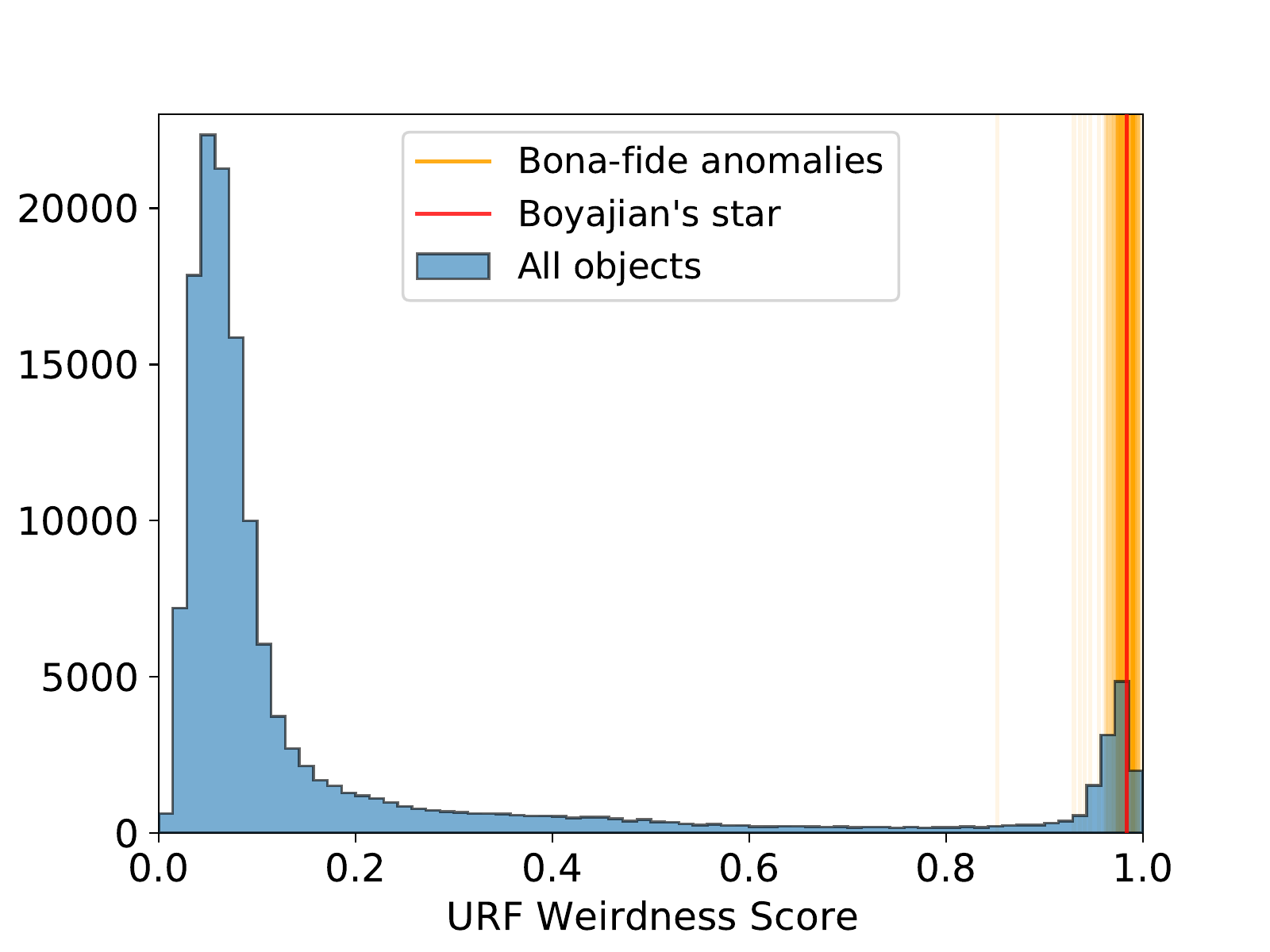}
\caption{Histograms of URF weirdness scores measured for dense (top) and sparse (bottom) Kepler Q16 dense light curves. Indicated in orange and red are the scores of the \emph{bona-fide} anomalies and Boyajian's star, respectively.}
\label{fig:weirdness_hist}
\end{figure}

\subsubsection{The identification of anomalous objects}\label{sec:URF_analysis}

In \autoref{fig:weirdness_hist} (top panel) we show the histogram of URF weirdness scores for all the 147,036 dense light curves. Note that the absolute range of URF score is not \emph{per-se} informative of the differences between objects, as it depends on the total number of trees and depth of the random forest, and we have therefore re-scaled the URF scores to the range $(0,1)$. The relative differences between objects are informative. We identify at least three groups of objects based solely on the values of the URF scores. There is a main core of ``normal'' objects, with normalized anomaly score centered at around 0.25, which includes about 60\% of all objects. Then there is a clear peak of objects that represent 18\% of the total, and that all have scores larger than about 0.85. We will take these to be anomalies. Finally, there is an excess of objects with scores centered at around 0.55, near the center of the range. This excess of objects, as we will see later in \autoref{sec:discussion}, mostly corresponds to members of the ``red clump'', a group of red giants that are slightly hotter than other red-giant-branch stars of the same luminosity, and have a degenerate helium core \citep{1999MNRAS.308..818G}.

Based on the distribution of the anomaly scores, we select a weirdness score of 0.85 as the URF anomaly threshold hereafter. At this score the distribution in the top panel of \autoref{fig:weirdness_hist} departs from a decreasing behavior and starts increasing, signaling an inflexion point beyond which there is an excess of objects with high anomaly scores. We note that the selection of this threshold is informed by the distribution, but it is to some extent arbitrary, because the transition between normal and anomalous objects is a continuous, rather than a discontinuity.

\begin{figure*}
\includegraphics[width=1.02\textwidth,,trim=3.8cm 1.6cm 3cm 2cm,clip=true]{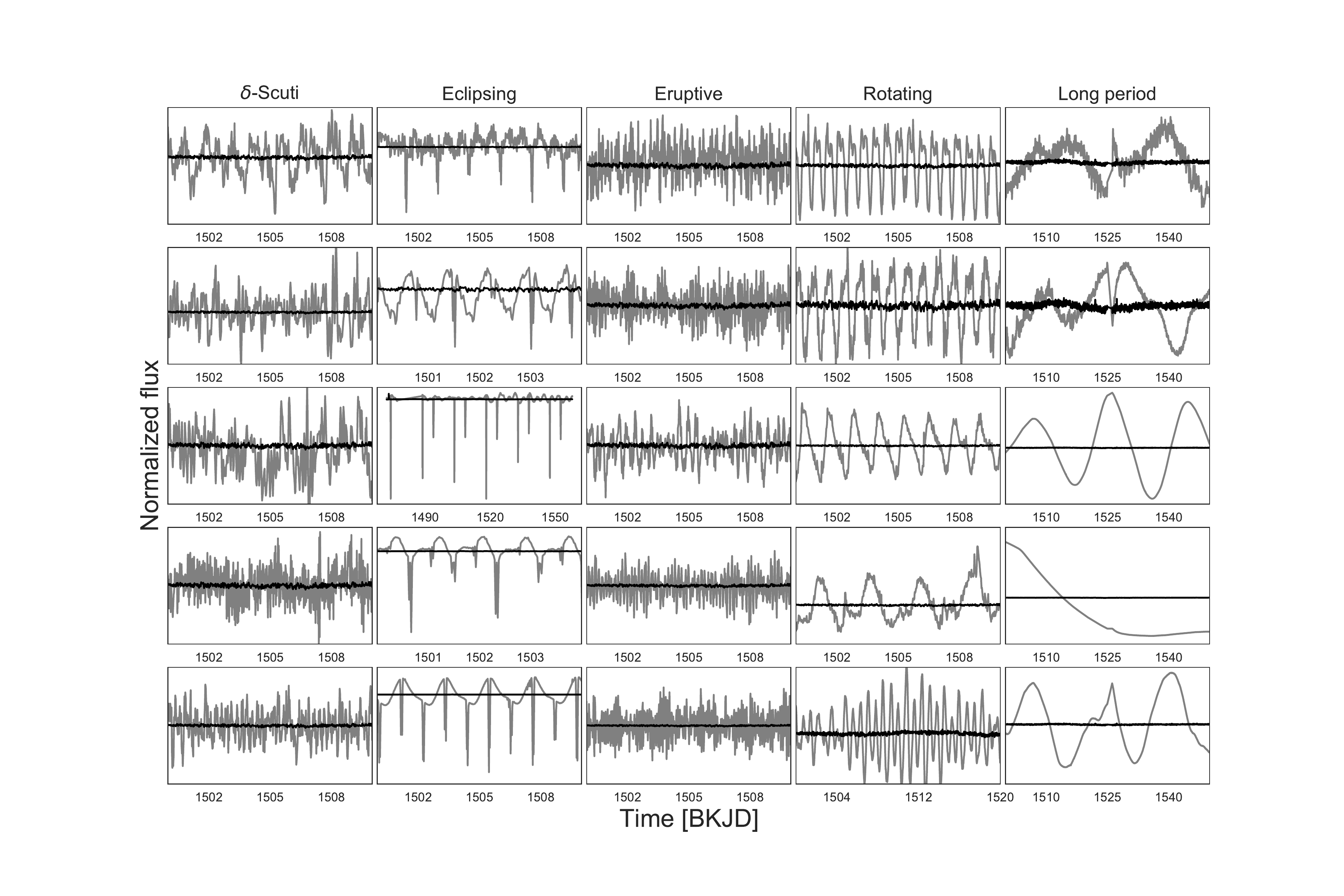}
\caption{A selection of URF anomalous light curves with an associated common class. Note that the ranges in both axes change from light curve to light curve. For comparison, each light curve is shown together with the light curve of low-URF score object KIC 8211660 (black line)}
\label{fig:anomalous_lc_known1}
\end{figure*}

\begin{figure*}
\includegraphics[width=1.02\textwidth,trim=3.8cm 1.6cm 3cm 2cm,clip=true]{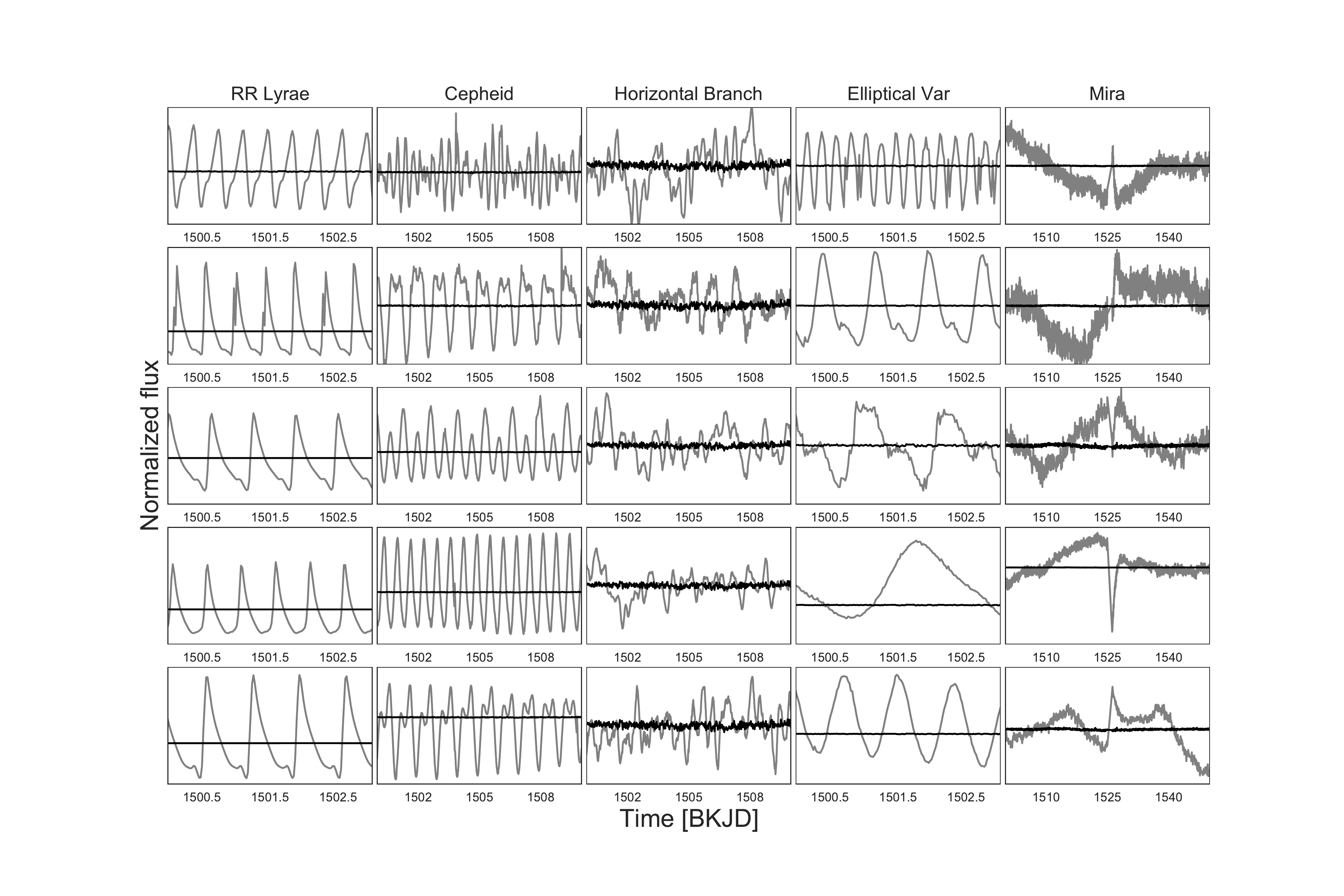}
\caption{A selection of URF anomalous light curves associated with less common know classes. Note that the ranges in both axes change from light curve to light curve. For comparison, each light curve is shown together with the light curve of low-URF score object KIC 8211660 (black line)}
\label{fig:anomalous_lc_known2}
\end{figure*}

\begin{figure*}
\includegraphics[width=1.02\textwidth,trim=3.8cm 1.6cm 3cm 2cm,clip=true]{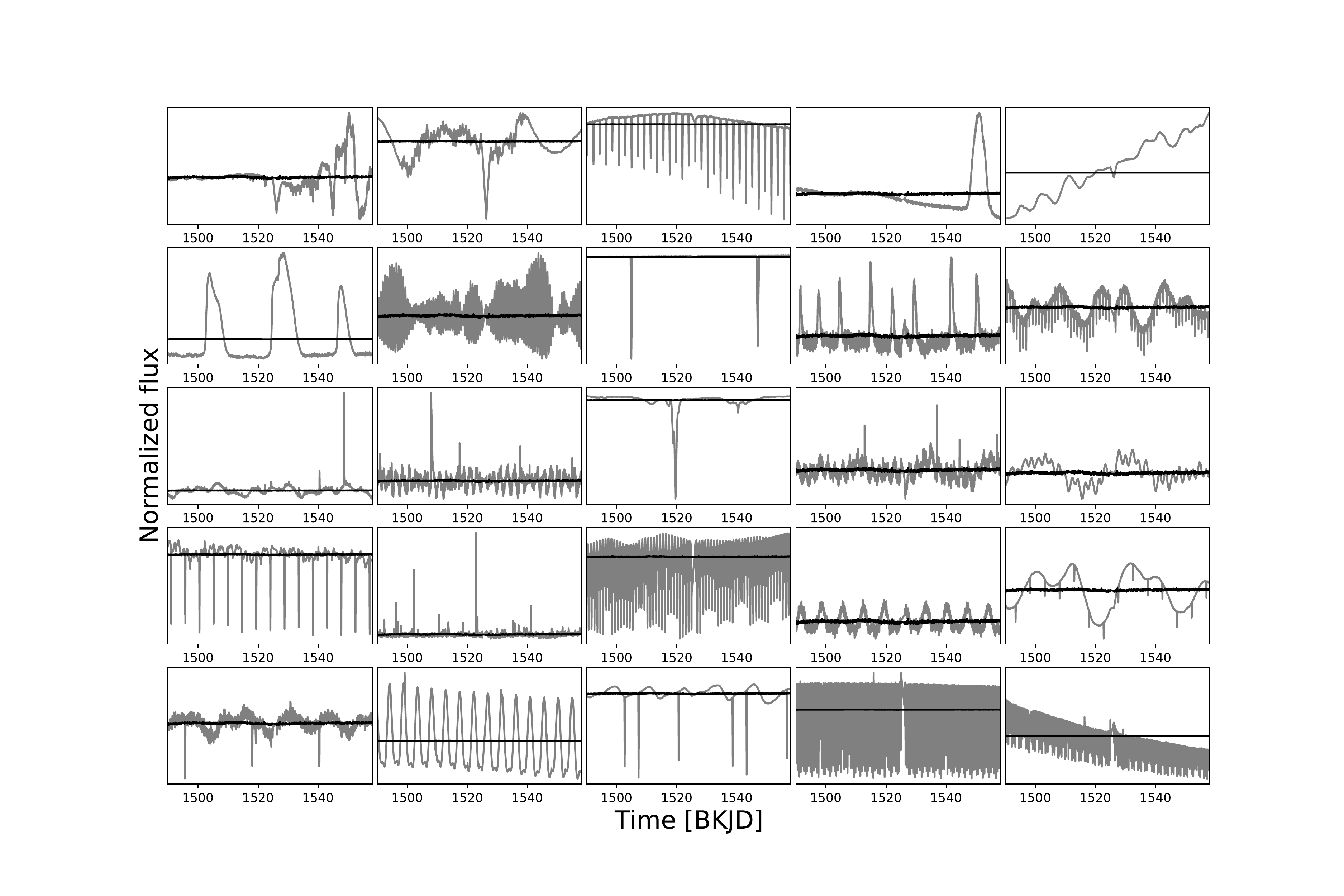}
\caption{A selection of URF anomalous light curves without a current classification in the SIMBAD database. Boyajian's star is shown in the center panel. Note that the y-axis range in changes from light curve to light curve, whereas the time axis is kept the same for all. For comparison, each light curve is shown together with the light curve of low-URF score object KIC 8211660 (black line)}
\label{fig:anomalous_lc_unknown}
\end{figure*}

We also indicate in \autoref{fig:weirdness_hist} the URF weirdness scores for all the \emph{bona-fide} anomalies of \autoref{tab:bona_fide}, shown as vertical orange lines. We observe that all of the anomalous objects of interest fall within the anomalous peak, \ie, they are all in the top 18\% of the weirdness scores (Boyajian's star is in the top 7\%). We determined the uncertainty in the anomaly score by running the full algorithm 10 times for a subset of about 20,000 objects, and measuring the standard deviation of the anomaly score for each object across these 10 runs. We find that the typical uncertainty depends on the mean value of the score, with high anomaly scores having smaller standard deviations. Specifically, the URF score is uncertain at the level of a few percent (2\%-3\%) for objects with mean URF score above 0.85, and at the level of about 10\% for objects in the low end of the URF score distribution. The extreme low-score end of the distribution is populated by objects whose features are consistent with being sampled from the distribution that describes the bulk of the objects, specifically near its mean. In this respect, they are unremarkable as anomalies. One example of such objects is KIC~8211660 (URF score 0.0015), which is shown as a black line in \autoref{fig:anomalous_lc_known1}, \autoref{fig:anomalous_lc_known2}, and \autoref{fig:anomalous_lc_unknown}.

As discussed above, the \emph{bona-fide} anomalious objects that we just demonstrated to be able to recover represent a few different astrophysical phenomena with different photometric properties. The anomaly scores are loosely related with ``anomaly type'', but there is a significant overlap between classes. We can quantify this by looking at the average scores of anomalies for which an independent classification is available in the SIMBAD database. For example, among the most anomalous objects are $\delta$-Scuti stars, for which the mode of the URF score is 0.99 (top 0.6\%). Eruptive stars have a similar mode, but have a larger fraction of members with much lower anomaly scores. Eclipsing binaries have a score mode of 0.975 (top 1.6\%). $\gamma$-Dor stars have a score mode of 0.96 (top 4\%). Rotating variables have a score mode of 0.91 (top 8\%), and represent the vast majority of the high anomaly score peak. Long period variables have a score mode of 0.88 (top 17\%), and so on. Other types of object are also found  (in smaller numbers) to have high score, including RRLyrae, cataclysmic variables, Cepheids, White Dwarfs, ellipsoidal variables, Mira variables, as well as active galactic nuclei, and quasars. Remarkably, a fraction as large as $\sim4\%$ of all the Q16 targets, or about 5000 objects, are both unclassified and within the anomalous peak. 

We show examples of light curves found by our method to have a high anomaly score in \autoref{fig:anomalous_lc_known1}, \autoref{fig:anomalous_lc_known2}, and \autoref{fig:anomalous_lc_unknown}. \autoref{fig:anomalous_lc_known1} shows known variability types (we remind the reader that these types are rare in the KIC), such as eclipsing binaries, long period variables, and $\delta$-Scuti stars. They all show significant amplitude variations with respect to the light curves of objects with low anomaly score (we have plotted the light curve of one of these normal objects, KIC~8211660, as a dark line for reference). They also have periodic or semi-periodic behaviors. \autoref{fig:anomalous_lc_known2} shows light curves of objects of known type that achieve a high URF anomaly score, but that belong to classes that are even rarer in our dataset. These include RR Lyrae stars (only a handful in the entire dataset), ellipsoidal variables, and Mira stars. Just as in the first group, these anomalies have wide amplitude variations and a tendency for periodic behavior. Finally, in \autoref{fig:anomalous_lc_unknown} we show objects found by our method to be anomalous and that are not classified in the SIMBAD database. Some of them appear in the literature as candidate dwarf novae, extrasolar planetary transits, and variable RGB stars, but the majority remain unclassified. Boyajian's star appears in this group, and it is shown in the center panel

The rich set of different anomaly types that we are able to identify indicates that the URF algorithm applied to the set of features constructed from the light curve points and their periodograms is an effective approach to identifying time-domain anomalies of varying astrophysical nature, including paradigm-changing objects such as Boyajian's star, but also a wide range of variability types that depart from the behavior of the majority of Kepler's sources. The complexity of the light curves result in significant dispersion and considerable overlaps between different groups in the distribution of URF scores. This implies that, given an unclassified anomaly, we cannot associate it with a specific class, or tell whether it belongs to a potentially novel class of objects, based on the URF score alone. In \autoref{sec:tsne_results} we propose a method to identify analogs.

We provide the full list of unclassified anomalies found with our method in the electronic version of the paper, together with their URF anomaly scores, low-dimensional manifold embedding features, and where available, also \emph{Gaia} absolute magnitudes and colors. 

\subsubsection{Sparse light curves}

Can we still identify anomalous objects when the amount of information is reduced? Time-domain surveys do not generally result in regular and well sampled light curves, like Kepler's. In fact, ground-based surveys, including the upcoming LSST, will have significantly more irregular and sparser cadences when compared to the dataset studied here so far.

We have applied our method to the set of sparse light curves generated by randomly sampling 10\% of the original light curve points. Just as with the original dataset, we have generated the periodograms by applying the Lomb-Scargle method to the sparse light curves, and concatenated them to the data arrays to form our feature vector. We have then adjusted the hyper-parameters of the URF algorithm to maximize accuracy in the classification step as described in \autoref{sec:URF}.

The resulting histogram of URF weirdness scores is shown in the bottom panel of \autoref{fig:weirdness_hist}. The sparser light curves have resulted in a much less structured distribution of the scores, but the general result holds: most of the \emph{bona-fide} anomalies, Boyajian's star included, remain within the peak of anomalous object, which is now much more separated from all the rest of the objects and only includes 8\% of all objects (the anomaly peak in the case of the full-sampled lightcurves contained 18\%). This suggests that information about the anomalous nature of objects is mostly contained at low frequencies. However we do not identify as anomalous all the original anomalies.

What information are we missing in the sparse light curves for the purposes of anomaly identification? One would be tempted to infer that the information lost is related to high-frequency oscillatory modes, to which we have lost sensitivity when under-sampling the original dataset. But the difference in the high frequency cuts of the periodograms for dense and sparse light curves is not very significant (1 hour versus 4 hours), and most of the 
\emph{bona-fide} anomalies have characteristics timescales larger than a few hours. In fact, what we are losing by only considering the sparse set is the ability to discern significant large-scale differences in the distribution of spectral power between high frequencies and low frequencies.

\begin{figure*}
\includegraphics[scale=0.64,angle=0,trim=1.5cm 1.5cm 1.5cm 1.5cm,clip=true]{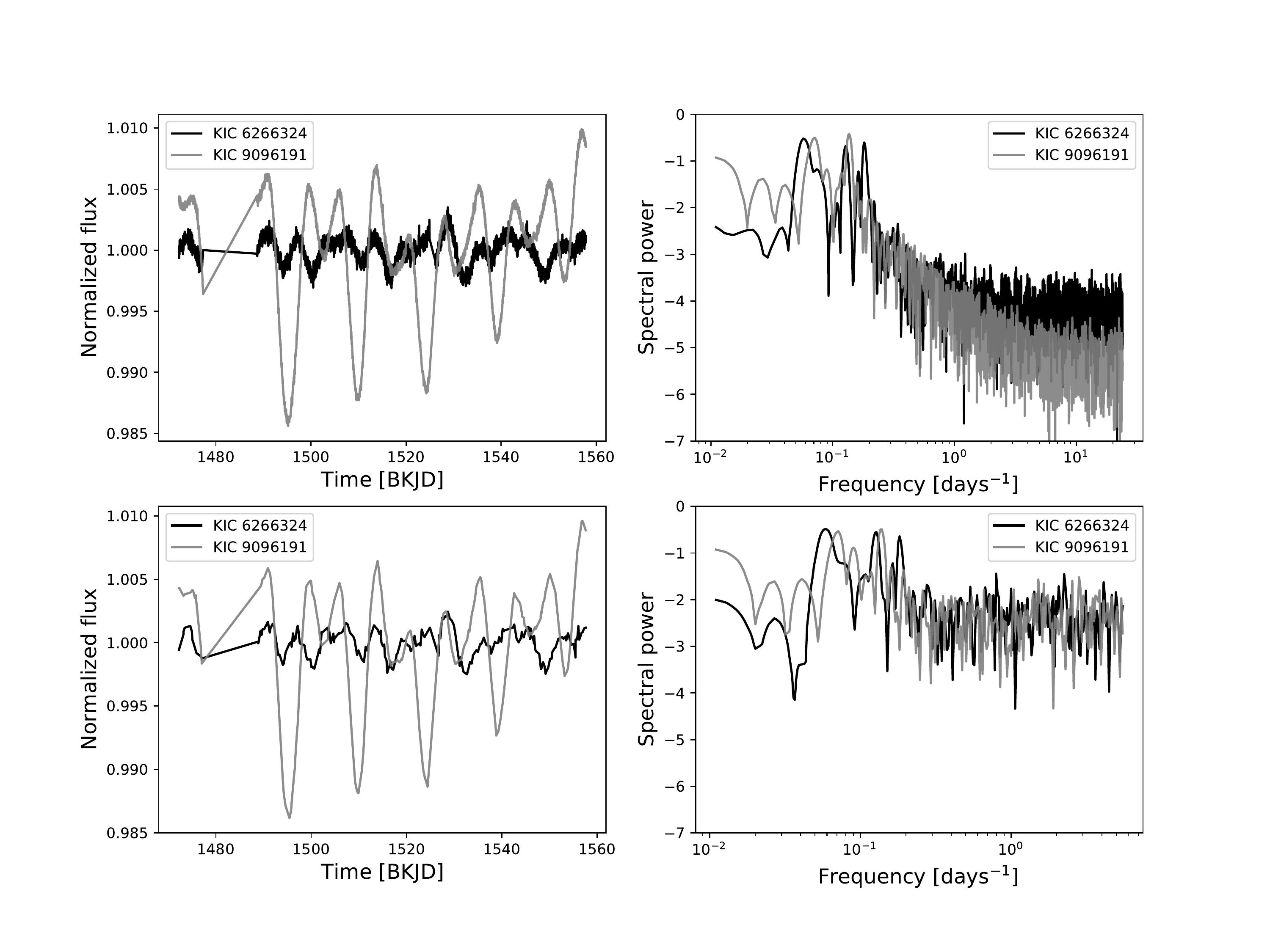}
\caption{Effect of information loss on anomaly detection. Shown are the light curves (left panels) and periodograms (right panels) for two rotation-variable objects, and for both the dense (top panels) and sparse (bottom panels) versions. KIC 9096191 (gray line) is identified as an anomaly in both the dense and sparse datasets. KIC 6266324 (black line) is identified as an anomaly only in the dense dataset. Note that the power spectrum is more uniform across frequencies for the sparse light curves.}
\label{fig:freq_separation}
\end{figure*}

To illustrate this, in \autoref{fig:freq_separation} we show the light curves (both dense and sparse) and the corresponding periodograms for two rotating variable stars with high anomaly scores. One of the two (KIC 9096191) is identified as an anomaly in both the dense and sparse datasets, whereas the other (KIC 6266324) is only identified in the dense dataset. We argue that the reason for this difference is that any relative low-frequency power excess in the power spectrum is damped by the loss of data points (\ie the overall shape of the power spectrum is flatter for the sparse light curves)\footnote{In noisy light curves of fainter or farther away objects, this excess is less pronounced because noise adds power primarily in high frequencies.}. This implies that anomalies selected mainly on the basis of an uneven spectral power distribution will only be spotted when the dense light curves are used. Some, of course, might still be selected as anomalies in the sparse light curve if, for example, the amplitude of the variability is unusually large, as is the case for KIC 9096191. 

\begin{figure}
\includegraphics[scale=0.3,angle=270,trim=0.0cm 0.0cm 0.0cm 0.0cm,clip=true]{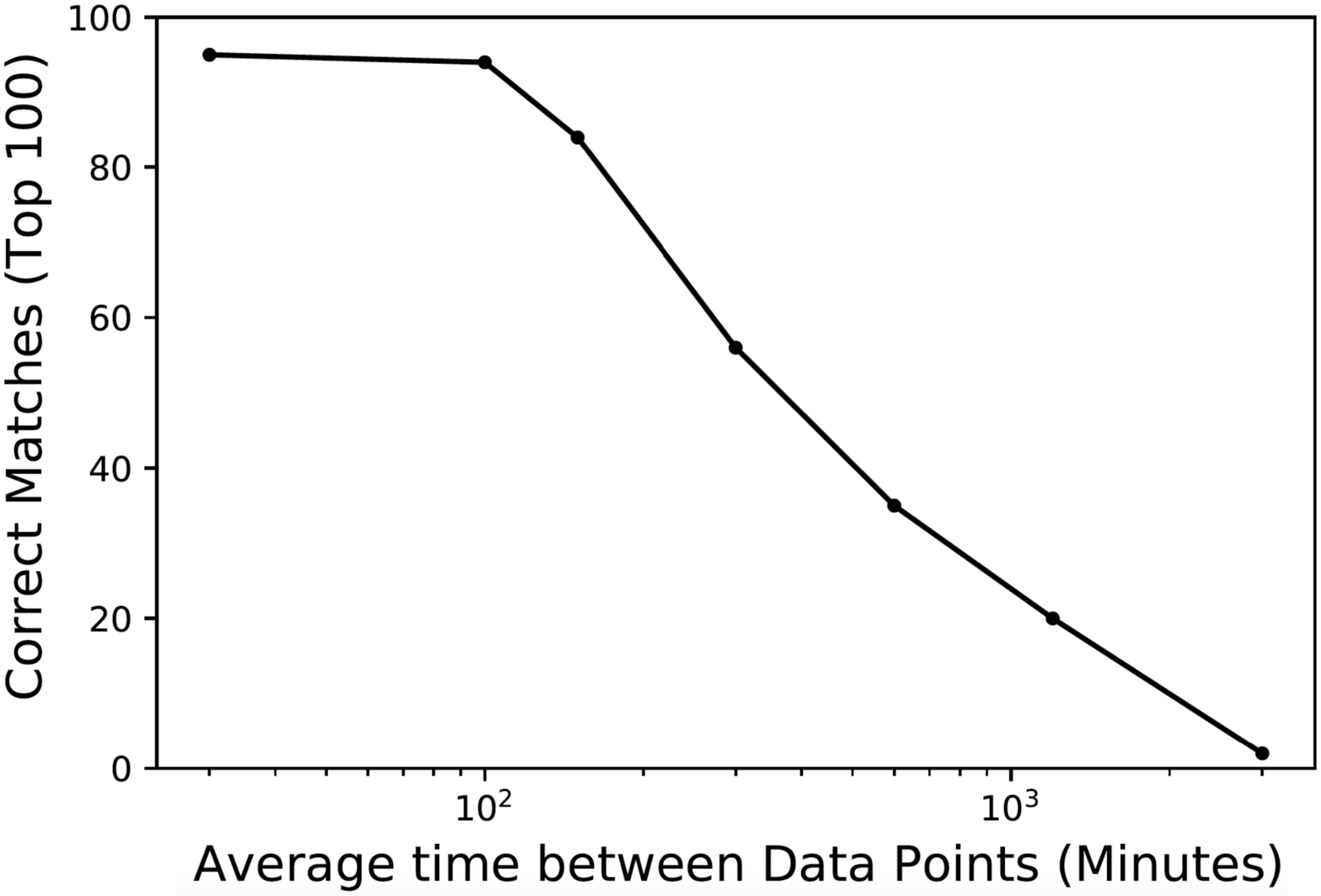}
\caption{The recovery rate of anomalies as a function of the average separation between points, for a subset of 2500 light curves.}
\label{fig:cadence_effect}
\end{figure}

A more general question is: how sparse can the cadence of a light curve get before one loses the ability to detect anomalies entirely? In \autoref{fig:cadence_effect} we show the fraction of the 100 most anomalous  objects found using the dense light curves that are still identified as anomalous in sparse light curves, for increasing separation between light curve points, for sub-sample of 2500 Q16 light curves. The average separation between points needs to increase by a factor of about 5 before the recovery rate drops below 80\%. This means that we could decrease the \emph{Kepler} Q16 cadence from 30 minutes to 2.5 hours and we would still detect 80\% of the anomalies that we have found using the full light curves.

\subsubsection{What drives the anomalous nature of light curves?}\label{sec:drives}

We now turn to the question: what makes a Kepler light curve anomalous? In this section we provide an answer in terms of the algorithm used and the features that were fed to it in computing the anomaly scores. 

\begin{figure*}
\includegraphics[scale=0.55,angle=0,trim=5.5cm 0.0cm 0cm 0.0cm,clip=true]{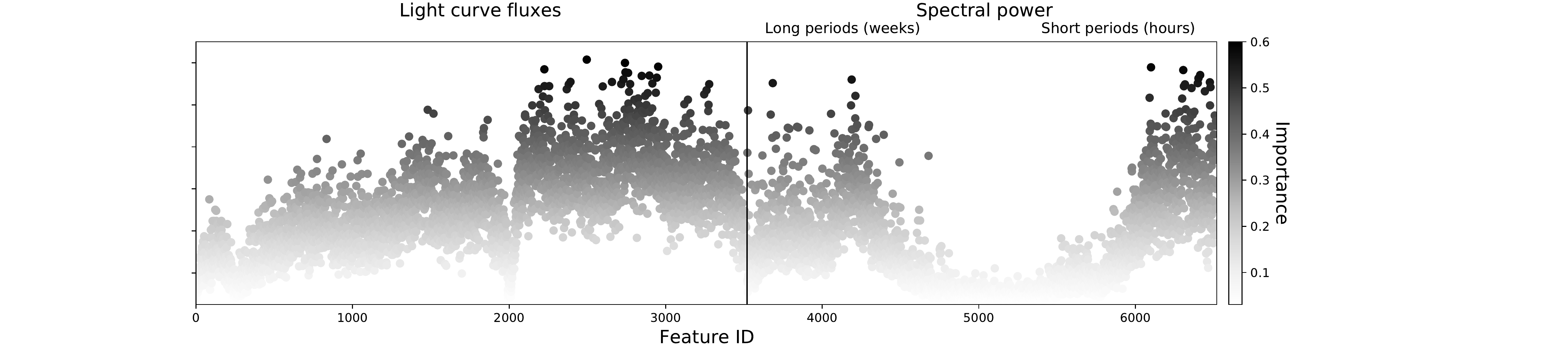}
\caption{Feature importance distribution for the URF. The vertical line marks the boundary between normalized fluxes (left) and frequencies (right). Each dot represents a feature, and they are color-coded by normalized importance. }
\label{fig:feature_imp}
\end{figure*}

In \autoref{fig:feature_imp} we show the distribution of feature importances derived during training of the URF classifier step. To estimate the importance of each feature, we have used the \verb+feature_importances_+ attribute of the random forest classifier in \verb+scikit-learn+, which assigns importances to features during training according to the accumulated decrease in impurity every time a feature is used in splitting. Important features are therefore those that, when used for splitting, result in a maximal isolation (less impurity) of anomalies. Leftwards of the vertical lines are the features associated with light curve points, rightwards of the line are the features associated with spectral power at different frequencies. There is relatively little variation in importance among features that represent light curve points, although there is a small increase of importance in the later half of the light curve. In fact, this upward trend follows quite closely the behavior of the \emph{average} light curve for the entire Kepler quarter, and might be the result of a residual in the de-trending process. The dip near the mid-point of the light curve features is related to a discontinuity due to lack of data around BKJD 1525. Something similar happens between BKJD 1477 and BKJD 1487. In general, light curve points are important in the classification step when they represent significant deviations from the mean flux. 

In the features associated with the power spectrum (rightward of the vertical line in \autoref{fig:feature_imp}) a more clear trend emerges. Spectral power values that contribute the most to the classification step, and therefore to the anomaly score, are either shorter than a few hours, or longer than about 10 days. The classification and anomaly detection are thus mostly affected by characteristic variability timescales of hours or weeks, whereas days-long characteristic timescales are less relevant for distinguishing between objects. 

We argue that features that are important in classification step of the URF (which, as a reminder, is the step in which the model learns to discriminate between true and synthetic data, see \autoref{sec:methods}) also dominate the anomaly detection. For example, anomalous light curves should have more points that deviate from the mean with respect to more common light curves. Similarly, anomalous light curves should have significant differences in the spectral power at those frequencies that were relevant to perform the classification during the training phase. This would explain why pulsating stars with short periods, such as $\delta$-Scuti stars, or long period variables are selected by the method, especially when they also have high amplitude variations. In particular, the combination of long (weeks/months) modulations with prominent dips and flares, as is the case of Boyajian's star (or certain types of eclipsing binaries), result in an anomalous behavior recognized by our method.

\begin{figure*}
\includegraphics[scale=0.55,angle=0,trim=2.5cm 0.3cm 2cm 1cm,clip=true]{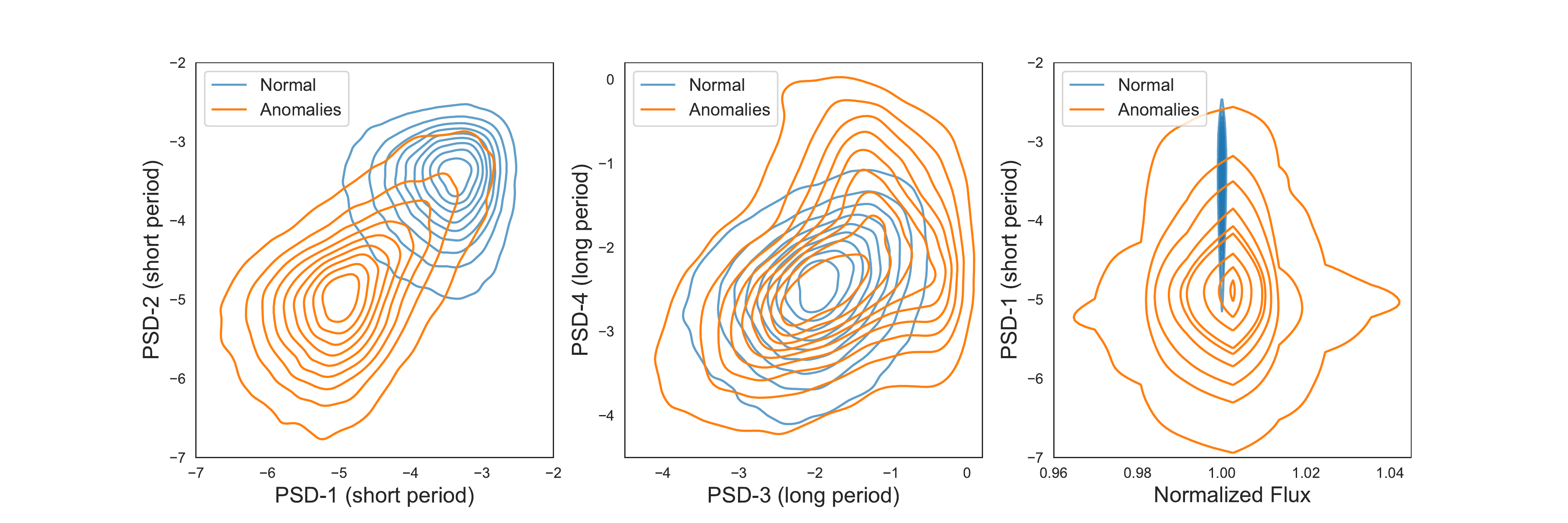}
\caption{Two-dimensional density distributions of several important features within the URF model for normal (blue) and anomalous (orange) light curves. Shown are for different power spectral densities (PSDs), and one normalized flux value.}
\label{fig:drives_weirdness}
\end{figure*}

In order to test our hypothesis, we look at how the most important features are distributed for both normal and anomalous light curves. We show the marginal distributions of a few pairs of features in \autoref{fig:drives_weirdness}. We selected the plotted features by ranking all features by importance, from more to less important, and then, for each group of high frequency power features, low frequency power features, and time-ordered light curve points, selecting from the top 5. The left panel shows the distribution of spectral power density (PSD) at two frequencies roughly corresponding to timescales of 1.5 hours (PSD-1) and 3 hours (PSD-2), for a representative group of normal objects (low URF score) and a representative group of anomalous objects (high URF score). For the same objects, the middle panel shows the distributions for low frequencies (15 days in the PSD-3 axis and 60 days in the PSD-4 axis), and the right panel shows the normalized flux at BKJD 1536 and the spectral power at 1.5 hours (PSD-1). 

We note that anomalies distribute quite differently in the joint space of features compared to normal objects. Of relevance is the fact that anomalous objects have less relative spectral power at high frequencies, but show a larger degree of correlation between the spectral power values at those frequencies. 
The relative variability amplitudes of anomalies are significantly enhanced with respect to the normal objects.

From this perspective, Boyajian's star is remarkable in more than one way. It shows pronounced dips seen in the light curve that deviate more than 20\% from the mean flux, but additionally it also shows shallower (0.5\%) modulations with characteristic timescales of several weeks. These modulations result in a complex and unique power spectrum dominated by the large timescales, some power bumps in short timescales, and a noticeable decrease in spectral power at frequencies corresponding to timescales of half a day, as well as higher frequency oscillations with similar characteristic periods. This timescale coincides with the typical duration of the main flux dips (see \autoref{fig:periodograms}).

\subsection{Finding analogs of light curves of interest}
\label{sec:tsne_results}
We have demonstrated that the URF score is a good measure of the anomalous nature of a light curve, and we have investigated how anomalous light curves differ from normal light curves in the space of features defined by the light curve points and the power spectrum. However, we have also pointed out that the URF score alone is insufficient for finding light curves that are similar between them, which is an essential task if we are to find analogs of specific light curves or truly unique objects. In this section we investigate whether manifold-learning methods that reduce the dimensionality of the light curves are suitable for the identification of groups containing anomalous light curves sharing similar properties. We show the results of this dimensionality reduction and explore its own value as an anomaly detection method, before we move on to combining the results with the URF scores in order to find groups of similar anomalous objects.

\subsubsection{Reduced dimensionality for finding analogs}

\begin{figure*}
\includegraphics[scale=0.45,angle=0,trim={0cm 0cm 0cm 0cm},clip]{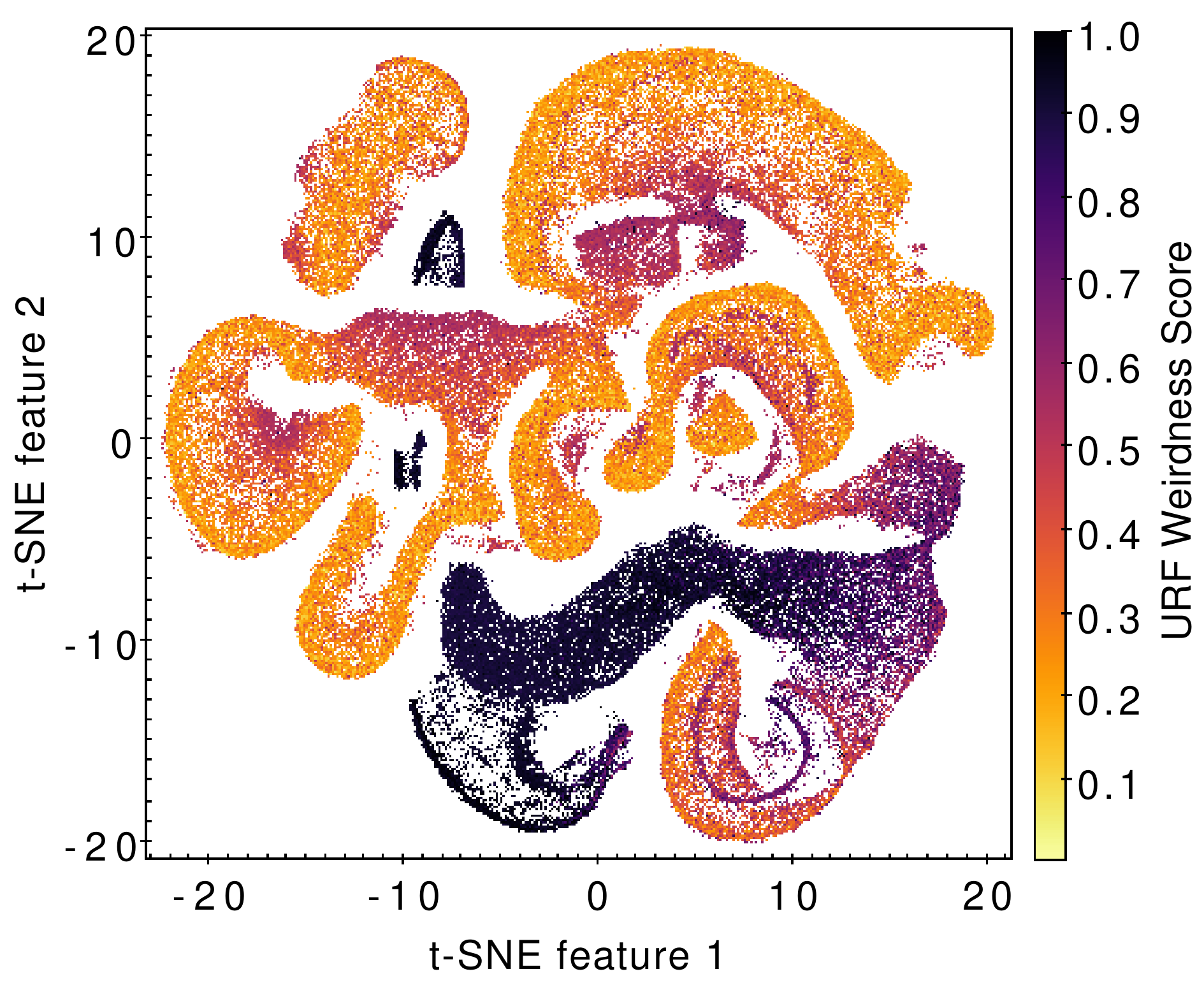}
\includegraphics[scale=0.45,angle=0,trim={0cm 0cm 0cm 0cm},clip]{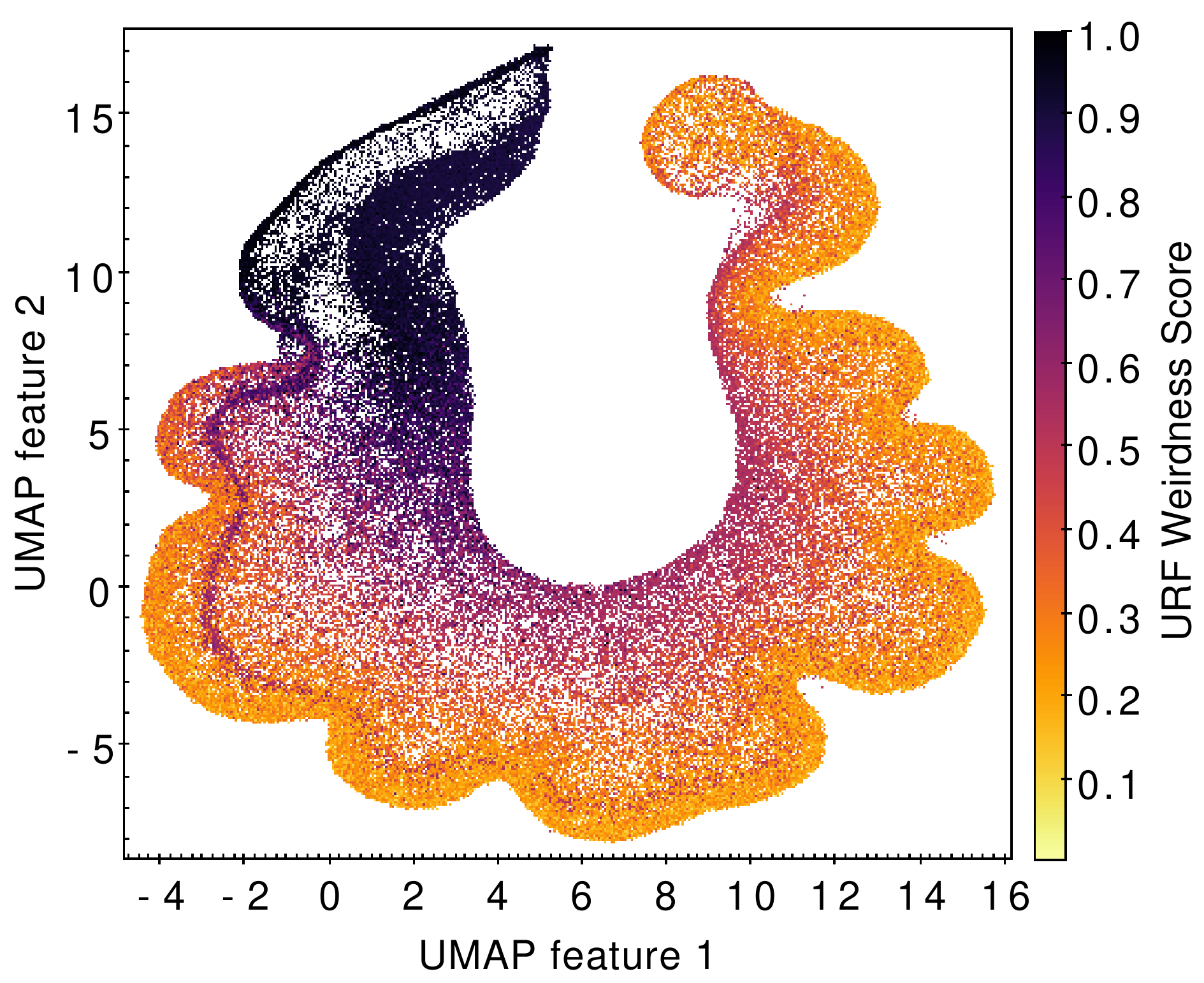}\\
\caption{\emph{Left}: The 2D embedding of the dense light curves using the \tsne~algorithm, color coded by URF weirdness score. \emph{Right}: The 2D embedding of the dense light curves using the UMAP algorithm, color coded by URF weirdness score.}
\label{fig:embeddings}
\end{figure*}

\tsne~and UMAP (as described in \autoref{sec:tsne} and \autoref{sec:umap} respectively) are separately applied to a representative feature data set, namely the \emph{DMDT} image representation of light curves introduced in \citealt{Mahabal17}, and described in \autoref{sec:dmdtmaps}. The hyperparameters used are summarized in \autoref{sec:manifolds}. Each of the two algorithms reduces the dimensionality of the data to only two dimensions, while trying to preserve similarity between light curves. We show the resulting \tsne~and UMAP embeddings for the dense light curves in \autoref{fig:embeddings}, where each point represents a light curve, and has been color-coded by its independently determined URF score. 

Despite the loss of information associated with the dimensionality reduction, the embedding maps show a complex structure that relates to the broad range of variability types present in the \emph{Kepler} dataset. Both maps show distint, differenciated regions.
The \tsne~embedding map is somewhat more complex, fragmented into different ``islands'' and showing  a less connected representation, whereas the UMAP embedding map shows a more continuous distribution and no fragmentation. Our analysis indicates that this distinction is due to the fact that \tsne~is more sensitive to baseline trends, \ie, similar objects might end up in different \tsne~islands if they look alike but have a different upward or downward trend. But both maps show a population of objects represented by dense filaments that span a considerable range of feature values.

Remarkably, both maps show a striking correlation between the location of a light curve in the 2D space of the map, and the independently determined URF score for the light curve. This suggests that the properties that make a light curve anomalous, \ie, increased amplitude variability, having certain characteristic variability timescales, or having truly unique light curve features, are to some extent \emph{embedded} in these maps. More specifically, members of the anomalous peak of the URF distribution occupy a well defined area of these two maps, roughly corresponding to the black areas in the figures. In the two-dimensional (four-dimensional if both maps are considered) space of the manifold embeddings, the degeneracies between different types of light curves having similar URF scores are now broken, allowing us to associate objects with similar characteristics. This implies that anomalies that are locally close to each other in the 4-dimensional space of these embeddings are similar between them, \ie, they have analog behaviors.

\begin{figure*}
\includegraphics[scale=0.43,angle=0,trim={0cm 0cm 0cm 0cm},clip]{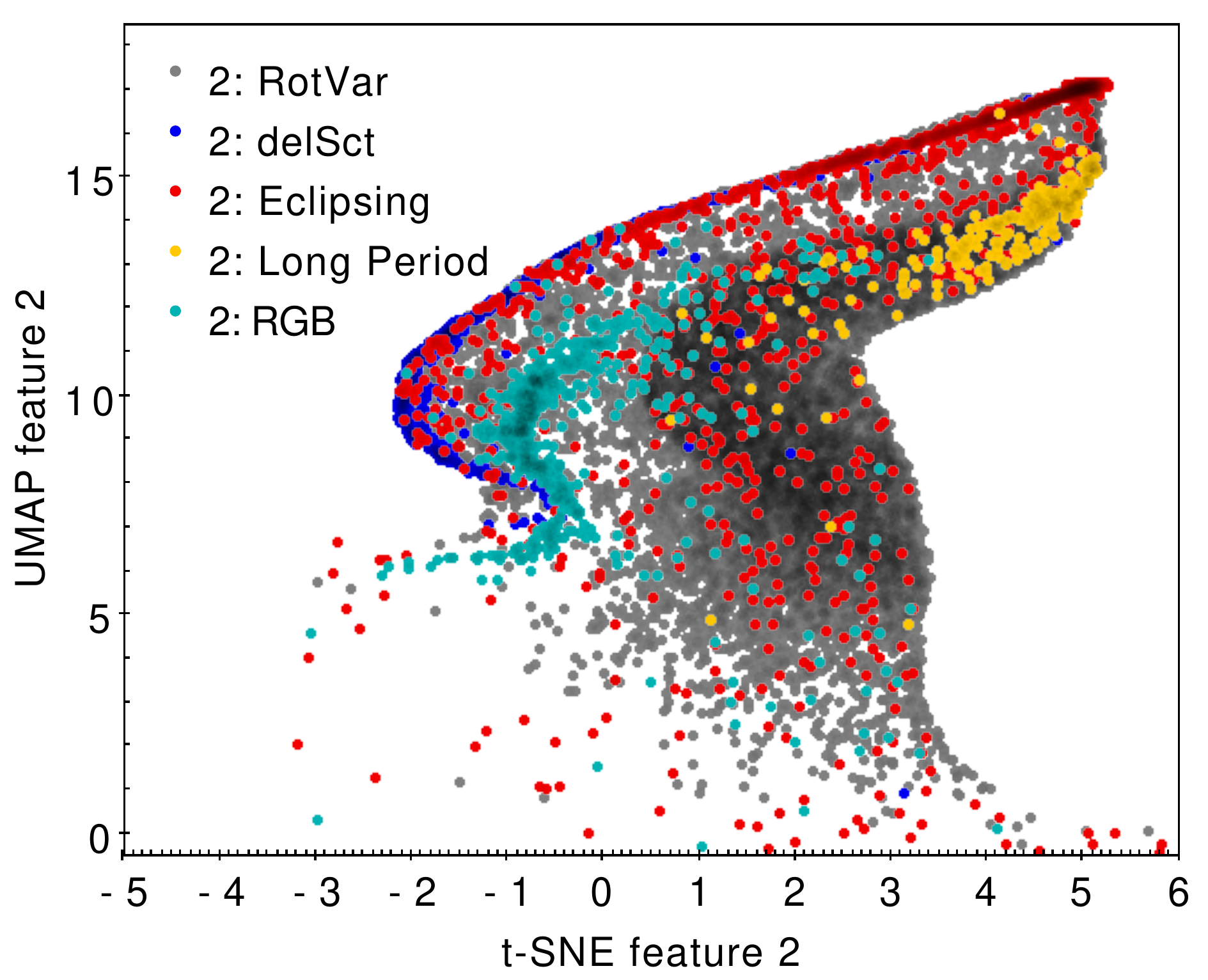}
\includegraphics[scale=0.44,angle=0,trim={0cm 0cm 0cm 0cm},clip]{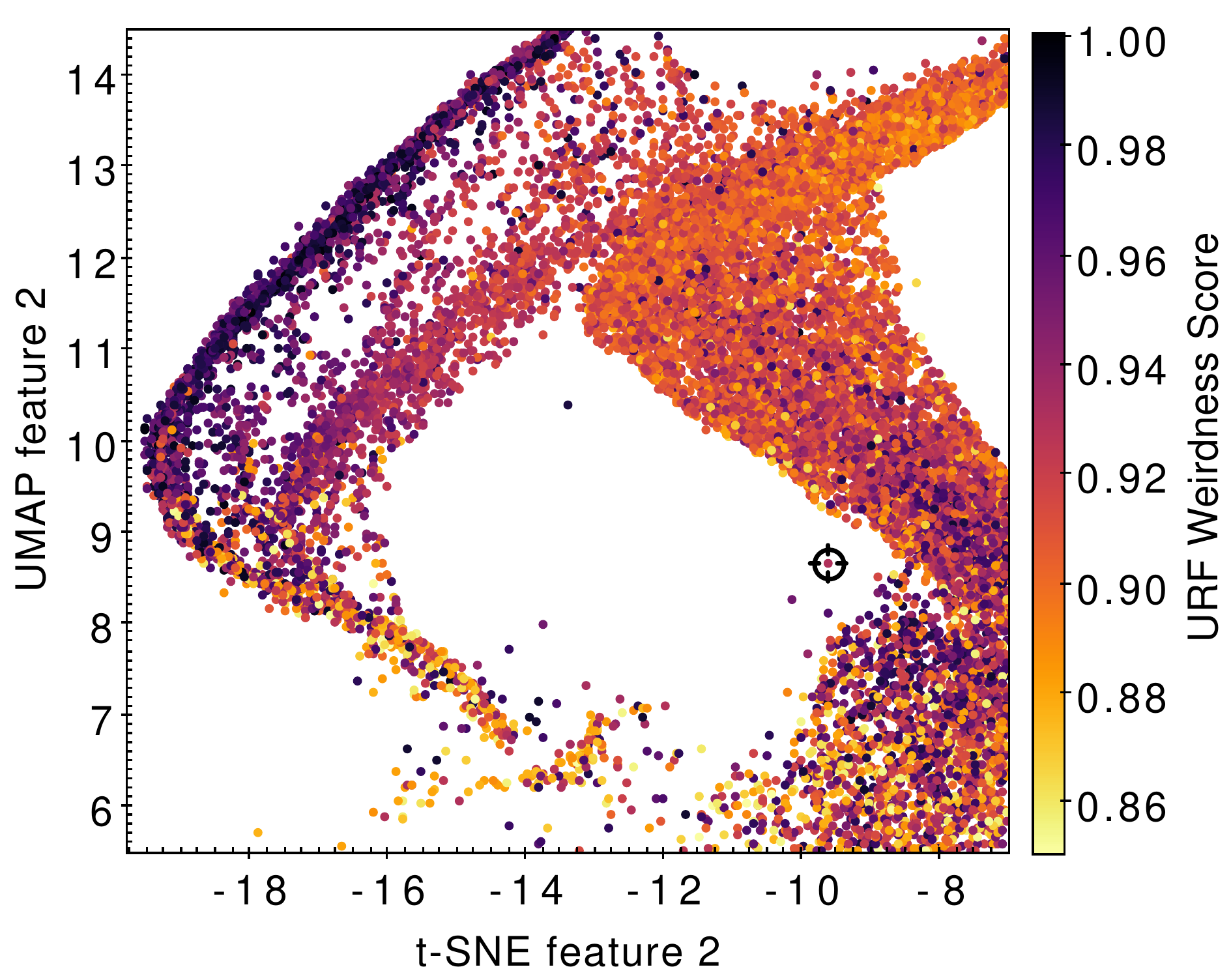}\\
\caption{\emph{Left}: The anomalous tip of the UMAP embedding map, with different types of anomalous objects (rotating variables, $\delta$-Scuti stars, eclipsing binaries, long period variables and oscillating RGB stars) indicated, from a crossmatch of our object list with the SIMBAD database. \emph{Right}: A particular projection of the 4-dimensional space of the four combined embeddings, in the region close to the location of Boyajian's star, which is indicated by the cross mark. Shown are only points corresponding to anomalies, and the color coding has been scaled to include only values in the anomalous peak of the distribution.}
\label{fig:embedded_anomalies}
\end{figure*}

To illustrate the last point, in the left panel of \autoref{fig:embedded_anomalies} we show the anomalous tip of the UMAP embedding map, this time explicitly indicating different types of known anomalies that we have identified by cross-matching our list of objects with the SIMBAD database. These classes closely match the types included in our list of \emph{bona-fide} anomalies. We note that objects of the same class tend to cluster together along well differentiated regions, or filaments of the map. The clustering is not exclusive, and there are relatively small overlaps between types, but considering the existing ambiguity in the classification of the sources this can be expected. Where, in this reduced space, do we find very unique anomalous objects, such as Boyajian's star? In both the \tsne~and UMAP embeddings of \autoref{fig:embeddings}, Boyajian's star falls within the group of anomalous objects dominated by rotating variables,  which in the \tsne~mapping corresponds to the elongated shape located at $y\sim-10$ and spanning the range $-8<x<-5$. In each individual map it is difficult to distinguish it from other, less remarkable anomalies. Its truly unique nature comes to light, however, if we move to the 4-dimensional space of the combined \tsne~and UMAP features. 

As illustrated in the right panel of \autoref{fig:embedded_anomalies}, Boyajian's star sits alone in an isolated region of the map, easily distinguishable from other anomalies, that in this region are mostly rotating variables. The two isolated anomalies closest  to Boyajian's star in this region are KIC 5385778, which just like Boyajian's star show a combination of dips and long-term variability (although the dips are regular and far less pronounced), and KIC 3544657, which displays long-term irregular variability and shallow, irregular dips. They are, however, too separated from each other to be considered true analogs. They are all unique in their own merits. Using a similar approach, anomalous objects that belong to a particular rare class (\eg, RRLyrae stars\footnote{RRLyrae stars sit at the very tip of the anomalous region in the UMAP embedding map. Whereas in the general context of all variable stars they are not rare, they are extremely rare in the dataset considered here.}) and that have not yet been classified, can be found.

In Appendix A, \autoref{tab:delScu}, \autoref{tab:oscBin}, \autoref{tab:erupRGB}, and \autoref{tab:LPV},  we provide small samples of the selected anomalies, that we have grouped by their probable class. In order to associate anomalies of unknown class to a specific labelled group, we first identified regions in the UMAP embedding with high density and high purity of a given class (\eg, Long Period Variables occupy a very specific region in the space of UMAP embeddings in \autoref{fig:embedded_anomalies}). We then manually selected objects in this region of the 2D map that had an anomaly score above 0.85, and that also did not have a previous classification according to the SIMBAD database. The labels we have assigned in the tables should be taken only as indicative, as no formal classification has been performed on these anomalies. 

\subsubsection{Distinguishing Euclidean anomalies in \tsne~and UMAP} \label{sec:manifold_anomalies}

The fact that unique anomalies such as Boyajian's star appear isolated in the four-dimensional space of the \tsne~and UMAP embeddings suggests that the Euclidean distance in this space could be used as an anomaly score, at least for some types of light curves. We now investigate what types of anomalies can be identified based on a distance metric in the embedding maps, and whether those anomalies are similar to those found with the URF method.

Since the \tsne~and UMAP methods work by closely simulating the probability distribution $q_{i|j}$ of point $i$ picking point $j$ as a neighbor in the high dimensional space, we can build a distribution based on the pairwise Euclidean distances between points in the 4-dimensional space of the combined \tsne~and UMAP embeddings. Specifically, for each point we consider its distance to the nearest neighbor. We then consider the distribution formed by normalizing these distances and take all points that are at least $3.4\sigma$'s from the mean in order to set a threshold for anomalous objects. The specific value of $3.4\sigma$ was chosen heuristically by inspecting the number of anomalies as a function of the threshold, and choosing the value at which the second derivative of this curve was closest to zero for both the \tsne\ and UMAP.  The points selected are those that are furthest from the cluster and, by the Euclidean metric, they are anomalies in the manifold-based method. 

\begin{figure}
\includegraphics[scale=0.58,angle=0,trim=0.2cm 0.2cm 0cm 0.9cm,clip=true]{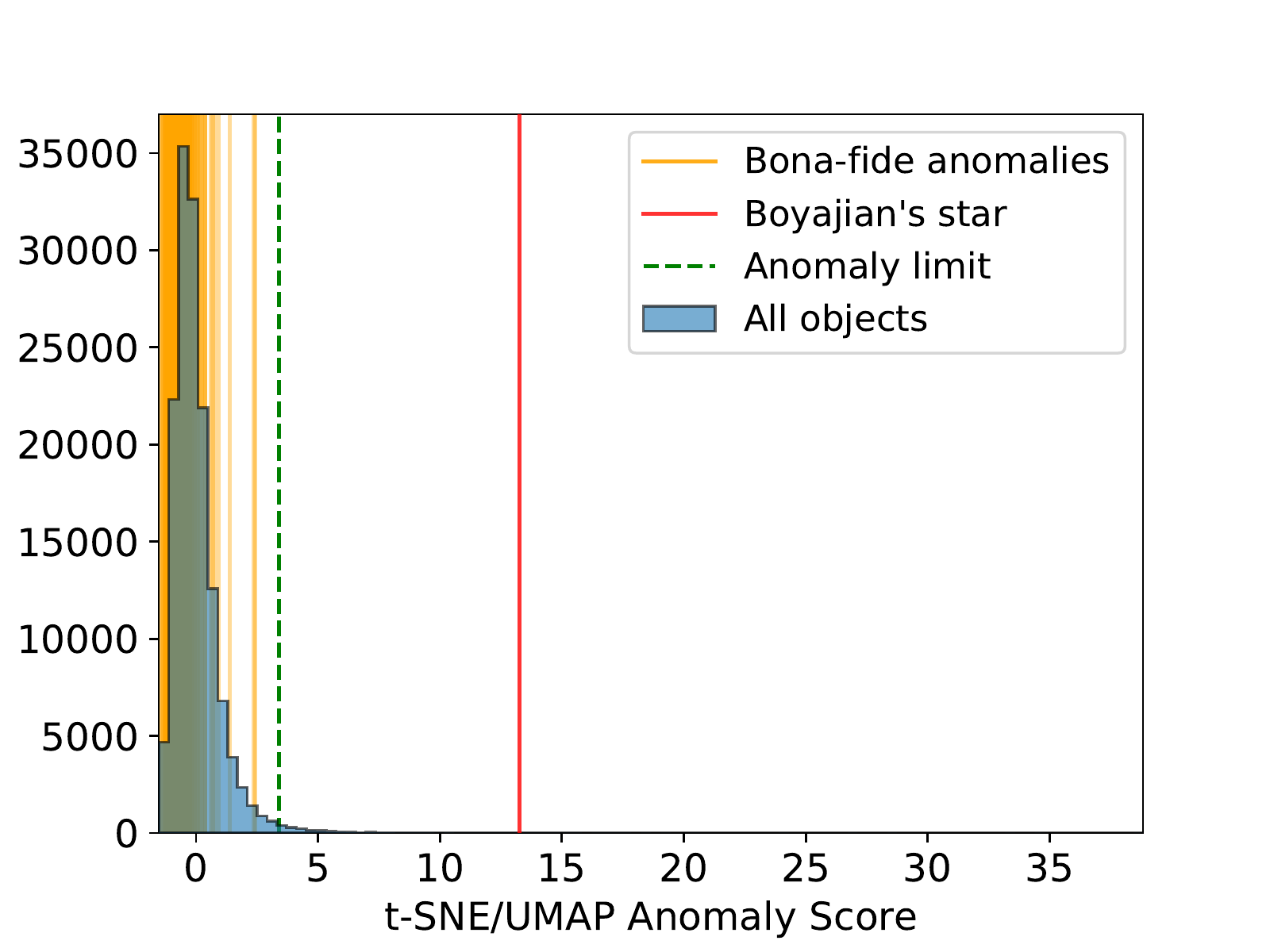}
\includegraphics[scale=0.58,angle=0,trim=0.2cm 0.2cm 0cm 0.9cm,clip=true]{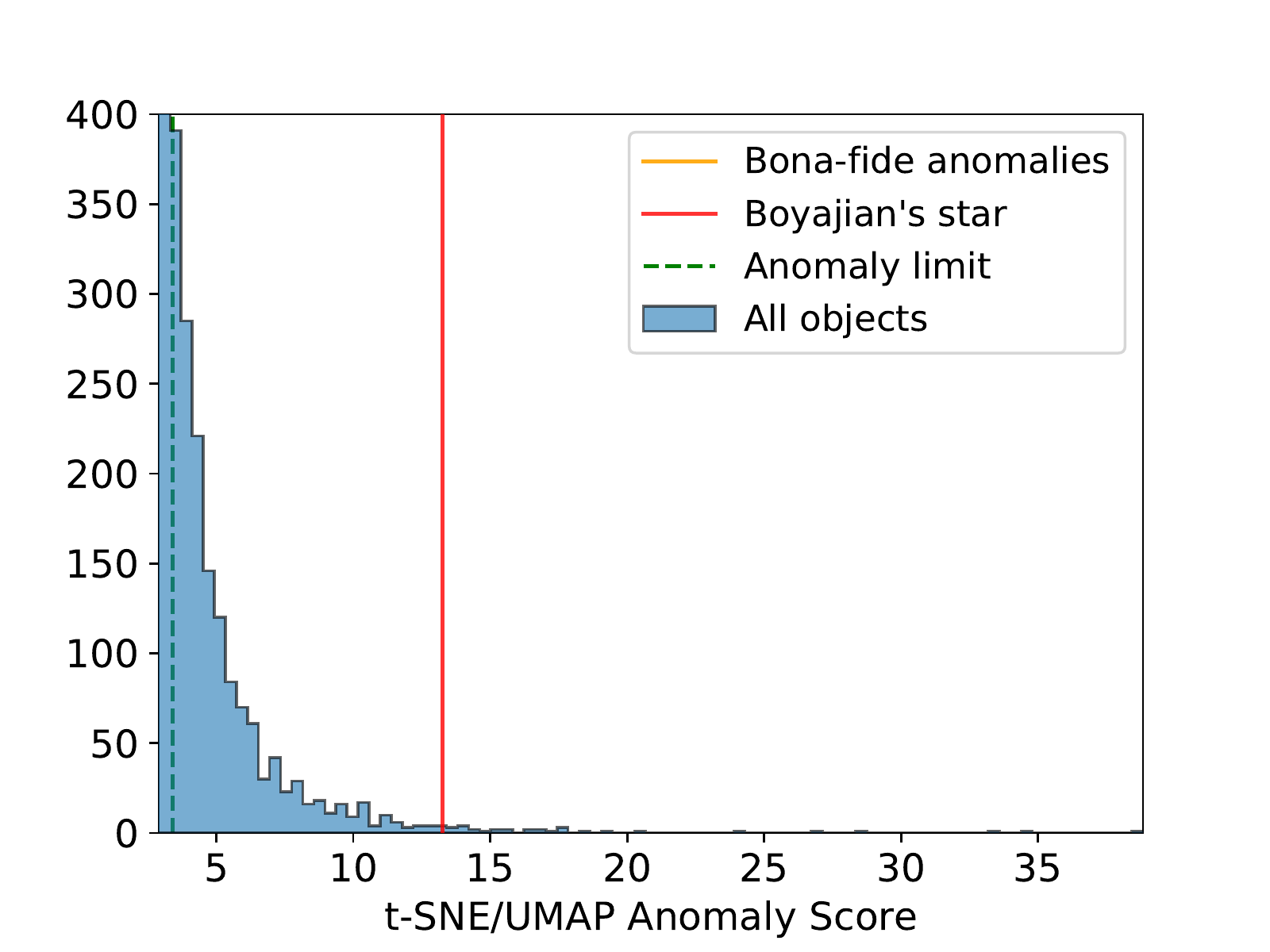}
\caption{Histograms of manifold anomaly scores for the dense Kepler Q16 light curves. Indicated in orange and red are the scores corresponding to \emph{bona-fide} anomalies, including Boyajian's star, as well as the anomaly limit (in green). The top panel shows all the objects, whereas the bottom panel shows the high anomaly score region of the distribution.}
\label{fig:manifold_scores}
\end{figure}

\begin{figure*}
\includegraphics[width=1.02\textwidth,trim=3.8cm 1.1cm 3cm 2cm,clip=true]{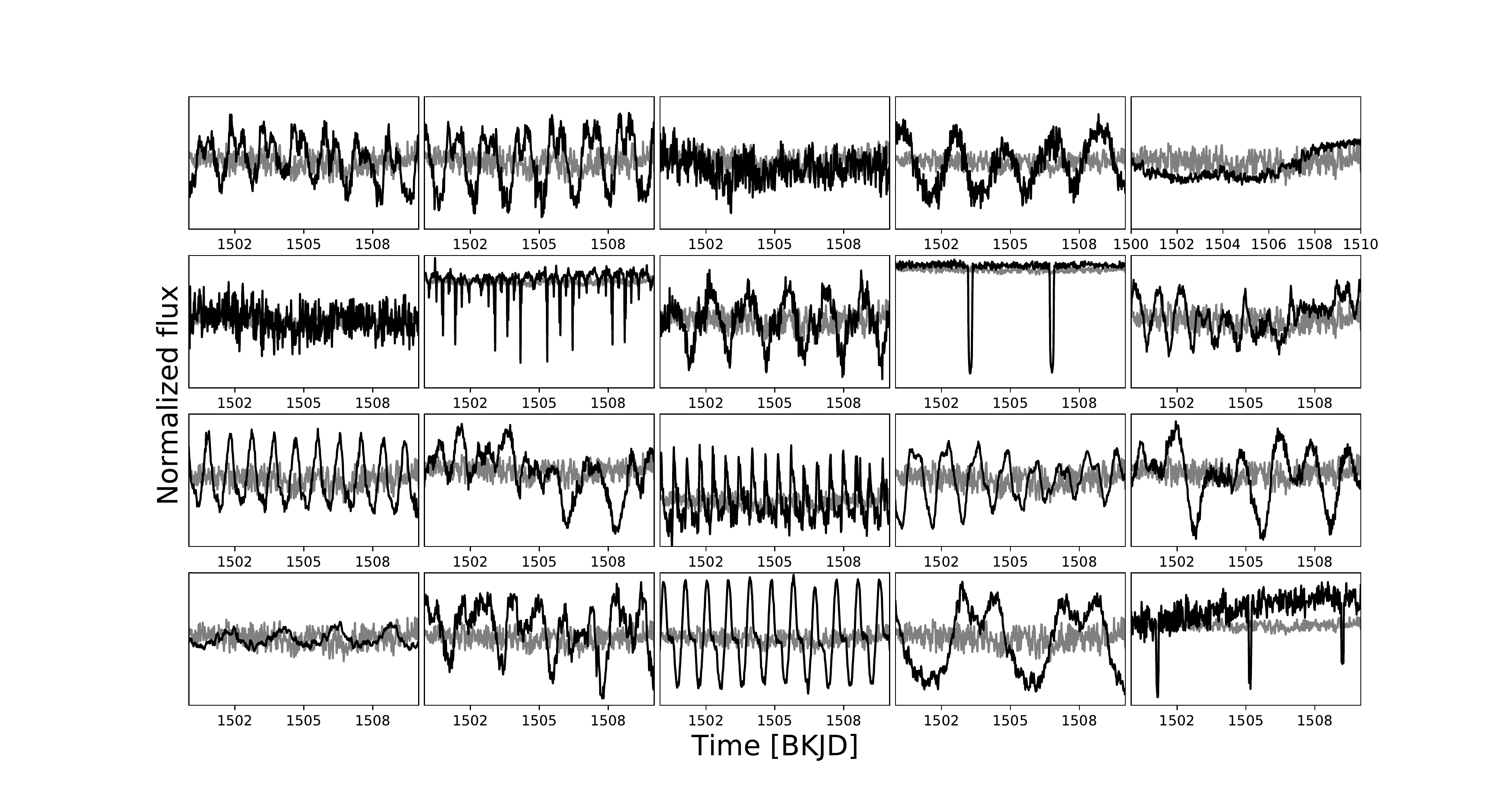}
\caption{Examples of anomalous light curves identified using the Euclidean distance in the manifold embeddings of the \emph{DMDT} maps. The anomalies are shown as black lines, whereas the gray line represents the reference object KIC 8211660}
\label{fig:manifold_anomalies}
\end{figure*}

In \autoref{fig:manifold_scores} we show the distribution of manifold anomaly scores computed as described for the dense light curves. In a similar manner as we did in \autoref{fig:weirdness_hist}, we indicate the scores of Boyajian's star and the rest of the \emph{bona-fide} anomalies with vertical lines. We note that this time the majority of our \emph{bona-fide} anomalies (in fact, all of them except for Boyajian's star) fall outside of the anomaly area, with many of them having fairly low scores. It is not too surprising that the methods may find different anomalies: \tsne\ and UMAP use the \emph{DMDT} image pixels as the features, which contains different information compared to the light curves-periodogram combinations. Second, as pointed out in the introduction, Euclidean distances become less meaningful as the dimension of the data grows, and thus when a high dimensional space is reduced to only a few dimensions with manifold approaches like the ones used here. Yet, Boyajian's star is clearly identified using this approach: Euclidean distances are still meaningful locally, and in fact we preferentially find certain types of isolated anomalous objects using distances between manifold features (see \autoref{fig:manifold_anomalies}). 

However, in the more general case, we are unable to know how many anomalies we are missing when we select them based on the Euclidean metric alone. We can understand this in terms of individual, or \emph{one-off} anomalies versus groups anomalies. The Euclidean distance identifies \emph{one-off} anomalies, isolated sources of which only one example exists, as well as anomalies that have only a few outlying features in an otherwise regular light curve (as in the case of rapid exoplanet transits), but it fails to identify groups of anomalies, such as RR Lyrae stars in the present Kepler dataset, whose anomalous behavior is more related to the overall variation of the light curve, taking into account many of the features at once. The URF method, on the other hand, successfully identifies both group anomalies with overall different feature distributions, and \emph{one-off} anomalies. A similar conclusion applies for the \tsne~and UMAP embeddings when the sparse light curves are considered. The correlation between the location of a light curve in the embedding map and its URF score is still present for these sparse light curves. But the general idea about using Euclidean metrics for identifying \emph{one-off} anomalies still holds. 

We find a total of 1552 anomalous dense light curves using the 4-dimensional Euclidean metric and the diagnostic described above. \autoref{tab:manifold_anomalies} list the first 20 when they are ordered by the Euclidean score. The manifold anomaly scores for those object do not correlate with the URF scores, with some of the manifold anomalies having low URF scores (and vice-versa). But the most remarkable manifold anomalies are also selected in the URF method. Boyajian's star, as expected, ranks high in the manifold Euclidean score, at position 33 from the top most anomalous object. In \autoref{fig:manifold_anomalies} we show some representative examples of anomalies being selected by the manifold method, but not selected in the URF method. They are all in the top 150 of Euclidean scores. We note that, unlike those identified with the URF method, these anomalies are not as clearly dominated by large amplitude variations. Deviations from the mean normalized flux are not as pronounced as in the URF anomalies of Figures \ref{fig:anomalous_lc_known1}-\ref{fig:anomalous_lc_unknown}. Instead, they appear to be dominated by complex periodic or semiperiodic behavior, either in the form of fast dips, such as those happening in transits with a typical duration comparable to or less than the Kepler cadence, or of periodic oscillations with more than one mode and characteristic timescales of a few days, which incidentally is the timescale that the URF method deems as less important (see \autoref{fig:feature_imp}). This $\sim$days-long periodicity, together with the fast transits, account for the vast majority of the manifold anomalies. The fact that these anomalies are being selected is probably due to the presence of features in the DMDT maps associated with these periodic behaviors, although it is difficult to interpret how those patterns reveal themselves in the 2D embeddings to which the images are eventually reduced. As a matter of fact, the \emph{DMDT} maps for these anomalies do not look very different from the randomly selected maps of \autoref{fig:tsne_dense_dmdt_images}. This suggests that the 4-dimensional Euclidean anomaly metric selects anomalies based on subtle \emph{DMDT} map differences between a given anomaly and the average map of the objects that are most like it. Sudden rapid dips in a light curve that is otherwise ordinary are a good example of this type of anomaly.

\begin{figure}
\includegraphics[scale=0.7,angle=0,trim={1.5cm 3.5cm 8cm 3.5cm},clip]{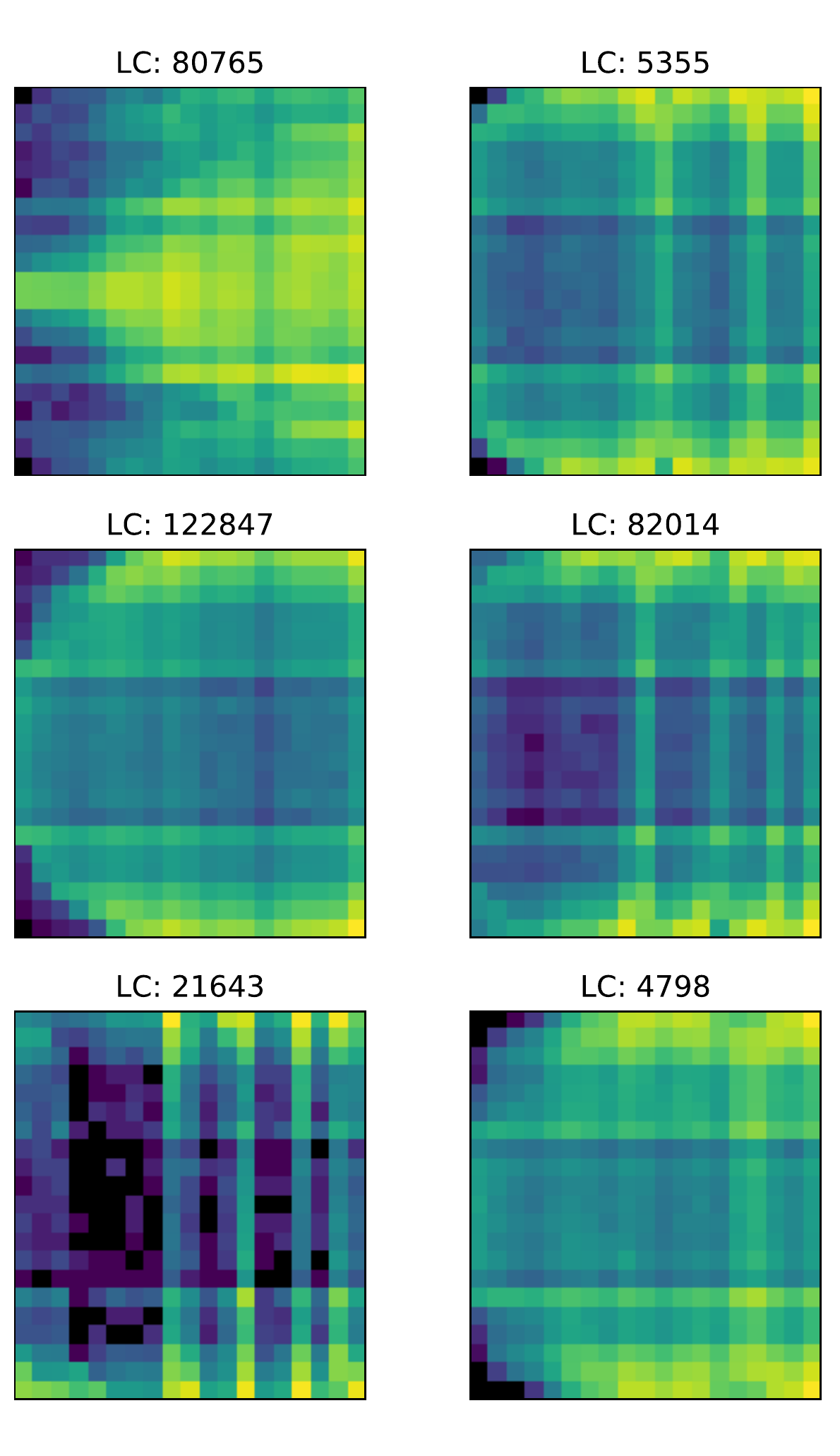}
\caption{DMDT maps for a sample of our \emph{bona-fide} anomalies. Shown clockwise from upper left are Boyajian's star, a classical Cepheid, a $\gamma$-Dor variable, a slow pulsating B star, an RRLyrae star,and an ellipsoidal binary.}
\label{fig:dmdt_anomalies}
\end{figure}

In contrast, the \emph{DMDT} maps of URF anomalies, of which we show a few in \autoref{fig:dmdt_anomalies} show a more complex structure: they are more widely populated in the vertical DM axis (which is explained by the fact that they are amplitude-dominated), and show vertical and horizontal stripes representing either periodicities or magnitude dips. The manifold dimensionality reduction using either \tsne~or UMAP is very successful at translating those features into groups of similar, or analog light curves (which is why we see a correlation between the URF anomaly score and the 2D embeddings), but not very successful at identifying those complex features as anomalies. The Euclidean metric, we argue, is a good indicator of \emph{local} anomalies (\ie, those with small deviations from an otherwise normal light curve). The additional information that we gain about light curve behavior from the spectral power features in the URF method appears to be fundamental in the identification of a broader range of anomaly types, including those selected \emph{a priori} as our \emph{bona-fide} anomalies (RR Lyrae stars, Cepheids, etc.), that we know to be rare in the KIC. We therefore conclude that the URF score that we have derived in \autoref{sec:urf_anomalies} for our set of objects is better suited for finding all types of anomalies, both \emph{one-off} cases with no counterparts (either because they show a few outlying features, or because they distribute differently in the joint space of all features), and groups of objects on known type that are rare in a given dataset.

\subsubsection{The full algorithm}
\begin{figure}
\begin{tikzpicture}[node distance=2cm]

\node (start) [startstop] {Start};
\node (in1) [io, below of=start] {Kepler};
\node (pro1) [process, right of=in1,xshift=3cm] {Normalize};
\node (pro2) [process, below of=pro1] {Spike Removal};
\node (pro3a) [process, left of=pro2,xshift=-3cm]{Periodograms};
\node (pro4) [process, below of=pro3a]{URF};
\node (out1) [io, right of=pro4,xshift=3cm] {Score};
\node (dec1) [decision, below of=out1,yshift=-0.3cm] {Scores only?};
\node (stop1) [startstop, below of=dec1,yshift=-0.3cm] {End};
\node (pro3b) [process, left of=dec1,xshift=-3cm]{DMDT maps};
\node (pro5) [process, below of=pro3b,yshift=-0.3cm]{UMAP/t-SNE};
\node (out2) [io, below of=pro5] {Embeddings};
\node (pro6) [process, below of=out2]{Search Neighbors};
\node (stop) [startstop, below of=pro6] {End};

\draw [arrow] (start) -- (in1);
\draw [arrow] (in1) -- (pro1);
\draw [arrow] (pro1) -- (pro2);
\draw [arrow] (pro2) -- (pro3a);
\draw [arrow] (pro3a) -- (pro4);
\draw [arrow] (pro4) -- (out1);
\draw [arrow] (out1) -- (dec1);
\draw [arrow] (dec1) -- node[anchor=east] {yes} (stop1);
\draw [arrow] (dec1) -- node[anchor=south] {no} (pro3b);
\draw [arrow] (pro3b) -- (pro5);
\draw [arrow] (pro5) -- (out2);
\draw [arrow] (out2) -- (pro6);
\draw [arrow] (pro6) -- (stop);

\end{tikzpicture}
\caption{A flow chart showing how our proposed algorithm works, both for finding anomaly scores only, and for finding analogs of light curves of interest.}
\label{fig:flow_chart}
\end{figure}
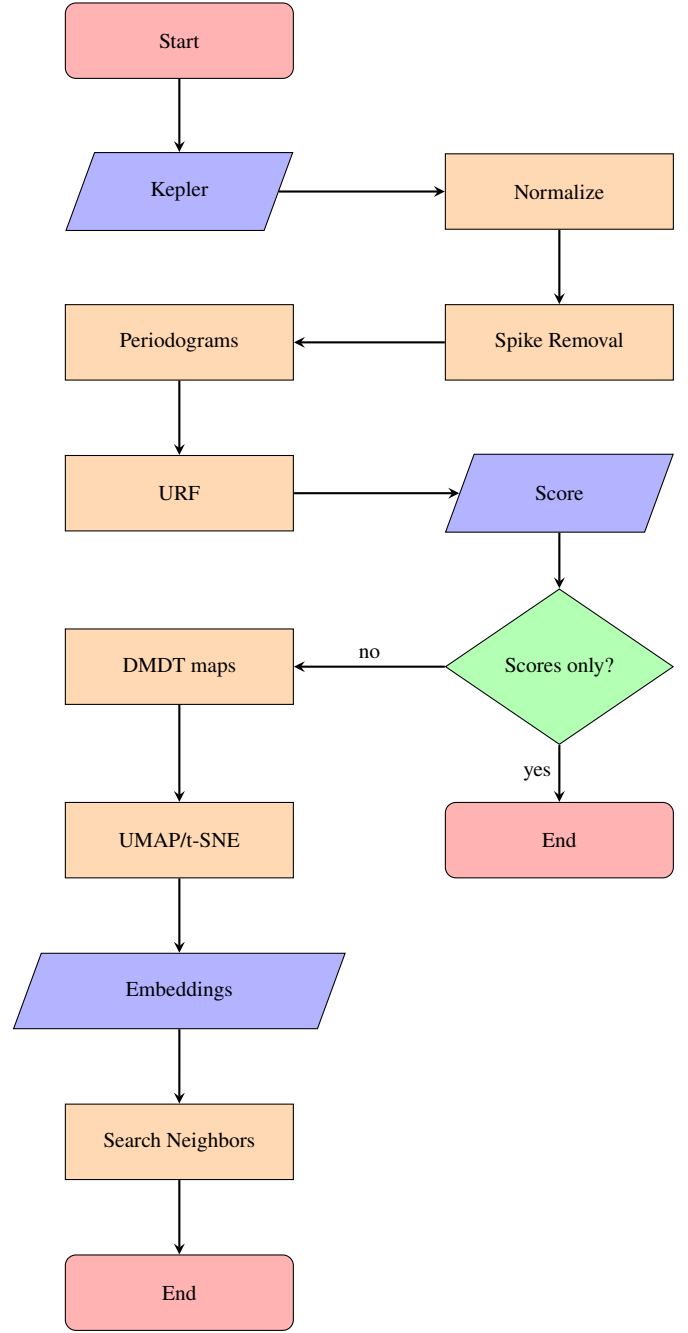

We now present an integrated view of our algorithm. Depending on the specific interest of the user, the method proposed here can be used in two modules in order to explore time-domain data. The user can choose to only find anomalies using the URF method applied to the light curve and periodogram points. But the user can in addition find analogs to anomalous light curves of interest by performing dimensionality reduction to the DMDT maps generated from the light curves, and then look for neighbors of the light curve of interest in the space of low dimensionality. In \autoref{fig:flow_chart} we show a flow chart that summarizes the method.

We could have adopted the perfectly viable option of combining all the features that we have derived, including light curve points, periodograms, and DMDT maps, and run the URF algorithm on the resulting embeddings. This would have resulted in different scores, and potentially different anomalies. However, we have demonstrated in \autoref{fig:embedded_anomalies} that the URF scores and the \tsne/UMAP embeddings are correlated, and therefore contain similar information. The advantage of splitting the analysis in two parts (anomaly identification with URF using the periodogram and light curve points, plus analog identification with the \tsne/UMAP embeddings using the DMDT maps) is that we avoid redundancy (given the correlation already mentioned) while allowing for modular analysis. For example, a user can decide to perform only the anomaly detection part using URF, for which calculating the periodograms alone is less computationally expensive than calculating both the periodograms and the DMDT maps. Similarly, a user can opt for getting the DMDT maps, run them through t-SNE/UMAP, and look for light curves that are similar based on their proximity in these embeddings.

\subsection{Astrophysical implications}
\label{sec:implications}

We now turn to the question of whether there is a correlation between anomalous variability of an object and its overall physical properties. Our results indicate that anomalous light curves can be associated to known astrophysical processes, such as intrinsic stellar pulsations of different types \citep[\eg\ ][]{arras06}, flares \citep{paduel18, 2014ApJ...797..122D}, or extrinsic binary eclipsing phenomena \citep{prsa13, prsa16}, but also to phenomena that are a challenge to current models, such as the case of Boyajian's star. Therefore, you would expect that there is some correlation between the URF score and the luminosity and temperatures of these objects.

\begin{figure*}
\includegraphics[scale=0.45,angle=0,trim={0cm 0cm 0cm 0cm},clip]{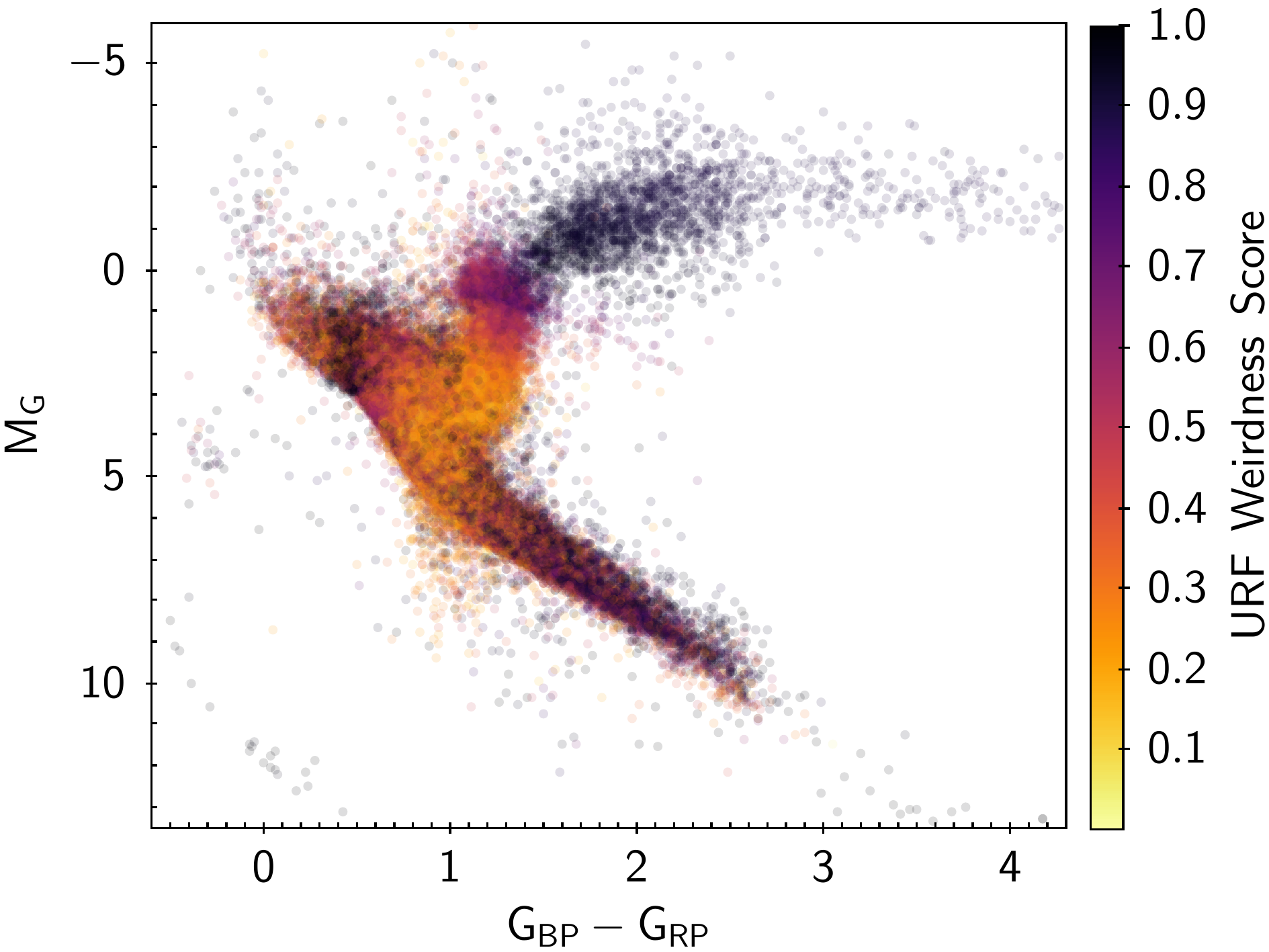}
\includegraphics[scale=0.45,angle=0,trim={0cm 0cm 0cm 0cm},clip]{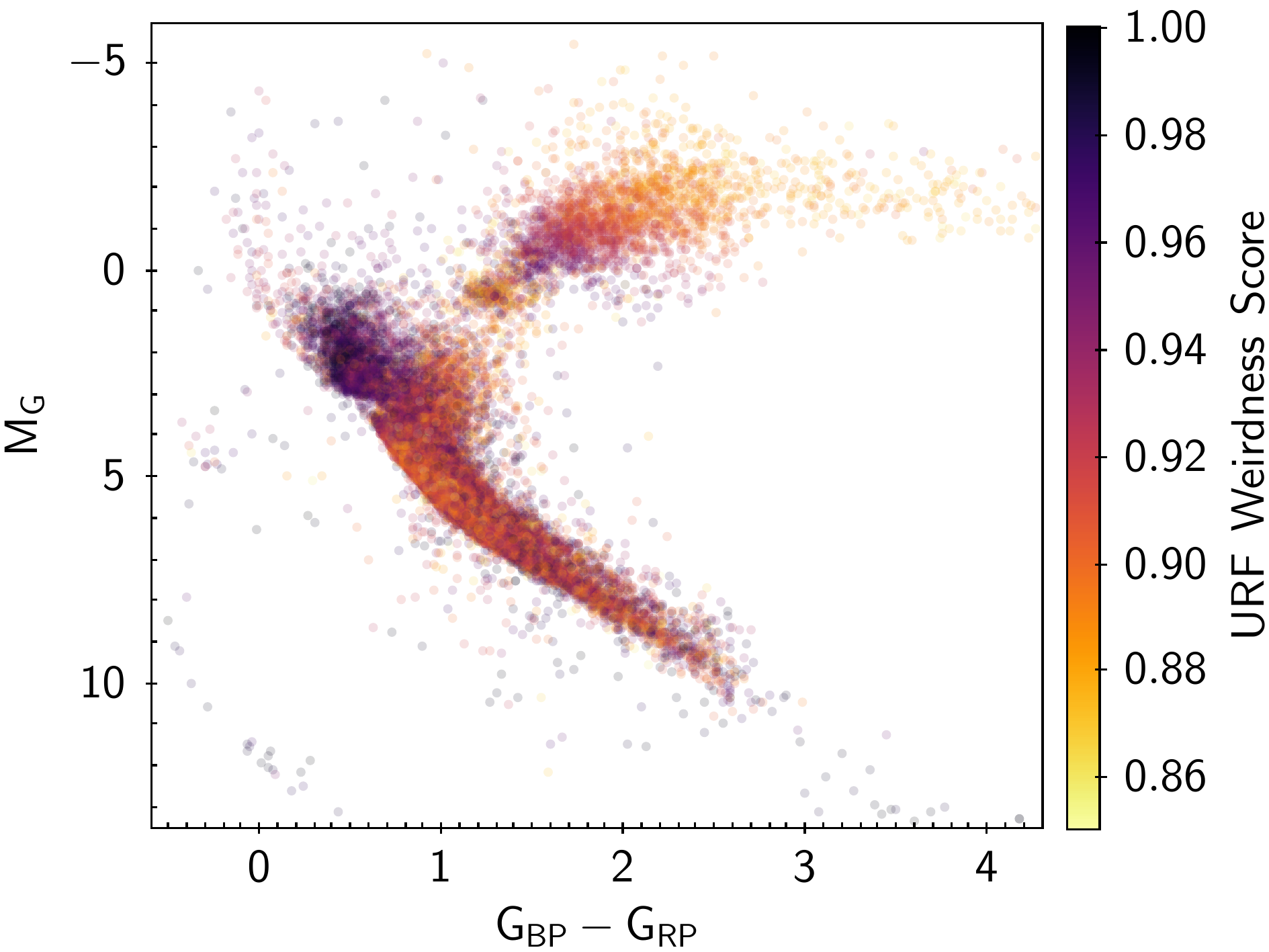}\\
\caption{\emph{Left}: The Hertzprung-Russell diagram for those stars in our set with measured \emph{Gaia} magnitudes and parallaxes, color-coded by the URF score for the dense light curves. \emph{Right}: The diagram for the anomalous objects, as determined using the dense light curves (URF score larger than 0.85). The colorbar has been renormalized to contain only values larger than 0.85.}
\label{fig:hr_diagram}
\end{figure*}

In order to investigate this, we have obtained \emph{Gaia} DR2 colors and derived G band magnitudes \citep{gaia1, gaia2} for all the objects in our sample with available \emph{Gaia} measurements, and produced a Hertzprung-Russell diagram that we have color-coded according to the URF anomaly score. We show this diagram in \autoref{fig:hr_diagram}. The left panel shows the entire set of objects analyzed in this paper, whereas the right panel shows those objects that we consider anomalous, with a URF score larger than 0.85. 

A clear correlation is revealed between astrophysical properties and anomalous variability. The majority of anomalies lie along the main sequence (MS), but those with the highest URF score are not preferentially located along the MS. Instead, pulsators in the instability strip such as $\delta$-Scuti stars, $\gamma$-Doradus stars, and RR Lyrae stars take the first place among the most anomalous objects, followed by eruptive stars and red dwarfs with high amplitude stochastic variability. White dwarfs and hot sub-dwarfs also belong to the group of the most anomalous objects, also due mostly to their high amplitude variations. Then come most of eclipsing binaries and rotational variables, with the latter being the group most represented along the main sequence. The last place among the anomalous is taken by irregular pulsating giants, long period variables, and Mira stars, in that order. The URF score gradient for the anomalies is clearly seen in the right panel of \autoref{fig:hr_diagram}.

The point density of objects is different between the two diagrams in \autoref{fig:hr_diagram}, which indicates that anomalies distribute differently in their physical properties with respect to the general population of KIC objects. With respect to the latter, there is an overdensity of anomalies in the instability strip, where pulsations due to instabilities in stars with ionized He atmospheres are common. In addition, anomalies cluster more strongly compared to the KIC at lower luminosities along the main sequence. This might be related to the larger incidence of eclipsing binaries and flaring young stars at late spectral types.

Perhaps the most unexpected anomalies are those that lie along the MS, and for which no particular strong pulsation modes or high amplitude variations are expected, and that are not rotational variables or eclipsing binaries. These objects, of which Boyajian's star is an example, are local anomalies that can be identified using an Euclidean metric in the space of manifold embeddings, as we have discussed. The correlation between URF anomaly score, embedded low-dimensionality features, and astrophysical properties indicates a possible path toward new discoveries. In particular, the application of a similar analysis to TESS light curves will allow us to identify scientifically compelling light curves for further analysis. We are currently performing this analysis on TESS data (Crake et al. in prep.). Note that this type of analysis is not limited to light curves with regular cadences. Our experimental setup allows us to adjust for differences in light curve lengths, cadences, and gaps between light curve points. Both the spectral decomposition and the \emph{DMDT} images can be produced for light curves of any length and cadence.

Red clump stars deserve a special mention. They are seen in the left panel of \autoref{fig:hr_diagram}, as the purple clump of objects at higher temperatures and lower luminosities compared to the anomalous giants. We have seen that they also stick out in the distribution of anomaly scores, having URF scores of about 0.55. These are RGB stars supported by helium fusion on their core. They tend to clump in the HR diagram because they end up having the same luminosity at their red giant stage regardless of their initial age or composition \citep{girardi16}. In terms of variability, they have amplitude variations that are less pronounced compared to other RGB stars (but still clearly seen in Kepler light curves), and these amplitude variations are also less regular compared to the RGB pulsations. In the UMAP embedding of the right panel in \autoref{fig:embeddings}, they live in the purple filament seen to the left, just below the anomalous filament of the instability strip pulsators.

\section{Discussion and conclusions}\label{sec:discussion}

Anomaly detection in time domain astronomy enables the discovery of unique light curves that could lead to the formulation of new hypotheses. Such is the case of the anomalous object Boyajian's star, whose \emph{Kepler} light curve has become the gold standard in the search for anomalous light curves. The search for anomalies is even more relevant in the era of \emph{TESS} and the Vera C. Rubin Observatory Legacy Survey of Space and Time (LSST), which will deliver thousands of light curves per night over a period of 10 years. Yet, it would seem a very difficult task to define an anomalous light curve, as the anomalous nature of an astronomical time series depends on the details of the methods employed, as well as on the features that are used to represent the light curve ensemble. Here we have proposed an approach for the detection of time-domain anomalies and of their analogs, and we have applied it to \emph{Kepler} data. We have accomplished the following:

\begin{enumerate}
    \item By comparing our candidates for anomalous light curves with a list of \emph{bona-fide} objects that we knew in advance to be rare in the context of the Kepler Input Catalog (KIC), we have shown that we can isolate rare and unique light curves, including Boyajian's star and a broad range of variable stars that are underrepresented in the KIC. 
     \item We provide an empirical definition: an anomaly is an object whose most relevant features (\ie\ those that most efficiently reduce impurity during the random forest classification), distribute differently compared to the corresponding features of non-anomalous objects. (\autoref{fig:feature_imp}, \autoref{fig:drives_weirdness}).
     \item Starting from a simple set of features, namely the light curve points and their power spectra, we provide a specific measure for anomalous behavior, the URF score, whose distribution over the entire dataset defines a clear set of anomalies, including at least 5000 anomalies that have not been previously classified in the KIC (\autoref{fig:weirdness_hist}).
     \item Using an image representation of the light curves and manifold methods for dimensionality reduction, we break the degeneracy in the one-dimensional URF score, and provide a nearest-neighbors method to group objects with similar characteristics, facilitating the identification of analogs to interesting anomalies (\autoref{fig:embeddings}, \autoref{fig:embedded_anomalies}).
     
     \item Using Euclidean distances in the four-dimensional space created by assembling the \tsne\ and UMAP projections, group anomalies (such as rare known stellar types) are not identified, but \emph{one-off} anomalies are clearly identified. The latter can be truly unique stars that, like Boyajian's star, that have no counter part in the sample, or otherwise regular objects with a few extremely outlying features.
\end{enumerate}

Along the way, we have also demonstrated the impact of pre-processing on anomaly detection, as observational or instrumental artifacts such as spurious spikes or bad pixels can have a significant impact on the anomaly score (\autoref{fig:ITscorecorrelation}, \autoref{fig:LCSNR}). We have also shown that our method can be applied to sparser, unevenly sampled light curves by deploying it on a version of the light curves with only 10\% of the points and demonstrated it is still possible to find some of the relevant anomalous objects, while missing anomalies that are selected based on large-scale differences in their power spectra (\autoref{fig:freq_separation}). This is relevant to extend our work to ground-based surveys.

Our work is complementary to previous methods that have attempted to identify time-domain anomalies in \emph{Kepler} data (\eg\  \citealt{Giles19}) and incorporates new features that allow for the identification of analogs.

The remarkable correlation between the location of a light curve in the space of \tsne~and UMAP embedded 2-dimensional features, and the independently determined URF weirdness score, indicates that the apparent complexity of the Kepler light curves, with a broad range of frequencies, amplitudes, transient behavior, and periodicity, can be represented by a handful of numbers, with anomalies occupying specific regions in these representations. In particular, we have shown that ``group'' anomalies such as RR Lyrae stars, instability strip pulsators, long period variables, live together in specific areas of this lower-dimensional space, whereas ``isolated'' anomalies, such as Boyajian's star, stick out locally as objects without neighbors in the vicinity of the embedded maps. 

Finally, we have demonstrated that there is a linkage between anomalous behavior and astrophysical properties (\autoref{fig:hr_diagram}). We have shown that, although \emph{Kepler} anomalies live in many different region of the HR diagram (most of them along the main sequence), the distribution of their astrophysical properties is different compared to the bulk of the objects. We have also shown that ``group'' anomalies live in physically meaningful regions of the HR diagram, such as the instability strip and the red giant branch, but that the most fertile ground for discovery lies within the realm of individually isolated anomalous light curves of objects that live in otherwise uneventful regions of the HR diagram.


\section*{Acknowledgements}

We thank the referee for a very detailed report that made this article significantly better. 

We thank the organizers and participants of the \emph{Detecting the Unexpected} workshop that took place at STScI in 2017. The ideas for this work came from a hack during that workshop and have produced also other papers. In particular, we thank Lucianne Walkovicz for a continuous exchange of ideas and for proposing the original hack. We also thank Dalya Baron for useful insight about the use of the URF method. We thank the original hackers' team which included Kelle Cruz, and Umaa Rebbapragada

The authors acknowledge the support of the Vera C. Rubin Observatory Legacy Survey of Space and Time Transient and Variable Stars Science Collaboration (TVS SC), of which most of the authors are member and that provided opportunities for collaboration and exchange of ideas and knowledge.

This paper includes data collected by the Kepler mission and obtained from the MAST data archive at the Space Telescope Science Institute (STScI). Funding for the Kepler mission is provided by the NASA Science Mission Directorate. STScI is operated by the Association of Universities for Research in Astronomy, Inc., under NASA contract NAS 5-26555.

This work made use of several \texttt{Python} modules including:  \begin{itemize}
   \item \texttt{numpy} \citep{harris2020array}
      \item \texttt{maplotlib} \citep{matplotlib}
      \item \texttt{scikit-learn} \citep{JMLR:v12:pedregosa11a}
            \item \texttt{seaborn} \citep{seaborn}
    
\end{itemize}







\medskip
\medskip
\textbf{DATA AVAILABILITY}\\

The data underlying this article were accessed from the Mikulski Archive for Space Telescopes (MAST), at \url{https://mast.stsci.edu/portal/Mashup/Clients/Mast/Portal.html}. The derived data generated in this research can be accessed from the \texttt{GitHub} repository \url{https://github.com/kushaltirumala/WaldoInSky}.

\bibliographystyle{mnras}

\bibliography{ref}

\begin{thebibliography}{}
\makeatletter
\relax
\def\mn@urlcharsother{\let\do\@makeother \do\$\do\&\do\#\do\^\do\_\do\%\do\~}
\def\mn@doi{\begingroup\mn@urlcharsother \@ifnextchar [ {\mn@doi@}
  {\mn@doi@[]}}
\def\mn@doi@[#1]#2{\def\@tempa{#1}\ifx\@tempa\@empty \href
  {http://dx.doi.org/#2} {doi:#2}\else \href {http://dx.doi.org/#2} {#1}\fi
  \endgroup}
\def\mn@eprint#1#2{\mn@eprint@#1:#2::\@nil}
\def\mn@eprint@arXiv#1{\href {http://arxiv.org/abs/#1} {{\tt arXiv:#1}}}
\def\mn@eprint@dblp#1{\href {http://dblp.uni-trier.de/rec/bibtex/#1.xml}
  {dblp:#1}}
\def\mn@eprint@#1:#2:#3:#4\@nil{\def\@tempa {#1}\def\@tempb {#2}\def\@tempc
  {#3}\ifx \@tempc \@empty \let \@tempc \@tempb \let \@tempb \@tempa \fi \ifx
  \@tempb \@empty \def\@tempb {arXiv}\fi \@ifundefined
  {mn@eprint@\@tempb}{\@tempb:\@tempc}{\expandafter \expandafter \csname
  mn@eprint@\@tempb\endcsname \expandafter{\@tempc}}}

\bibitem[\protect\citeauthoryear{Aggarwal \& Yu}{Aggarwal \&
  Yu}{2001}]{aggarwal2001outlier}
Aggarwal C.~C.,  Yu P.~S.,  2001, in Proceedings of the 2001 ACM SIGMOD
  international conference on Management of data. pp 37--46

\bibitem[\protect\citeauthoryear{{Aleo} et~al.,}{{Aleo} et~al.}{2020}]{Aleo20}
{Aleo} P.~D.,  et~al., 2020, \mn@doi [Research Notes of the American
  Astronomical Society] {10.3847/2515-5172/aba6e8}, \href
  {https://ui.adsabs.harvard.edu/abs/2020RNAAS...4..112A} {4, 112}

\bibitem[\protect\citeauthoryear{{Arras}, {Townsley}  \& {Bildsten}}{{Arras}
  et~al.}{2006}]{arras06}
{Arras} P.,  {Townsley} D.~M.,   {Bildsten} L.,  2006, \mn@doi [\apjl]
  {10.1086/505178}, \href
  {https://ui.adsabs.harvard.edu/abs/2006ApJ...643L.119A} {643, L119}

\bibitem[\protect\citeauthoryear{{Baron} \& {Poznanski}}{{Baron} \&
  {Poznanski}}{2017a}]{2017ascl.soft05015B}
{Baron} D.,  {Poznanski} D.,  2017a, {WeirdestGalaxies: Outlier Detection
  Algorithm on Galaxy Spectra}, Astrophysics Source Code Library (\mn@eprint
  {ascl} {1611.07526})

\bibitem[\protect\citeauthoryear{Baron \& Poznanski}{Baron \&
  Poznanski}{2017b}]{Baron17}
Baron D.,  Poznanski D.,  2017b, Monthly Notices of the Royal Astronomical
  Society, 465, 4530

\bibitem[\protect\citeauthoryear{{Bellm} et~al.,}{{Bellm} et~al.}{2019}]{ztf}
{Bellm} E.~C.,  et~al., 2019, \mn@doi [\pasp] {10.1088/1538-3873/aaecbe}, \href
  {https://ui.adsabs.harvard.edu/abs/2019PASP..131a8002B} {131, 018002}

\bibitem[\protect\citeauthoryear{Bengio, Courville  \& Vincent}{Bengio
  et~al.}{2013}]{Bengio14}
Bengio Y.,  Courville A.,   Vincent P.,  2013, \mn@doi [IEEE Transactions on
  Pattern Analysis and Machine Intelligence] {10.1109/TPAMI.2013.50}, 35, 1798

\bibitem[\protect\citeauthoryear{{Bianco} et~al.}{{Bianco}
  et~al.}{2021}]{Bianco21}
{Bianco} F.,  et~al., 2021, ApJS, submitted

\bibitem[\protect\citeauthoryear{Biau}{Biau}{2012}]{Biau12}
Biau G.,  2012, Journal of Machine Learning Research, 13, 1063

\bibitem[\protect\citeauthoryear{{Bl{\'a}zquez-Garc{\'\i}a}, {Conde}, {Mori}
  \& {Lozano}}{{Bl{\'a}zquez-Garc{\'\i}a} et~al.}{2020}]{blazquez20}
{Bl{\'a}zquez-Garc{\'\i}a} A.,  {Conde} A.,  {Mori} U.,   {Lozano} J.~A.,
  2020, arXiv e-prints, \href
  {https://ui.adsabs.harvard.edu/abs/2020arXiv200204236B} {p. arXiv:2002.04236}

\bibitem[\protect\citeauthoryear{Boyajian et~al.,}{Boyajian
  et~al.}{2016}]{Boyajian16}
Boyajian T.~S.,  et~al., 2016, \mn@doi [Monthly Notices of the Royal
  Astronomical Society] {10.1093/mnras/stw218}, 457, 3988

\bibitem[\protect\citeauthoryear{Breiman}{Breiman}{2001}]{Breiman01}
Breiman L.,  2001, \mn@doi [Machine Learning] {10.1023/A:1010933404324}, 45, 5

\bibitem[\protect\citeauthoryear{Buitinck et~al.,}{Buitinck
  et~al.}{2013}]{Buitinck13}
Buitinck L.,  et~al., 2013, arXiv preprint arXiv:1309.0238

\bibitem[\protect\citeauthoryear{Che, Purushotham, Cho, Sontag  \& Liu}{Che
  et~al.}{2018}]{Che18}
Che Z.,  Purushotham S.,  Cho K.,  Sontag D.,   Liu Y.,  2018, \mn@doi
  [Scientific Reports] {10.1038/s41598-018-24271-9}, 8

\bibitem[\protect\citeauthoryear{Chen, Diethe, Twomey  \& Flach}{Chen
  et~al.}{2018}]{Chen18}
Chen H.,  Diethe T.,  Twomey N.,   Flach P.,  2018.

\bibitem[\protect\citeauthoryear{{Conroy} et~al.,}{{Conroy}
  et~al.}{2018}]{Conroy18}
{Conroy} C.,  et~al., 2018, \mn@doi [\apj] {10.3847/1538-4357/aad460}, \href
  {https://ui.adsabs.harvard.edu/abs/2018ApJ...864..111C} {864, 111}

\bibitem[\protect\citeauthoryear{{Davenport} et~al.,}{{Davenport}
  et~al.}{2014}]{2014ApJ...797..122D}
{Davenport} J. R.~A.,  et~al., 2014, \mn@doi [\apj]
  {10.1088/0004-637X/797/2/122}, \href
  {https://ui.adsabs.harvard.edu/abs/2014ApJ...797..122D} {797, 122}

\bibitem[\protect\citeauthoryear{{Debosscher}, {Sarro}, {Aerts}, {Cuypers},
  {Vandenbussche}, {Garrido}  \& {Solano}}{{Debosscher}
  et~al.}{2007}]{debosscher07}
{Debosscher} J.,  {Sarro} L.~M.,  {Aerts} C.,  {Cuypers} J.,  {Vandenbussche}
  B.,  {Garrido} R.,   {Solano} E.,  2007, \mn@doi [\aap]
  {10.1051/0004-6361:20077638}, \href
  {https://ui.adsabs.harvard.edu/abs/2007A&A...475.1159D} {475, 1159}

\bibitem[\protect\citeauthoryear{{Drake} et~al.,}{{Drake}
  et~al.}{2012}]{catalina}
{Drake} A.~J.,  et~al., 2012, in {Griffin} E.,  {Hanisch} R.,   {Seaman} R.,
  eds,  IAU Symposium Vol. 285, New Horizons in Time Domain Astronomy. pp
  306--308 (\mn@eprint {arXiv} {1111.2566}), \mn@doi{10.1017/S1743921312000889}

\bibitem[\protect\citeauthoryear{Druetto, Roberti, Cancelliere, Cavagnino  \&
  Gai}{Druetto et~al.}{2019}]{Druetto19}
Druetto A.,  Roberti M.,  Cancelliere R.,  Cavagnino D.,   Gai M.,  2019, in
  International Work-Conference on Artificial Neural Networks. pp 390--401

\bibitem[\protect\citeauthoryear{Dubath et~al.,}{Dubath
  et~al.}{2011}]{Dubath11}
Dubath P.,  et~al., 2011, \mn@doi [Monthly Notices of the Royal Astronomical
  Society] {10.1111/j.1365-2966.2011.18575.x}, 414, 2602

\bibitem[\protect\citeauthoryear{Dutta, Giannella, Borne  \& Kargupta}{Dutta
  et~al.}{2007}]{Dutta07}
Dutta H.,  Giannella C.,  Borne K.,   Kargupta H.,  2007, in: Proceedings of
  the 2007 SIAM International Conference on Data Mining.
SIAM

\bibitem[\protect\citeauthoryear{Emmott, Das, Dietterich, Fern  \& Wong}{Emmott
  et~al.}{2013}]{10.1145/2500853.2500858}
Emmott A.~F.,  Das S.,  Dietterich T.,  Fern A.,   Wong W.-K.,  2013, in
  Proceedings of the ACM SIGKDD Workshop on Outlier Detection and Description.
  ODD '13.
Association for Computing Machinery, New York, NY, USA, p. 16–21,
  \mn@doi{10.1145/2500853.2500858}, \url
  {https://doi.org/10.1145/2500853.2500858}

\bibitem[\protect\citeauthoryear{{Eyer} \& {Mowlavi}}{{Eyer} \&
  {Mowlavi}}{2008}]{Eyer08}
{Eyer} L.,  {Mowlavi} N.,  2008, in Journal of Physics Conference Series. p.
  012010 (\mn@eprint {arXiv} {0712.3797}),
  \mn@doi{10.1088/1742-6596/118/1/012010}

\bibitem[\protect\citeauthoryear{{Eyer}, {S{\"u}veges}, {De Ridder}, {Regibo},
  {Mowlavi}, {Holl}, {Rimoldini}  \& {Bouchy}}{{Eyer} et~al.}{2019}]{Eyer19}
{Eyer} L.,  {S{\"u}veges} M.,  {De Ridder} J.,  {Regibo} S.,  {Mowlavi} N.,
  {Holl} B.,  {Rimoldini} L.,   {Bouchy} F.,  2019, \mn@doi [\pasp]
  {10.1088/1538-3873/ab2511}, \href
  {https://ui.adsabs.harvard.edu/abs/2019PASP..131h8001E} {131, 088001}

\bibitem[\protect\citeauthoryear{Fulcher}{Fulcher}{2017}]{Fulcher17}
Fulcher B.,  2017, Feature-based time-series analysis

\bibitem[\protect\citeauthoryear{{Gaia Collaboration} et~al.,}{{Gaia
  Collaboration} et~al.}{2016}]{gaia1}
{Gaia Collaboration} et~al., 2016, \mn@doi [\aap]
  {10.1051/0004-6361/201629272}, \href
  {https://ui.adsabs.harvard.edu/abs/2016A&A...595A...1G} {595, A1}

\bibitem[\protect\citeauthoryear{{Gaia Collaboration} et~al.,}{{Gaia
  Collaboration} et~al.}{2018}]{gaia2}
{Gaia Collaboration} et~al., 2018, \mn@doi [\aap]
  {10.1051/0004-6361/201833051}, \href
  {https://ui.adsabs.harvard.edu/abs/2018A&A...616A...1G} {616, A1}

\bibitem[\protect\citeauthoryear{Giles \& Walkowicz}{Giles \&
  Walkowicz}{2019}]{Giles19}
Giles D.,  Walkowicz L.,  2019, Monthly Notices of the Royal Astronomical
  Society, 484, 834

\bibitem[\protect\citeauthoryear{{Giles} \& {Walkowicz}}{{Giles} \&
  {Walkowicz}}{2020}]{Giles20}
{Giles} D.~K.,  {Walkowicz} L.,  2020, \mn@doi [\mnras]
  {10.1093/mnras/staa2736}, \href
  {https://ui.adsabs.harvard.edu/abs/2020MNRAS.499..524G} {499, 524}

\bibitem[\protect\citeauthoryear{{Girardi}}{{Girardi}}{1999}]{1999MNRAS.308..818G}
{Girardi} L.,  1999, \mn@doi [\mnras] {10.1046/j.1365-8711.1999.02746.x}, \href
  {https://ui.adsabs.harvard.edu/abs/1999MNRAS.308..818G} {308, 818}

\bibitem[\protect\citeauthoryear{{Girardi}}{{Girardi}}{2016}]{girardi16}
{Girardi} L.,  2016, \mn@doi [\araa] {10.1146/annurev-astro-081915-023354},
  \href {https://ui.adsabs.harvard.edu/abs/2016ARA&A..54...95G} {54, 95}

\bibitem[\protect\citeauthoryear{Goldstein \& Uchida}{Goldstein \&
  Uchida}{2016}]{goldstein16}
Goldstein M.,  Uchida S.,  2016, \mn@doi [PLoS One]
  {10.1371/journal.pone.0152173}, 11

\bibitem[\protect\citeauthoryear{Graham, Drake, Djorgovski, Mahabal, Donalek,
  Duan  \& Maker}{Graham et~al.}{2013}]{Graham13}
Graham M.~J.,  Drake A.~J.,  Djorgovski S.~G.,  Mahabal A.~A.,  Donalek C.,
  Duan V.,   Maker A.,  2013, \mn@doi [Monthly Notices of the Royal
  Astronomical Society] {10.1093/mnras/stt1264}, 434, 3423

\bibitem[\protect\citeauthoryear{Harris et~al.,}{Harris
  et~al.}{2020}]{harris2020array}
Harris C.~R.,  et~al., 2020, \mn@doi [Nature] {10.1038/s41586-020-2649-2}, 585,
  357

\bibitem[\protect\citeauthoryear{Henrion, Hand, Gandy  \& Mortlock}{Henrion
  et~al.}{2013}]{Henrion13}
Henrion M.,  Hand D.~J.,  Gandy A.,   Mortlock D.~J.,  2013, Statistical
  Analysis and Data Mining: The ASA Data Science Journal, 6, 53

\bibitem[\protect\citeauthoryear{Hinton \& Roweis}{Hinton \&
  Roweis}{2003}]{Hinton03}
Hinton G.~E.,  Roweis S.~T.,  2003, in Advances in neural information
  processing systems. pp 857--864

\bibitem[\protect\citeauthoryear{Hunter}{Hunter}{2007}]{matplotlib}
Hunter J.~D.,  2007, \mn@doi [Computing in Science Engineering]
  {10.1109/MCSE.2007.55}, 9, 90

\bibitem[\protect\citeauthoryear{{Ishida} et~al.,}{{Ishida}
  et~al.}{2019}]{Ishida19}
{Ishida} E. E.~O.,  et~al., 2019, arXiv e-prints, \href
  {https://ui.adsabs.harvard.edu/abs/2019arXiv190913260I} {p. arXiv:1909.13260}

\bibitem[\protect\citeauthoryear{{Ivezi{\'c}} et~al.,}{{Ivezi{\'c}}
  et~al.}{2019}]{Ivezic19}
{Ivezi{\'c}} {\v{Z}}.,  et~al., 2019, \mn@doi [\apj]
  {10.3847/1538-4357/ab042c}, \href
  {https://ui.adsabs.harvard.edu/abs/2019ApJ...873..111I} {873, 111}

\bibitem[\protect\citeauthoryear{{Jamal} \& {Bloom}}{{Jamal} \&
  {Bloom}}{2020}]{jamal20}
{Jamal} S.,  {Bloom} J.~S.,  2020, \mn@doi [\apjs] {10.3847/1538-4365/aba8ff},
  \href {https://ui.adsabs.harvard.edu/abs/2020ApJS..250...30J} {250, 30}

\bibitem[\protect\citeauthoryear{{Jenkins}}{{Jenkins}}{2017}]{Jenkins17}
{Jenkins} J.~M.,  2017, {Kepler Data Processing Handbook: Philosophy and
  Scope}, Kepler Science Document KSCI-19081-002

\bibitem[\protect\citeauthoryear{{Johnston} \& {Oluseyi}}{{Johnston} \&
  {Oluseyi}}{2017}]{Johnston17}
{Johnston} K.~B.,  {Oluseyi} H.~M.,  2017, \mn@doi [\na]
  {10.1016/j.newast.2016.10.004}, \href
  {https://ui.adsabs.harvard.edu/abs/2017NewA...52...35J} {52, 35}

\bibitem[\protect\citeauthoryear{{Johnston}, {Caballero-Nieves}, {Peter},
  {Petit}  \& {Haber}}{{Johnston} et~al.}{2019}]{Johnston19}
{Johnston} K.~B.,  {Caballero-Nieves} S.~M.,  {Peter} A.~M.,  {Petit} V.,
  {Haber} R.,  2019, in {Teuben} P.~J.,  {Pound} M.~W.,  {Thomas} B.~A.,
  {Warner} E.~M.,  eds,  Astronomical Society of the Pacific Conference Series
  Vol. 523, Astronomical Data Analysis Software and Systems XXVII. p.~83

\bibitem[\protect\citeauthoryear{{Kessler} et~al.,}{{Kessler}
  et~al.}{2019}]{Kessler19}
{Kessler} R.,  et~al., 2019, \mn@doi [\pasp] {10.1088/1538-3873/ab26f1}, \href
  {https://ui.adsabs.harvard.edu/abs/2019PASP..131i4501K} {131, 094501}

\bibitem[\protect\citeauthoryear{{Kochanek} et~al.,}{{Kochanek}
  et~al.}{2017}]{asassn}
{Kochanek} C.~S.,  et~al., 2017, \mn@doi [\pasp] {10.1088/1538-3873/aa80d9},
  \href {https://ui.adsabs.harvard.edu/abs/2017PASP..129j4502K} {129, 104502}

\bibitem[\protect\citeauthoryear{Kullback \& Leibler}{Kullback \&
  Leibler}{1951}]{Kullback51}
Kullback S.,  Leibler R.~A.,  1951, \mn@doi [Ann. Math. Statist.]
  {10.1214/aoms/1177729694}, 22, 79

\bibitem[\protect\citeauthoryear{{Li}, {Ragosta}, {Clarkson}  \& {Bianco}}{{Li}
  et~al.}{2021}]{Li21}
{Li} X.,  {Ragosta} F.,  {Clarkson} W.~I.,   {Bianco} F.~B.,  2021, arXiv
  e-prints, \href {https://ui.adsabs.harvard.edu/abs/2021arXiv210710281L} {p.
  arXiv:2107.10281}

\bibitem[\protect\citeauthoryear{Liu, Ting  \& Zhou}{Liu et~al.}{2012}]{Liu12}
Liu F.~T.,  Ting K.~M.,   Zhou Z.-H.,  2012, ACM Transactions on Knowledge
  Discovery from Data (TKDD), 6, 1

\bibitem[\protect\citeauthoryear{{Lochner} \& {Bassett}}{{Lochner} \&
  {Bassett}}{2021}]{2021A&C....3600481L}
{Lochner} M.,  {Bassett} B.~A.,  2021, \mn@doi [Astronomy and Computing]
  {10.1016/j.ascom.2021.100481}, \href
  {https://ui.adsabs.harvard.edu/abs/2021A&C....3600481L} {36, 100481}

\bibitem[\protect\citeauthoryear{Lomb}{Lomb}{1976}]{Lomb76}
Lomb N.~R.,  1976, Astrophysics and space science, 39, 447

\bibitem[\protect\citeauthoryear{Maaten \& Hinton}{Maaten \&
  Hinton}{2008}]{Maaten08}
Maaten L. v.~d.,  Hinton G.,  2008, Journal of machine learning research, 9,
  2579

\bibitem[\protect\citeauthoryear{Mahabal, Sheth, Gieseke, Pai, Djorgovski,
  Drake  \& Graham}{Mahabal et~al.}{2017}]{Mahabal17}
Mahabal A.,  Sheth K.,  Gieseke F.,  Pai A.,  Djorgovski S.~G.,  Drake A.~J.,
  Graham M.~J.,  2017, in 2017 IEEE Symposium Series on Computational
  Intelligence (SSCI). pp~1--8

\bibitem[\protect\citeauthoryear{{Malanchev} et~al.,}{{Malanchev}
  et~al.}{2021}]{2021MNRAS.502.5147M}
{Malanchev} K.~L.,  et~al., 2021, \mn@doi [\mnras] {10.1093/mnras/stab316},
  \href {https://ui.adsabs.harvard.edu/abs/2021MNRAS.502.5147M} {502, 5147}

\bibitem[\protect\citeauthoryear{Margalef-Bentabol, Huertas-Company, Charnock,
  Margalef-Bentabol, Bernardi, Dubois, Storey-Fisher  \&
  Zanis}{Margalef-Bentabol et~al.}{2020}]{Margalef20}
Margalef-Bentabol B.,  Huertas-Company M.,  Charnock T.,  Margalef-Bentabol C.,
   Bernardi M.,  Dubois Y.,  Storey-Fisher K.,   Zanis L.,  2020, arXiv
  preprint arXiv:2003.08263

\bibitem[\protect\citeauthoryear{{McInnes}, {Healy}  \& {Melville}}{{McInnes}
  et~al.}{2018}]{McInnes18}
{McInnes} L.,  {Healy} J.,   {Melville} J.,  2018, arXiv e-prints, \href
  {https://ui.adsabs.harvard.edu/abs/2018arXiv180203426M} {p. arXiv:1802.03426}

\bibitem[\protect\citeauthoryear{{Meech} et~al.,}{{Meech}
  et~al.}{2017}]{Meech17}
{Meech} K.~J.,  et~al., 2017, \mn@doi [\nat] {10.1038/nature25020}, \href
  {https://ui.adsabs.harvard.edu/abs/2017Natur.552..378M} {552, 378}

\bibitem[\protect\citeauthoryear{{Miniutti} et~al.,}{{Miniutti}
  et~al.}{2019}]{Miniutti19}
{Miniutti} G.,  et~al., 2019, \mn@doi [\nat] {10.1038/s41586-019-1556-x}, \href
  {https://ui.adsabs.harvard.edu/abs/2019Natur.573..381M} {573, 381}

\bibitem[\protect\citeauthoryear{{Nun}, {Protopapas}, {Sim}, {Zhu}, {Dave},
  {Castro}  \& {Pichara}}{{Nun} et~al.}{2015}]{Nun15}
{Nun} I.,  {Protopapas} P.,  {Sim} B.,  {Zhu} M.,  {Dave} R.,  {Castro} N.,
  {Pichara} K.,  2015, arXiv e-prints, \href
  {https://ui.adsabs.harvard.edu/abs/2015arXiv150600010N} {p. arXiv:1506.00010}

\bibitem[\protect\citeauthoryear{{Nun}, {Protopapas}, {Sim}  \& {Chen}}{{Nun}
  et~al.}{2016}]{Nun16}
{Nun} I.,  {Protopapas} P.,  {Sim} B.,   {Chen} W.,  2016, \mn@doi [\aj]
  {10.3847/0004-6256/152/3/71}, \href
  {https://ui.adsabs.harvard.edu/abs/2016AJ....152...71N} {152, 71}

\bibitem[\protect\citeauthoryear{{Paudel}, {Gizis}, {Mullan}, {Schmidt},
  {Burgasser}, {Williams}  \& {Berger}}{{Paudel} et~al.}{2018}]{paduel18}
{Paudel} R.~R.,  {Gizis} J.~E.,  {Mullan} D.~J.,  {Schmidt} S.~J.,  {Burgasser}
  A.~J.,  {Williams} P. K.~G.,   {Berger} E.,  2018, \mn@doi [\apj]
  {10.3847/1538-4357/aac8e0}, \href
  {https://ui.adsabs.harvard.edu/abs/2018ApJ...861...76P} {861, 76}

\bibitem[\protect\citeauthoryear{Paudel, Gizis, Mullan, Schmidt, Burgasser,
  Williams, Youngblood  \& Stassun}{Paudel et~al.}{2019}]{Paudel19}
Paudel R.~R.,  Gizis J.~E.,  Mullan D.,  Schmidt S.~J.,  Burgasser A.~J.,
  Williams P.~K.,  Youngblood A.,   Stassun K.,  2019, Monthly Notices of the
  Royal Astronomical Society, 486, 1438

\bibitem[\protect\citeauthoryear{Pedregosa et~al.,}{Pedregosa
  et~al.}{2011a}]{scikit-learn}
Pedregosa F.,  et~al., 2011a, Journal of Machine Learning Research, 12, 2825

\bibitem[\protect\citeauthoryear{Pedregosa et~al.,}{Pedregosa
  et~al.}{2011b}]{JMLR:v12:pedregosa11a}
Pedregosa F.,  et~al., 2011b, Journal of Machine Learning Research, 12, 2825

\bibitem[\protect\citeauthoryear{{Prsa} \& {Eclipsing Binary Working
  Group}}{{Prsa} \& {Eclipsing Binary Working Group}}{2013}]{prsa13}
{Prsa} A.,  {Eclipsing Binary Working Group} 2013, in Giants of Eclipse. p.
  40102

\bibitem[\protect\citeauthoryear{{Pruzhinskaya}, {Malanchev}, {Kornilov},
  {Ishida}, {Mondon}, {Volnova}  \& {Korolev}}{{Pruzhinskaya}
  et~al.}{2019}]{Pruzhinskaya19}
{Pruzhinskaya} M.~V.,  {Malanchev} K.~L.,  {Kornilov} M.~V.,  {Ishida}
  E.~E.~O.,  {Mondon} F.,  {Volnova} A.~A.,   {Korolev} V.~S.,  2019, \mn@doi
  [\mnras] {10.1093/mnras/stz2362}, \href
  {https://ui.adsabs.harvard.edu/abs/2019MNRAS.489.3591P} {489, 3591}

\bibitem[\protect\citeauthoryear{{Pr{\v{s}}a} et~al.,}{{Pr{\v{s}}a}
  et~al.}{2016}]{prsa16}
{Pr{\v{s}}a} A.,  et~al., 2016, \mn@doi [\apjs] {10.3847/1538-4365/227/2/29},
  \href {https://ui.adsabs.harvard.edu/abs/2016ApJS..227...29P} {227, 29}

\bibitem[\protect\citeauthoryear{{Rebbapragada}, {Protopapas}, {Brodley}  \&
  {Alcock}}{{Rebbapragada} et~al.}{2009}]{Rebbapragada09}
{Rebbapragada} U.,  {Protopapas} P.,  {Brodley} C.~E.,   {Alcock} C.,  2009,
  arXiv e-prints, \href {https://ui.adsabs.harvard.edu/abs/2009arXiv0905.3428R}
  {p. arXiv:0905.3428}

\bibitem[\protect\citeauthoryear{{Reis}, {Poznanski}, {Baron}, {Zasowski}  \&
  {Shahaf}}{{Reis} et~al.}{2018}]{Reis18}
{Reis} I.,  {Poznanski} D.,  {Baron} D.,  {Zasowski} G.,   {Shahaf} S.,  2018,
  \mn@doi [\mnras] {10.1093/mnras/sty348}, \href
  {https://ui.adsabs.harvard.edu/abs/2018MNRAS.476.2117R} {476, 2117}

\bibitem[\protect\citeauthoryear{{Richards} et~al.,}{{Richards}
  et~al.}{2011}]{Richards11}
{Richards} J.~W.,  et~al., 2011, \mn@doi [\apj] {10.1088/0004-637X/733/1/10},
  \href {https://ui.adsabs.harvard.edu/abs/2011ApJ...733...10R} {733, 10}

\bibitem[\protect\citeauthoryear{{Scargle}}{{Scargle}}{1982}]{Scargle82}
{Scargle} J.~D.,  1982, \mn@doi [\apj] {10.1086/160554}, \href
  {https://ui.adsabs.harvard.edu/abs/1982ApJ...263..835S} {263, 835}

\bibitem[\protect\citeauthoryear{{Schmidt}}{{Schmidt}}{2019}]{Schmidt19}
{Schmidt} M.,  2019, arXiv e-prints, \href
  {https://ui.adsabs.harvard.edu/abs/2019arXiv190702574S} {p. arXiv:1907.02574}

\bibitem[\protect\citeauthoryear{Shi \& Horvath}{Shi \& Horvath}{2006}]{Shi06}
Shi T.,  Horvath S.,  2006, \mn@doi [Journal of Computational and Graphical
  Statistics] {10.1198/106186006X94072}, 15, 118

\bibitem[\protect\citeauthoryear{{Storey-Fisher}, {Huertas-Company},
  {Ramachandra}, {Lanusse}, {Leauthaud}, {Luo}, {Huang}  \&
  {Prochaska}}{{Storey-Fisher} et~al.}{2021}]{storey21}
{Storey-Fisher} K.,  {Huertas-Company} M.,  {Ramachandra} N.,  {Lanusse} F.,
  {Leauthaud} A.,  {Luo} Y.,  {Huang} S.,   {Prochaska} J.~X.,  2021, arXiv
  e-prints, \href {https://ui.adsabs.harvard.edu/abs/2021arXiv210502434S} {p.
  arXiv:2105.02434}

\bibitem[\protect\citeauthoryear{{Szklen{\'a}r}, {B{\'o}di},
  {Tarczay-Neh{\'e}z}, {Vida}, {Marton}, {Mez{\H{o}}}, {Forr{\'o}}  \&
  {Szab{\'o}}}{{Szklen{\'a}r} et~al.}{2020}]{szklenar20}
{Szklen{\'a}r} T.,  {B{\'o}di} A.,  {Tarczay-Neh{\'e}z} D.,  {Vida} K.,
  {Marton} G.,  {Mez{\H{o}}} G.,  {Forr{\'o}} A.,   {Szab{\'o}} R.,  2020,
  \mn@doi [\apjl] {10.3847/2041-8213/ab9ca4}, \href
  {https://ui.adsabs.harvard.edu/abs/2020ApJ...897L..12S} {897, L12}

\bibitem[\protect\citeauthoryear{{VanderPlas}}{{VanderPlas}}{2018}]{2018ApJS..236...16V}
{VanderPlas} J.~T.,  2018, \mn@doi [\apjs] {10.3847/1538-4365/aab766}, \href
  {https://ui.adsabs.harvard.edu/abs/2018ApJS..236...16V} {236, 16}

\bibitem[\protect\citeauthoryear{Waskom et~al.,}{Waskom et~al.}{2017}]{seaborn}
Waskom M.,  et~al., 2017, mwaskom/seaborn: v0.8.1 (September 2017),
  \mn@doi{10.5281/zenodo.883859}, \url {https://doi.org/10.5281/zenodo.883859}

\bibitem[\protect\citeauthoryear{{York} et~al.,}{{York} et~al.}{2000}]{SDSS}
{York} D.~G.,  et~al., 2000, \mn@doi [\aj] {10.1086/301513}, \href
  {https://ui.adsabs.harvard.edu/abs/2000AJ....120.1579Y} {120, 1579}

\bibitem[\protect\citeauthoryear{{{\v{S}}koda}, {Podsztavek}  \&
  {Tvrd{\'\i}k}}{{{\v{S}}koda} et~al.}{2020}]{Skoda20}
{{\v{S}}koda} P.,  {Podsztavek} O.,   {Tvrd{\'\i}k} P.,  2020, arXiv e-prints,
  \href {https://ui.adsabs.harvard.edu/abs/2020arXiv200903219S} {p.
  arXiv:2009.03219}

\makeatother
\end{thebibliography}

\appendix
\section{Tables of anomalies} \label{ap:tables}

Listed below are the anomalous objects that were selected from the study in \citet{debosscher07}, as well as sample sets of anomalies selected with out method.

\begin{table*}
\centering
\caption{\emph{Bona-fide} anomalies from \citet{debosscher07}}
\label{tab:bona_fide}
\begin{tabular}{ccccc}
\hline
KIC ID & Class 1 & Class 2 & Class 3 & Score \\
\hline
12601939 &	CLCEP &	MISC &	ECL &	0.008786218632 \\
12406908  &	CLCEP &	ELL &	ROT &	0.008273873742 \\ 
3561372	& CLCEP &	MISC &	SR &	0.01282938125 \\
8481420	& CLCEP &	ROT &	MISC &	0.006202438995 \\
5095098	& CLCEP &	MISC &	ROT &	0.005815868967 \\
8259835	& CLCEP &	ELL &	ROT &	0.005275752976 \\
1573138	& CLCEP &	ELL &	ROT &	0.007246326707 \\
5217688	& CLCEP &	ROT &	MISC &	0.0086037896 \\
7957881 &	CLCEP &	ROT &	ELL &	0.006562796276 \\
4357272 &	CLCEP &	MISC &	GDOR &	0.005871296046 \\
6423857 &	CLCEP &	ELL &	ROT &	0.006206039431 \\
9413885 &	ELL &	CLCEP &	ROT &	0.006566994353 \\
7800087 &	ELL &	BCEP &	ECL &	0.009391201802 \\
3240305 &	ELL &	BCEP &	RRC &	0.008396707286 \\
11246163 &	ELL &	BCEP &	RVTAU &	0.009316268909 \\
5460828 &	ELL &	MISC &	ROT &	0.005373922226 \\
9117123 &	ELL &	BCEP &	CLCEP &	0.006926996579 \\
8396230 &	ELL &	MISC &	GDOR &	0.005702632788 \\
10815379 &	ELL &	ECL &	RRD &	0.005149065953 \\
4175618 &	ELL &	BCEP &	RRC &	0.005983051588 \\
5301955 &	ELL &	MISC &	BCEP &	0.01535287946 \\
3859213 &	ELL &	CLCEP &	MISC &	0.005018452801 \\
11091336 &	ELL &	ROT &	ECL &	0.00522920081 \\
2159783 &	ELL &	BCEP &	RRC &	0.0162872493 \\
9606106 &	ELL &	ROT &	ECL &	0.008558313951 \\
11666309 &	ELL &	MISC &	GDOR &	0.006768599231 \\
2285420 &	ELL &	ROT &	 ECL &	0.005544164958 \\
9450669 &	ELL &	ROT &	CLCEP &	0.008378505168 \\
6035535 &	ELL &	BCEP &	ECL &	0.007001462287 \\
7433177 &	ELL &	ECL &	ROT &	0.005067273826 \\
6719893 &	ELL &	CLCEP &	BCEP &	0.008011748077 \\
3848042 &	ELL &	BCEP &	ECL &	0.005772841488 \\
3222369 &	ELL &	ROT &	CLCEP &	0.005141579964 \\
5474812 &	ELL &	ROT &	ECL &	0.007563596577 \\
7036755 &	ELL &	ECL &	RRD &	0.007398642537 \\
1570924 &	ELL &	ROT &	ECL &	0.01289010831 \\
5181824 &	ELL &	ECL &	RRD &	0.006513773867 \\
9306085 &	ELL &	BCEP &	MISC &	0.007914872094 \\
10646009 &	ELL &	CLCEP &	ROT &	0.01089612709 \\
5956051 &	ELL &	BCEP &	RRC &	0.007217262209 \\
3528198 &	ELL &	ROT &	MISC &	0.005084600456 \\
10275197 &	ELL &	ECL &	BCEP &	0.00510990445 \\
10618253 &	ELL &	BCEP &	RRC &	0.01247752439 \\
9761113 &	ELL &	ROT &	ECL &	0.005797477873 \\
5446821 &	ELL &	MISC &	ECL &	0.005348593006 \\
11717798 &	ELL &	BCEP &	RRC &	0.006313991487 \\
11862915 &	ELL &	CLCEP &	ECL &	0.006677180058 \\
8682921 &	ELL &	BCEP &	ECL &	0.006591784762 \\
5646176 &	ELL &	MISC &	ROT &	0.008414125691 \\
4059007 &	ELL &	ROT &	CLCEP &	0.01003724288 \\
8012943 &	ELL &	ROT &	BCEP &	0.00629319905 \\
10515986 &	ELL &	ROT &	ECL &	0.006091237395 \\
11087095 &	ELL &	ECL &	RRD &	0.005589115642 \\
12365015 &	ELL &	ECL &	RRD &	0.007136313026 \\
6675318 &	ELL &	BCEP &	CLCEP &	0.006399049821 \\
10320278 &	ELL &	MISC &	ECL &	0.007443713072 \\
9812607 &	ELL &	ROT &	ECL &	0.006178859492 \\
6129451 &	ELL &	ROT &	CLCEP &	0.005499182815 \\
4819564 &	ELL &	BCEP &	RRC &	0.00621635076 \\
10711646 &	GDOR &	MISC &	ECL &	0.01130325381 \\
6779613 &	GDOR &	ROT &	MISC &	0.005053367934 \\
4932691 &	GDOR &	MISC &	ROT &	0.007419884684 \\
6953103 &	GDOR &	SPB &	MISC &	0.008369353744 \\
3759394 &	GDOR &	MISC &	ROT &	0.005154493111 \\
\hline
\end{tabular}

\end{table*}

\begin{table*}
\centering
\contcaption{\emph{Bona-fide} anomalies from \citet{debosscher07}}
\label{tab:bona_fide2}
\begin{tabular}{ccccc}
\hline
KIC ID & Class 1 & Class 2 & Class 3 & Score \\
\hline
10292465 &	GDOR &	MISC &	ECL &	0.007451421701 \\
8097825 &	GDOR &	MISC &	ECL &	0.01075275798 \\
8590527 &	GDOR &	MISC &	ECL &	0.008553672218 \\
9210828 &	GDOR &	MISC &	ECL &	0.006667395564 \\
7671594 &	GDOR &	MISC &	ECL &	0.01277598394 \\
3847822 &	GDOR &	MISC &	SPB &	0.006330828654 \\
4752731 &	GDOR &	BCEP &	RRC &	0.007465211686 \\
7798259 &	GDOR &	MISC &	ECL &	0.01133006546 \\
6025466 &	GDOR &	MISC &	SPB &	0.006922493469 \\
4077442 &	GDOR &	MISC &	ECL &	0.008473826794 \\
7286410 &	GDOR &	MISC &	DMCEP &	0.005183427211 \\
5166136 &	GDOR &	BCEP &	MISC &	0.005201605104 \\
11922782 &	GDOR &	MISC &	ECL &	0.008167659534 \\
6836589 &	GDOR &	MISC &	DSCUT &	0.00738269835 \\
7672492 &	GDOR &	SPB &	ROT &	0.005770436504 \\
6863840 &	GDOR &	MISC &	ECL &	0.0108957491 \\
4358206 &	GDOR &	MISC &	ECL &	0.00998718986 \\
11455795 &	GDOR &	MISC &	ECL &	0.009661442003 \\
9945280 &	GDOR &	MISC &	ROT &	0.006896913873 \\
12258225 &	GDOR &	SPB &	ROT &	0.005384756292 \\
10490282 &	GDOR &	MISC &	SPB &	0.005621403494 \\
12350399 &	GDOR &	MISC &	SPB &	0.005130130217 \\
5510843 &	GDOR &	MISC &	ECL &	0.008097610133 \\
3455094 &	GDOR &	SPB &	MISC &	0.01119237607 \\
7021689 &	GDOR &	BCEP &	ELL &	0.006029030581 \\
11134079 &	GDOR &	MISC &	ECL &	0.01004903642 \\
6870327 &	GDOR &	MISC &	ECL &	0.005929492017 \\
5822633 &	GDOR &	MISC &	SPB &	0.007496501304 \\
8288719 &	GDOR &	MISC &	ECL &	0.007163174849 \\
7021124 &	RRAB &	RVTAU &	ECL &	0.03710292832 \\
6619830 &	RRAB &	CLCEP &	RVTAU &	0.01475008223 \\
6936115 &	RRAB &	RVTAU &	ECL &	0.03079985113 \\
5559631 &	RRAB &	RVTAU &	CLCEP &	0.02317068883 \\
4473355 &	RRC &	BCEP &	ROT &	0.00577276251 \\
5520878 &	RRC &	BCEP &	ELL &	0.014310101 \\
8265951 &	RRC &	BCEP &	ELL &	0.006168061003 \\
4064484 &	RRC &	BCEP &	ELL &	0.01581693962 \\
7698650 &	RRC &	BCEP &	ELL &	0.01541358106 \\
10229723 &	RRC &	BCEP &	ELL &	0.005765092672 \\
9453114 &	RRC &	BCEP &	ELL &	0.01515723456 \\
9350889 &	RRC &	BCEP &	ELL &	0.006120988373 \\
3104113 &	RRC &	BCEP &	ELL &	0.009539970725 \\
1572802 &	RRC &	ROT &	ECL &	0.007025044558 \\
2985366 &	RRC &	BCEP &	ECL &	0.008408515106 \\
10063343 &	RRC &	BCEP &	ECL &	0.006657052264 \\
11097678 &	RRC &	ELL &	BCEP &	0.006830221169 \\
12055014 &	RRC &	BCEP &	ELL &	0.009629194717 \\
5950759 &	RVTAU &	MISC &	BCEP &	0.03469484644 \\
6044064 &	RVTAU &	MISC &	GDOR &	0.02049138251 \\
9848000 &	SPB	& MISC &	GDOR &	0.005500802133 \\
8247608 &	SPB	 &  GDOR &	ROT &	0.006217369242 \\
4248763 &	SPB	 &  GDOR &	ROT &	0.008198738717 \\
3443221 &	SPB &	GDOR &	ECL &	0.009316805485 \\
3654076 &	SPB &	MISC &	ROT &	0.01123835044 \\
3001695 &	SPB &	GDOR &	MISC &	0.005912971376 \\
6425437 &	SPB &	GDOR &	ECL &	0.006307296393 \\
7339348 &	SPB &	ROT &	ELL &	0.006906215906 \\
8719419 &	SPB &	GDOR &	MISC &	0.005727435485 \\
3326917 &	SPB &	ROT &	MISC &	0.006382340192 \\
4932663 &	SPB &	GDOR &	ECL &	0.005664085207 \\
8041249 &	SPB &	MISC &	ROT &	0.006618825146 \\
3748748 &	SPB &	GDOR &	ECL &	0.005784702967 \\
6681516 &	SPB &	GDOR &	MISC &	0.005094068603 \\
2141387 &	SPB &	GDOR &	MISC &	0.007014599889 \\
\hline
\end{tabular}	
\end{table*}

\begin{table*}
\centering
\contcaption{\emph{Bona-fide} anomalies from \citet{debosscher07}}
\label{tab:bona_fide3}
\begin{tabular}{ccccc}
\hline
KIC ID & Class 1 & Class 2 & Class 3 & Score \\
\hline
7039421 &	SPB &	GDOR &	MISC &	0.006928340026 \\
5288646 &	SPB &	GDOR &	MISC &	0.005199013066 \\
6449081 &	SPB &	GDOR &	ECL &	0.005312343615 \\
8332664 &	SPB &	GDOR &	MISC &	0.005078668855 \\
6939772 &	SPB &	GDOR &	MISC &	0.005217383926 \\
7581697 &	SPB &	MISC &	GDOR &	0.006368475132 \\
8127495 &	SPB &	GDOR &	ROT &	0.005187879249 \\
9007322 &	SPB &	GDOR &	MISC &	0.005837423991 \\
4175707 &	SPB &	DSCUT &	GDOR &	0.008218706274 \\
7449844 &	SPB &	ACT &	ECL &	0.006529076219 \\
7448050 &	SPB &	GDOR &	MISC &	0.005403935735 \\
8113425 &	SPB &	MISC &	ROT &	0.008590883498 \\
5450166 &	SPB &	GDOR &	MISC &	0.008490744245 \\
3331147 &	SPB &	ECL &	RRD &	0.005492894475 \\
8355134 &	SPB &	GDOR &	MISC &	0.01831054767 \\
8264617 &	SPB &	GDOR &	MISC &	0.005602077117 \\
8180361 &	SPB &	GDOR &	ECL &	0.005657907911 \\
5021374 &	SPB &	GDOR &	MISC &	0.005309879497 \\
10669516 &	SPB &	GDOR &	MISC &	0.005520919223 \\
5000454 &	SPB &	GDOR &	MISC &	0.005089999299 \\
9267997 &	SPB &	MISC &	ROT &	0.008132484944 \\
6367159 &	SPB &	MISC &	ROT &	0.008734156537 \\
8712174 &	SPB &	GDOR &	ECL &	0.00718564492 \\
\hline
\end{tabular}
\begin{flushleft}
\emph{Note:} The KIC number corresponds to the Kepler ID. Class 1, Class 2 and Class3 correspond to the most probable classes according to \citet{debosscher07}. CLCEP: Classical Cepheid; ECL: Eclipsing Binary; ROT: Rotating variable; SR: Semi-regular star; ELL: Ellipsoidal binary; BCEP: $\beta$-Cephei star; RRAB: RR-Lyrae star RRab; RRC: RR-Lyrae star RRc; RRD: RR-Lyrae star RRd; RVTAU: RV-Tauri star; SPB: Slowly-pulsating B stars; GDOR: $\gamma$-Doradus type variables; DMCEP: Double-mode Cepheid; DSCUT: $\delta$-Scuti star; MISC: Miscellaneous. The last column is the anomaly score assigned in \citet{Giles19} .
\end{flushleft}
\end{table*}

\begin{table*}
\centering
\caption{Sample of unclassified anomalies similar to $\delta$-Scuti stars}
\label{tab:delScu}
\begin{tabular}{lccccccc}
\hline
  \multicolumn{1}{|c|}{KIC} &
  \multicolumn{1}{c|}{$G$ mag} &
  \multicolumn{1}{c|}{$G_{\rm{BP}}-G_{\rm{rp}}$} &
  \multicolumn{1}{c|}{t-SNE f1} &
  \multicolumn{1}{c|}{t-SNE f2} &
  \multicolumn{1}{c|}{UMAP f1} &
  \multicolumn{1}{c|}{UMAP f2} &
  \multicolumn{1}{c|}{URF score} \\
\hline
  100003119 &  &  & -6.2473454 & -18.123388 & -1.6541581 & 11.322543 & 0.9995\\
  100003307 &  &  & -6.74224 & -17.711239 & -1.4385592 & 11.609716 & 0.9984\\
  100003116 &  &  & -6.2032733 & -18.155602 & -1.7096267 & 11.327872 & 0.9977\\
  8093353 &  &  & -5.7698956 & -18.45632 & -1.8460397 & 11.076324 & 0.9976\\
  100003338 &  &  & -5.306527 & -18.780773 & -1.8841666 & 10.772886 & 0.9958\\
  10678547 & 0.776 & 0.341 & -3.2465124 & -19.316246 & -1.9911314 & 9.82429 & 0.9955\\
  10612592 &  &  & -6.1158776 & -18.221844 & -1.6524252 & 11.293662 & 0.9953\\
  2695999 & 4.145 & 0.994 & -0.5316093 & -18.305187 & -1.6424927 & 8.72731 & 0.9952\\
  5450881 & 0.860 & 0.046 & -0.66870624 & -18.537977 & -1.7074523 & 8.74083 & 0.9948\\
  10203328 &  &  & -6.723156 & -17.734406 & -1.4937031 & 11.62777 & 0.9946\\
  12603159 & 1.807 & 0.257 & -4.511584 & -19.094446 & -2.0708036 & 10.403458 & 0.9942\\
  100003285 &  &  & -5.8177295 & -18.417446 & -1.8020984 & 11.044083 & 0.9941\\
  8640132 & 2.492 & 0.548 & -5.997529 & -18.306383 & -1.8289222 & 11.231339 & 0.9936\\
  11657371 & 2.914 & 0.522 & -3.3878977 & -19.384365 & -2.045817 & 9.851587 & 0.9936\\
  7467547 & 3.773 & 0.827 & -2.9663498 & -19.302572 & -1.9664931 & 9.712989 & 0.9935\\
  9410674 & 6.803 & 1.383 & -6.272415 & -18.086332 & -1.5418396 & 11.323483 & 0.9933\\
  10975463 &  &  & -6.053544 & -18.263033 & -1.6969683 & 11.229567 & 0.9931\\
  100003367 &  &  & -3.3317416 & -19.339626 & -2.0612345 & 9.846666 & 0.9931\\
  9897683 &  &  & -5.782284 & -18.436483 & -1.7665132 & 10.997628 & 0.9930\\
  6311520 & 2.722 & 0.807 & -1.522426 & -19.018957 & -1.8995813 & 9.110288 & 0.9928\\
\hline\end{tabular}
\begin{flushleft}
\emph{Note:} The KIC number corresponds to the Kepler ID. The G magnitude and colors have been obtained from the \emph{Gaia} archive.
\end{flushleft}
\end{table*}

\begin{table*}
\centering
\caption{Sample of unclassified anomalies similar to oscillating binary stars.}
\label{tab:oscBin}
\begin{tabular}{lccccccc}
\hline
  \multicolumn{1}{|c|}{KIC} &
  \multicolumn{1}{c|}{$G$ mag} &
  \multicolumn{1}{c|}{$G_{\rm{BP}}-G_{\rm{rp}}$} &
  \multicolumn{1}{c|}{t-SNE f1} &
  \multicolumn{1}{c|}{t-SNE f2} &
  \multicolumn{1}{c|}{UMAP f1} &
  \multicolumn{1}{c|}{UMAP f2} &
  \multicolumn{1}{c|}{URF score} \\
\hline
 9594654 &  &  & -7.5821147 & 10.907919 & 5.2125616 & 17.08151 & 0.9994\\
 9776888 &  &  & -7.6291323 & 10.846095 & 5.149217 & 17.092026 & 0.9987\\
100003189 &  &  & -7.66666 & 10.798131 & 5.09078 & 17.053328 & 0.9971\\
9594468 &  &  & -7.6492314 & 10.820229 & 5.103816 & 17.09373 & 0.9968\\
9956596 &  &  & -7.705231 & 10.74775 & 4.9843044 & 16.999191 & 0.9925\\
100004178 &  &  & -7.581282 & 10.904561 & 5.238691 & 17.095936 & 0.9924\\
9471705 &  &  & -7.697963 & 10.7570915 & 5.02821 & 16.99908 & 0.9915\\
4049858 &  &  & -7.610809 & 10.868222 & 5.1962314 & 17.09528 & 0.9892\\
100004179 &  &  & -7.636067 & 10.837953 & 5.125358 & 17.076881 & 0.98509\\
9836795 & 11.232 & 2.449 & -7.718672 & 10.730212 & 5.061275 & 17.073288 & 0.9849\\
8719524 & 8.991 & 1.969 & -7.6057296 & 10.87642 & 5.2227488 & 17.093218 & 0.9846\\
7025613 & 13.028 & 3.764 & -7.5926914 & 10.893344 & 5.1981487 & 17.079847 & 0.9841\\
9716337 &  &  & -7.75843 & 10.673654 & 4.9066124 & 16.935802 & 0.9834\\
9541127 & 2.8402 & 0.594 & -7.64089 & 10.888142 & 5.0242763 & 17.040018 & 0.9797\\
9722737 & 2.187 & 0.707 & -7.716939 & 10.721025 & 4.959805 & 17.000145 & 0.9723\\
10284901 & 2.987 & 0.459 & -7.684295 & 10.778942 & 5.0390916 & 17.039068 & 0.9657\\
8087649 & 2.528 & 0.555 & -7.710859 & 10.722375 & 5.018523 & 16.984755 & 0.9613\\
10350225 & 8.473 &  & -7.694458 & 10.737052 & 4.968204 & 16.916359 & 0.9589\\
8963394 & 3.674 & 0.827 & -7.6731224 & 10.825514 & 4.927709 & 16.953476 & 0.9583\\
10454962 & 6.719 & 1.467 & -7.739664 & 10.872207 & 4.967913 & 17.029408 & 0.9570\\
\hline\end{tabular}
\begin{flushleft}
\emph{Note:} The KIC number corresponds to the Kepler ID. The G magnitude and colors have been obtained from the \emph{Gaia} archive.
\end{flushleft}
\end{table*}

\begin{table*}
\centering
\caption{Sample of unclassified anomalies similar to eruptive RGB stars}
\label{tab:erupRGB}
\begin{tabular}{lccccccc}
\hline
  \multicolumn{1}{|c|}{KIC} &
  \multicolumn{1}{c|}{$G$ mag} &
  \multicolumn{1}{c|}{$G_{\rm{BP}}-G_{\rm{rp}}$} &
  \multicolumn{1}{c|}{t-SNE f1} &
  \multicolumn{1}{c|}{t-SNE f2} &
  \multicolumn{1}{c|}{UMAP f1} &
  \multicolumn{1}{c|}{UMAP f2} &
  \multicolumn{1}{c|}{URF score} \\
\hline
5014753 & 5.441 &  & -4.153925 & -16.68397 & -0.90287846 & 10.464219 & 0.9862\\
8195444 & 3.473 & 1.687 & -4.506603 & -14.705019 & -0.102418125 & 11.566384 & 0.9802\\
8649496 & 0.358 & 2.009 & -1.8385679 & -17.647367 & -0.981706 & 9.115791 & 0.9754\\
8450468 & 3.916 & 0.783 & -2.5616813 & -17.238077 & -1.2301917 & 9.658302 & 0.9751\\
7200934 & -0.375 & 1.600 & -2.406651 & -17.494953 & -0.9304391 & 9.426681 & 0.9730\\
10416390 &  &  & -4.1185384 & -15.529406 & -0.19089723 & 10.818488 & 0.9717\\
8836489 & -0.861 & 1.801 & -3.1784277 & -17.400663 & -0.9736563 & 9.807226 & 0.9695\\
5219663 & -1.139 & 1.770 & -2.8299842 & -17.493786 & -1.0102772 & 9.604672 & 0.9679\\
1865744 & 6.551 &  & -3.6125305 & -17.09649 & -0.9249465 & 10.136411 & 0.9675\\
7984243 & 0.079 & 1.796 & -3.227355 & -17.062319 & -0.76908594 & 9.880368 & 0.9664\\
8145759 & -0.065 & 1.474 & -1.8224608 & -17.744188 & -1.053977 & 9.12446 & 0.9647\\
4587051 & -0.069 & 1.703 & -1.940113 & -17.528873 & -0.8508138 & 9.15528 & 0.9641\\
10991892 & -0.085 & 1.530 & -2.4526129 & -17.553347 & -0.965016 & 9.401397 & 0.9631\\
6849861 & -0.866 & 1.669 & -4.1245103 & -16.036352 & -0.50729203 & 10.685133 & 0.9626\\
9301126 & 0.484 & 1.965 & -3.4143038 & -17.166023 & -0.9480698 & 10.008888 & 0.9619\\
6025983 & -0.401 & 1.588 & -3.542263 & -16.890402 & -0.75655204 & 10.126622 & 0.9614\\
10447681 & 0.008 & 1.569 & -3.0860567 & -17.45611 & -1.0238457 & 9.715024 & 0.9614\\
7336419 & -0.140 & 1.504 & -3.1959655 & -16.872692 & -0.4300632 & 9.743351 & 0.9607\\
7779434 & 0.594 & 1.978 & -3.455784 & -17.093006 & -0.8364878 & 10.057132 & 0.9599\\
6020718 & 0.078 & 1.487 & -2.774681 & -17.514397 & -0.9522696 & 9.567208 & 0.9599\\
\hline\end{tabular}
\begin{flushleft}
\emph{Note:} The KIC number corresponds to the Kepler ID. The G magnitude and colors have been obtained from the \emph{Gaia} archive.
\end{flushleft}
\end{table*}

\begin{table*}
\centering
\caption{Sample of unclassified anomalies similar to Long Period Variable stars}
\label{tab:LPV}
\begin{tabular}{lccccccc}
\hline
  \multicolumn{1}{|c|}{KIC} &
  \multicolumn{1}{c|}{$G$ mag} &
  \multicolumn{1}{c|}{$G_{\rm{BP}}-G_{\rm{rp}}$} &
  \multicolumn{1}{c|}{t-SNE f1} &
  \multicolumn{1}{c|}{t-SNE f2} &
  \multicolumn{1}{c|}{UMAP f1} &
  \multicolumn{1}{c|}{UMAP f2} &
  \multicolumn{1}{c|}{URF score} \\
\hline
11082175 & 4.482 &  & -9.065751 & 0.0927 & 4.8437257 & 14.526902 & 0.9116\\
11554998 & 1.394 & 1.415 & -9.076439 & 0.16205257 & 4.858607 & 14.579996 & 0.9077\\
10483262 & -0.907 & 3.141 & -8.948204 & -0.19253835 & 4.7898993 & 14.154463 & 0.9052\\
4677837 & 0.025 & 1.568 & -9.091283 & 0.15311472 & 4.678336 & 14.575614 & 0.8992\\
10157826 & -1.709 & 2.390 & -9.011099 & -0.030076377 & 4.8064322 & 14.33454 & 0.8976\\
7909956 & 1.326 & 1.520 & -9.025238 & 0.0027910993 & 4.6902266 & 14.395064 & 0.8963\\
8376357 & -1.945 & 3.278 & -9.051126 & 0.22992167 & 4.945882 & 14.60285 & 0.8929\\
11122913 & -5.435 & 1.729 & -9.005106 & -0.0457747 & 4.8686867 & 14.345256 & 0.8919\\
12072767 & -1.326 & 3.220 & -9.067909 & 0.25513262 & 4.7925344 & 14.666139 & 0.8917\\
5733729 & -2.055 & 2.574 & -9.075015 & 0.3217304 & 5.001181 & 14.735213 & 0.8916\\
11033884 & -2.470 & 2.379 & -8.947156 & -0.16587107 & 4.861857 & 14.168224 & 0.8912\\
6020264 & -1.178 & 4.010 & -9.033311 & 0.003955113 & 4.690762 & 14.4270735 & 0.8909\\
7739645 & -2.122 & 2.060 & -9.041659 & 0.07632352 & 4.8761516 & 14.502619 & 0.8892\\
100004284 &  &  & -9.133263 & 0.33165124 & 4.618789 & 14.779659 & 0.8880\\
5640488 & -1.360 & 3.255 & -9.111134 & 0.32123733 & 4.7750044 & 14.762475 & 0.8879\\
9820825 & -2.267 & 2.667 & -9.112118 & 0.29049492 & 4.6843486 & 14.686565 & 0.8874\\
5219922 & -2.209 & 2.738 & -9.084163 & 0.15421346 & 4.6843576 & 14.578069 & 0.8873\\
4551712 & -1.890 & 2.276 & -8.941083 & -0.19517083 & 4.9198 & 14.1704035 & 0.8835\\
4919121 & -1.423 & 3.215 & -9.044839 & 0.026868027 & 4.773669 & 14.416367 & 0.8802\\
6605787 & -1.644 & 3.006 & -9.054727 & 0.14816612 & 4.9234815 & 14.552588 & 0.8799\\
\hline\end{tabular}
\begin{flushleft}
\emph{Note:} The KIC number corresponds to the Kepler ID. The G magnitude and colors have been obtained from the \emph{Gaia} archive.
\end{flushleft}
\end{table*}

\begin{table*}
\centering
\caption{Sample of manifold anomalies selected with the Euclidean score.}
\label{tab:manifold_anomalies}
\begin{tabular}{lccccccc}
\hline
  \multicolumn{1}{|c|}{KIC} &
  \multicolumn{1}{c|}{$G$ mag} &
  \multicolumn{1}{c|}{$G_{\rm{BP}}-G_{\rm{rp}}$} &
  \multicolumn{1}{c|}{t-SNE f1} &
  \multicolumn{1}{c|}{t-SNE f2} &
  \multicolumn{1}{c|}{UMAP f1} &
  \multicolumn{1}{c|}{UMAP f2} &
  \multicolumn{1}{c|}{Euclidean score} \\
  \hline
5168334 & 5.0244 & 1.080 & 3.2576716 & 4.429501 & 9.336882 & -2.5593443 & 38.828\\
5774442 & 2.880 & 0.632 & 4.844674 & -2.3890922 & 12.410824 & -1.5254657 & 34.779\\
9395222 & 2.861 & 0.498 & 4.8298926 & -2.3726006 & 2.0440662 & -4.773456 & 33.520\\
6129761 & 3.420 & 0.937 & -10.421026 & -3.44677 & 13.182329 & 3.4518988 & 28.460\\
6205481 & 6.054 & 1.095 & 0.15788801 & -13.37177 & 0.56017363 & 10.386182 & 27.023\\
9896606 & 2.712 & 0.611 & 8.722893 & -14.658413 & -3.344982 & 4.769703 & 24.044\\
8984949 & 3.826 & 0.816 & 1.0454575 & 14.84347 & 2.3673918 & -4.06511 & 20.270\\
5943622 & 4.736 & 0.827 & -14.95123 & 2.6781847 & 9.335588 & 6.3301888 & 19.137\\
9900027 & 5.091 & 0.901 & 1.7817745 & 2.506942 & 1.3364209 & -4.290253 & 18.315\\
4755287 & 3.405 & 0.636 & 0.06469339 & 15.6223955 & 5.067639 & -5.823458 & 17.808\\
5385792 & 3.829 & 0.746 & 3.401131 & -10.21556 & 0.7265468 & 7.2443156 & 17.578\\
7886329 & 4.113 & 0.775 & 10.268521 & 9.163061 & 9.72615 & -3.3141055 & 17.494\\
5192090 & 2.854 & 0.613 & -2.9854202 & -5.561847 & 14.04739 & -0.5290367 & 17.279\\
8038609 & 1.611 & 0.444 & -8.257905 & -8.0289545 & 2.2189136 & 14.65101 & 17.009\\
6706543 & 3.533 & 0.746 & 4.219155 & 8.535086 & 3.3016827 & -3.20653 & 16.656\\
9162503 & 2.991 & 0.612 & 0.94035065 & 15.112005 & 7.6065702 & -4.6533947 & 16.609\\
8191672 & 3.449 & 0.872 & -17.43241 & -2.7113748 & 10.702633 & 8.467638 & 16.571\\
8313059 & 3.624 & 0.736 & 1.5142245 & 14.066471 & 4.740681 & -5.021357 & 15.538\\
8197220 & 2.595 & 0.541 & 16.444843 & 3.6578653 & 10.613494 & -6.0641556 & 15.445\\
6707691 & 0.529 & 1.043 & 16.341946 & 9.632014 & 8.303985 & -6.536313 & 15.170\\
  \hline\end{tabular}
\begin{flushleft}
\emph{Note:} The KIC number corresponds to the Kepler ID. The G magnitude and colors have been obtained from the \emph{Gaia} archive.
\end{flushleft}
\end{table*}

\bsp	
\label{lastpage}
\end{document}